\documentclass{jfm}

\usepackage{graphicx}
\usepackage{newtxtext}
\usepackage{newtxmath}
\usepackage{amsmath}
\usepackage{natbib}
\usepackage{hyperref}
\usepackage{physics}
\usepackage{subcaption}
\usepackage{float}
\usepackage{graphicx}
\usepackage[absolute,overlay]{textpos}
\usepackage{overpic}
\usepackage{svg}
\newtheorem{theorem}{Theorem}

\usepackage{dirtytalk}
\usepackage{soul}

\hypersetup{
 colorlinks = true,
 urlcolor = blue,
 citecolor = black,
}

\newcommand{\RomanNumeralCaps}[1]
\linenumbers

\newcommand{\vc}[1]{\ensuremath{\mathbf{#1}}}

\title{Stability analysis of transitional flows based on disturbance magnitude}

\author{Ofek Frank-Shapir\aff{1}\corresp{\email{ofekfr@campus.technion.ac.il}},
 Igal Gluzman\aff{1}}

\affiliation{\aff{1}The Stephen B. Klein Faculty of Aerospace Engineering, Technion – Israel Institute of Technology,
Haifa 32000, Israel}

\begin{document}
\maketitle

\begin{abstract}
We propose a novel stability criterion for incompressible shear flows by combining input–output analysis and the small-gain theorem. The criterion yields an explicit threshold on the magnitude of velocity perturbations about a given base flow that guarantees stability. If this threshold is crossed---either due to nonmodal growth, exponential growth, or a bypass transition scenario---our analysis predicts a loss of stability that may lead to transition to turbulence. We consider three approximated models for nonlinearity: unstructured, structured with non-repeated blocks, and structured with repeated blocks. 
We show that the imposed threshold obtained by these three methods complies with a hierarchical relationship, where the unstructured case is the most conservative, imposing the lowest bound on disturbance magnitude. We apply this approach to three canonical and well-studied base flows: Couette, plane Poiseuille, and Blasius. For these three base flows, we compare our results with experiments, direct numerical simulation results, nonmodal nonlinear stability results, and linear stability theory (LST). In the limit of infinitesimally small perturbation magnitude, our stability criterion for the unstructured case recovers the results of LST. For finite perturbations, the structured cases that account for nonlinear interactions provided stability thresholds that are consistent with experimental observations and simulation results of transition at both subcritical and post-critical Reynolds numbers for the considered base flows in our study. In particular, we utilize our stability criterion to demonstrate that Couette flow can become unstable and transition can be triggered at different Reynolds numbers, which is consistent with past experimental observations.

\end{abstract}


\section{Introduction}
\label{sec:1}

In transitional flow stability studies, two main approaches are typically considered. (i) Modal analysis \citep{taira2020modal}---has been a leading approach in formulating linear stability theory (LST). This approach focuses on eigenvalue analysis of the flow response to initial conditions and involves infinitesimally small disturbances superimposed on a base flow. In LST, stability is defined in the infinite horizon sense, while disregarding short-time perturbation dynamics \citep{schmid2007nonmodal} and the effect of perturbation magnitude on flow stability. (ii) Nonmodal analysis---which allows focusing on the response to external forcing, and the study of flow behavior in the finite horizon sense, such as the study of transient growth due to the non-normality of the linearized Navier-Stokes (LNS) operator \citep{schmid2007nonmodal,karp2017secondary,lozano2021cause}. 
In particular, in a nonmodal approach using an input-output formulation, the governing equations are formulated as state equations. This allows for the computation of the system's frequency response, which contains a great deal of information and enables a detailed analysis of the response to external forcing. One field that employs such methodology is resolvent analysis \citep{schmid2007nonmodal,mckeon2010critical,mckeon2017engine,jovanovic2021bypass}. In these studies, the resolvent operator (which is analogous to the frequency response) is formulated to examine the input–output dynamical response of the linearized system to harmonic forcing about a certain base flow via associated resolvent modes. The main difference between input-output works and resolvent analysis is that in resolvent analysis, the input-output dynamics is transformed into analyses of the resolvent modes, via a Schmidt decomposition of the transfer operator, taking a singular value associated with the particular input–output mode pair. In this way, the analyses of the input–output dynamics is transformed into analyses of the resolvent modes and the associated energy amplification. 
In the current study, we place less emphasis on specific mode shapes and focus on amplification characteristics over a range of wave numbers. This type of analysis aligns more with previous input-output works \citep{jovanovic2001spatio,jovanovic2005componentwise,liu2021}.
Such tools have been shown to be both computationally efficient and to provide insight into the response of different base flows to actuation. Examples include studies that focused on the externally forced LNS equations to examine dynamic processes, structural features, and energy pathways for a variety of base flows \citep{farrell1993stochastic,bamieh2001energy,jovanovic2004unstable,jovanovic2005componentwise,HC10,mckeon2010critical,mckeon2013experimental,mckeon2017engine,madhusudanan2019coherent,liu2020input,symon2021energy}. Additional examples include our recent studies, where {the} dynamics of large-scale structures in a turbulent boundary layer are studied by analyzing the flow response to external periodic perturbations \citep{liu2022spatial}, and utilization of the input-output approach for modeling different actuation geometry patterns \citep{GG21,frank2025choosing}.

In \cite{schmid2007nonmodal}, it was stated: \say{despite these remarkable accomplishments, many questions were left unanswered, including the discrepancy between the computed critical Reynolds number and the observed transitional Reynolds number in many wall-bounded shear flows, the interplay between linear energy amplification and nonlinear energy conservation in the subcritical regime, the nonexistence of finite critical Reynolds numbers for pipe or plane Couette flow, and a frequent failure to observe theoretically predicted structures in unforced experiments}. Many of these issues highlighted by \cite{schmid2007nonmodal} continue to be valid nowadays. In particular, this study focuses on addressing the discrepancies between modal and nonmodal approaches for identifying the governing flow structures responsible for the onset of instability and predicting the critical Reynolds number, which we review in detail next.

Linear nonmodal analysis methods, such as linear input-output analysis \citep{jovanovic2005componentwise}, emphasize different predictions about the dominant flow structures. Classical results from linear modal stability analysis predict the prevalence of {Tollmien–Schlichting} ({TS}) waves as the modes that cause the loss of stability in shear flows. 
On the other hand, linear nonmodal analysis emphasizes streaky structures that are linked to the transient growth process \citep{reddy1993energy,schmid1994optimal,schmid2000linear,bamieh2001energy,jovanovic2004modeling,jovanovic2004unstable,jovanovic2005componentwise,schmid2007nonmodal}. 
Both approaches are validated via experimental observations and numerical simulations, where TS-waves are observed to lead to transition when the inflow is hardly noisy \citep{klebanoff1962three,laurien1989numerical,rist1995direct}, and streaky structures lead to transition when the input boundary layer is noisy \citep{roach1990influence,jacobs2001simulations,hernon2007experimental}.

When considering the critical Reynolds number, one key result of LST is that Couette flow is stable for all Reynolds numbers \citep{romanov1973stability}. However, experimental evidence \citep{tillmark1992experiments,dauchot1995finite}, stochastic viewpoint at stability theory \citep{ko2011effects}, and simulation results \citep{barkley2005computational,dou2012direct}, have observed transition in Couette flow at different Reynolds numbers, typically occurring in the range $\Rey=320-370$. The same holds for other flow geometries (e.g., plane Couette-Poiseuille, plane Poiseuille, and Blasius flows), for which there is plentiful literature, where instability is observed at sub-critical Reynolds numbers \citep{nishioka1985some,asai1995boundary,burin2012subcritical,sano2016universal,chefranov2016linear,klotz2017couette,Huang_Gao_Gao_Xi_2024} or remained stable even at post-critical Reynolds numbers  \citep[e.g,][]{pfenninger1961transition,nishioka1975experimental,avila2023transition}.   
This disagreement of linear analysis methods with experimental and simulation results can be attributed to the omission of finite-size disturbances and nonlinear interactions in the stability analysis methods
\citep{MULLIN_2010} that are evident from experiments \citep[e.g,][]{cohen2009aspects,sano2016universal} and direct numerical simulations (DNS) \citep[e.g.,][]{he2015transition}. 

Different approaches can be considered to represent the nonlinear advection term in the Navier-Stokes equations (NSE) and account for finite-size perturbations. Examples include using quadratic constraints and employing linear matrix inequalities (LMIs) to compute bounds on permissible perturbation magnitudes, ensuring the flow field remains laminar \citep{liu2020input_LMI,kalur2020stability,kalur2021estimating,kalur2021nonlinear,heide2025optimization}.
In addition, optimal perturbation analysis using nonmodal nonlinear methods \citep[e.g.,][]{duguet2013minimal,kerswell2018nonlinear,parente2022minimal} successfully addresses many shortcomings of linear analysis methods. These nonmodal nonlinear approaches are based on computing the evolution of the entire flow field and optimizing over the initial conditions. This process is very numerically demanding, limiting the ability of such works to study the effect of many perturbation magnitudes for a wide range of Reynolds numbers and wave-numbers. The framework proposed in this work for flow stability analysis is based on computing frequency-response operators within an input-output paradigm rather than simulating the entire flow-field evolution. This makes our approach more computationally efficient than adjoint-based methods, allowing us to examine large ranges of Reynolds numbers, revealing the complex interplay between dominant flow structures that determine the stability threshold.
In particular, we want to emphasize the study of \cite{liu2021}, where the authors proposed to recover the key flow features that are associated with nonlinear effects in input-output formulation by using the concept of {\emph{structured uncertainty}} \citep{packard1993,Zhou1995-dl} from the field of robust control theory. This approach is denoted as structured input-output analysis, in which the nonlinear term in the NSE is modeled as a fixed structure uncertainty that is interconnected to the linear frequency response operator obtained from the LNS equations.
Different structures for the uncertainty are proposed in this and subsequent studies: repeated block uncertainty \citep{mushtaq2023,bhattacharjee2023structured,mushtaq2024structured} and non-repeated block uncertainty \citep{liu2021,shuai2023structured}  to incorporate the nonlinearity into the system. Within the proposed structured input-output formulation, the small gain theorem is used to define the stability of a structured feedback interconnection. 
This theorem is also used in  \citet{jovanovic2004modeling} to show that the robust stability analysis is equivalent to the input-output analysis for a certain part of the system.

This study builds upon the mentioned studies above, where we utilize input-output analysis and the small gain theorem to derive disturbance-based criteria for analyzing the stability of flows. In our framework, we rewrite the LNS equations as a state-space based on the work of \citet{jovanovic2005componentwise}. Next, the framewrok is expanded to incorporate nonlinear interactions by introducing an uncertainty block,  using structured input-output analysis \citep{liu2021,mushtaq2023}. This formulation is used to derive disturbance-based criteria for analyzing the stability of the flows.
This is achieved by relating the inverse of the largest structured singular value (SSV) to the magnitude of velocity perturbations in the flow, and utilizing the small-gain theorem to establish a threshold on velocity disturbance magnitude that triggers flow instability. We conduct a detailed study of applying this criterion to three canonical flows that are fundamental in stability studies, over a wide range of Reynolds numbers.
In detail, we propose a novel stability analysis approach, based on determining a threshold on the disturbance magnitude present in the flow that triggers instability rather than examining the growth rate of modes in the flow through linear stability analysis. The threshold criterion provides an upper bound on the magnitude of velocity perturbations that ensures stability. This threshold can be reached via different bypass transition routes, such as through the existence of a finite-size perturbation in the
flow (like in noisy environments and realistic experimental conditions) or via a transient
growth scenario where a set of infinitesimal perturbations leads to a perturbation amplitude
amplification that can surpass the estimated threshold by our approach. Therefore, our criterion can be applied to study a wide range of physical routes to transition.

In~{\S}\ref{sec:2} we provide a mathematical framework to formulate the stability criterion with details on numerical implementation, {suitable for high-order flow models}, provided in~{\S}\ref{sec:3}. We use this criterion to analyze the transition of Couette, plane Poiseuille, and Blasius flows for a wide range of Reynolds numbers. {For each base flow,  we consider three input-output formulations: unstructured input-output, structured input-output with non-repeating uncertainty, and structured input-output with repeating uncertainty. We demonstrate that our threshold would comply with a hierarchical relationship, where the unstructured case is most conservative, imposing the lowest bound for transition, whereas structured cases are closer to the true value, by accounting for nonlinearity. In~{\S}\ref{sec:4}, we also identify the dominant structures in transitional flows that determine the stability threshold bound for a wide range of Reynolds numbers and show a complex interplay between the structures that set the stability threshold in sub-critical, near-critical, and post-critical Reynolds number ranges.

We demonstrate that our criterion allows us to address the issues mentioned in \cite{schmid2007nonmodal}, regarding the discrepancy in transitional Reynolds number in many wall-bounded shear flows, and the dominant flow structures that govern different stages of transition. As would be detailed in~{\S}\ref{sec:5}, our stability approach is in agreement with previous experimental work, which found that finite-amplitude perturbations can cause transition, and the critical amplitude depends on the Reynolds number \citep{dauchot1995finite,philip2007scaling,peixinho2007finite}. We also use the analysis presented in this paper to obtain stability diagrams, compare them to stability diagrams from LST, and compute the critical perturbation energy required to cause flow instability, which shows a good agreement with DNS and nonmodal nonlinear approaches \citep{reddy1998stability,duguet2010towards,duguet2013minimal,parente2022minimal}. 
Similar to other works using nonmodal approaches, our novel stability criterion overcomes a shortcoming of LST, which shows the nonexistence of a finite critical Reynolds number for plane Couette flow. We use our novel stability criterion to show that Couette flow is indeed unstable and can transition to turbulence at various Reynolds numbers, as supported by experimental results \citep{nishioka1985some,klotz2017couette,burin2012subcritical}. 
Lastly, conclusions are discussed in~{\S}\ref{sec:6}.

\section{Theoretical Background}
\label{sec:2}

We consider incompressible shear flow, for which the LNS equations for perturbations in fluid velocity and pressure $(\mathbf{u}, p)$ about the base flow $(\boldsymbol{\overline{u}}, \overline{p})$ are as follows: 
\begin{equation}
\label{eq:2.1}
\begin{matrix}
 \frac{\partial \mathbf{u}}{\partial t} = -\mathbf{\bar{u}} \cdot \nabla \mathbf{u} - \mathbf{u} \cdot \nabla \mathbf{\bar{u}} - \nabla p + \frac{1}{\Rey} \Delta \mathbf{u} + \boldsymbol{d}, \\
\nabla \cdot \mathbf{u} = 0.
\end{matrix}
\end{equation}
Here, $\mathbf{u}=\begin{bmatrix} u & v & w\end{bmatrix}^T$ is the perturbation velocity vector, corresponding to the streamwise $x$, wall normal $y$, and spanwise $z$ directions, respectively. Throughout this work, we consider base velocity profiles of the form $\mathbf{\bar{u}} = \begin{bmatrix} U(y) & 0 & 0 \end{bmatrix}^T$. Additionally, $\boldsymbol{d}=\begin{bmatrix} d_x & d_y & d_z\end{bmatrix}^T$ is the term representing body forcing. $\nabla$ is the del operator,  and $\Delta\equiv\nabla^2$ is the Laplacian operator.
 
In addition, we assume spatial invariance of the parallel flow field in the horizontal directions to allow Fourier transforms in the $x$ and $z$ directions, providing a system of two partial differential equations depending on spatial wave numbers $k_x$, $k_z$, and $y$, the position in the wall-normal direction \citep{kim1987turbulence}. Such an assumption is valid for Couette, {plane Poiseuille}, and Blasius flow under a quasi-parallel assumption.
This process allows us to write the LNS equations in a state-space form, which simplifies performing input-output analysis, via a formulation we adopt from \citet{jovanovic2005componentwise}:
 
\begin{equation}
\label{eq:2.2}
\begin{matrix}
\frac{\partial \boldsymbol{\psi}}{\partial t}(k_x,y,k_z,t) = [\mathscr{A}(k_x,k_z)\boldsymbol{\psi}(k_x,k_z,t)](y) + [\mathscr{B}(k_x,k_z)\boldsymbol{d}(k_x,k_z,t)](y), \\
\boldsymbol{\phi}(k_x,y,k_z,t) = [\mathscr{C}(k_x,k_z)\boldsymbol{\psi}(k_x,k_z,t)](y).
\end{matrix}
\end{equation}
Here $\boldsymbol{\psi} \equiv \begin{bmatrix} v & \omega_y \end{bmatrix} ^T$ is the state vector, comprised of the wall-normal velocity perturbation $v$, and the wall-normal vorticity perturbation $\omega_y = {\partial u}/{\partial z} - {\partial w}/{\partial x}$.  The output vector $\boldsymbol{\phi}$ in our analysis is set to be $\boldsymbol{\phi} \equiv \mathbf{u} = \begin{bmatrix} u & v & w\end{bmatrix}^T$.
Lastly, $\boldsymbol{d}$ is the body forcing term applied to the flow system. 
The operators denoted as $\mathscr{A}$, $\mathscr{B}$ and $\mathscr{C}$ are defined as \citep{jovanovic2005componentwise}

\begin{equation}
\label{eq:2.3}
\begin{aligned}
\mathscr{A} \equiv 
\begin{bmatrix}
 \mathscr{A}_{11} & 0 \\
 \mathscr{A}_{21} & \mathscr{A}_{22}
\end{bmatrix}
\equiv
\begin{bmatrix}
 -ik_x\Delta^{-1}U\Delta+ik_x\Delta^{-1}U''+\frac{1}{Re}\Delta^{-1}\Delta^2 & 0 \\
 -ik_zU' & -ik_xU+\frac{1}{Re}\Delta
\end{bmatrix},
\end{aligned}
\end{equation}

\begin{equation}
\label{eq:2.4}
\begin{aligned}
\mathscr{B} \equiv 
\begin{bmatrix}
 \mathscr{B}_{x} & \mathscr{B}_{y} & \mathscr{B}_{z}
\end{bmatrix}
\equiv
\begin{bmatrix}
 \Delta^{-1} & 0 \\
 0 & I
\end{bmatrix}
\begin{bmatrix}
 -ik_x\frac{\partial}{\partial y} & -(k_x^2 + k_z^2) & -ik_z\frac{\partial}{\partial y} \\
 ik_z & 0 & -ik_x
\end{bmatrix},
\end{aligned}
\end{equation}

\begin{equation}
\label{eq:2.5}
\begin{aligned}
\mathscr{C} \equiv 
\begin{bmatrix}
 \mathscr{C}_{u} \\ 
 \mathscr{C}_{v} \\ 
 \mathscr{C}_{w}
\end{bmatrix}
\equiv
\frac{1}{k_x^2+k_z^2}
\begin{bmatrix}
 ik_x\frac{\partial}{\partial y} & -ik_z \\
 k_x^2+k_z^2 & 0 \\
 ik_z\frac{\partial}{\partial y} & ik_x
\end{bmatrix},
\end{aligned}
\end{equation}
where $U'=\frac{dU(y)}{dy}$ and $U''=\frac{d^{2}U(y)}{dy^{2}}$. The Laplacian operator is defined as $\Delta=\frac{\partial^2}{\partial y^2}-(k_x^{2}+k_z^{2})$ because of the Fourier transforms done in directions $x$ and $z$. Here $\mathscr{A}$ is the operator representing the dynamics of the system and is comprised of the Orr-Sommerfeld operator ($\mathscr{A}_{11}$), the coupling operator  ($\mathscr{A}_{21}$), and the Squire operator ($\mathscr{A}_{22}$); $\mathscr{B}$ is the forcing operator, representing the effect of forcing input on the flow; $\mathscr{C}$ is the operator relating the state variables---wall-normal velocity and vorticity perturbations---to the three velocity perturbations ($u,v,w$). We note that the operator $\mathscr{C}$ can be modified to acquire the desired flow variables of interest as output of the system. Each of the operators $\mathscr{A}$, $\mathscr{B}$ and $\mathscr{C}$ implicitly depends on the wall-normal coordinate $y$.

Herein, the boundary conditions on $v$ and $\omega_y$ depend on the specific base flow. For Couette and plane Poiseuille flows, the boundary conditions are:

\begin{equation}
\label{eq:2.6}
 v(y= \pm 1) = \frac{\partial v}{\partial y}(y= \pm 1) = \omega_{y}(y= \pm 1) =0, \forall k_x,k_z \in \mathbb{R}, t \geq 0.
\end{equation}
For Blasius base flow, the boundary conditions are:

\begin{equation}
\label{eq:2.7}
 v(y= 0,\infty) = \frac{\partial v}{\partial y}(y= 0,\infty) = \omega_{y}(y= 0,\infty) =0, \forall k_x,k_z \in \mathbb{R}, t \geq 0.
\end{equation}
Lastly, the Reynolds number is defined as $\Rey = \mathcal{U}L / \nu$, where $\mathcal{U}$ is the reference velocity, $L$ is the reference height, and $\nu$ is the kinematic viscosity of the fluid. These parameters are also used to render the variables in this work dimensionless. In particular, for Couette flow, $\mathcal{U}$ is half the difference between base flow velocity values at the lower and upper walls; for plane Poiseuille flow, it is the centerline velocity; for Blasius, it is the freestream velocity. Here $L$ is the channel half-height for both Couette and plane Poiseuille flows, and for Blasius flow, it is the displacement thickness.

\subsection{Linear frequency Response}
In this section, we consider the frequency response associated with the LNS system. The derivations in this section will be used as the foundation for the more complex nonlinear analysis, which we employ to obtain a stability criterion that is based on perturbation magnitude. 
Here, we consider a forcing term $\boldsymbol{d}$, which is harmonic in time and in the $x$ and $z$ directions, i.e., 
$\boldsymbol{d}(t,x,y,z) = \overline{\boldsymbol{d}}(y)\exp[i(k_x x + k_z  z + \omega t)]$,
where $\omega \in \mathbb{R}$ is the temporal frequency and $k_x,k_z\in\mathbb{R}$ are wavenumbers of the streamwise and spanwise flow directions, respectively, and $\overline{\boldsymbol{d}}(y)$ is some function of $y$, following the formulation in \citet{jovanovic2005componentwise}. In our study, we do not consider any specific form of  $\overline{\boldsymbol{d}}(y)$, instead the forcing component is viewed as the feedback in the interconnected system in Fig.~\ref{fig:1} to account for the effect of the nonlinear advection term of the full NS system as will be detailed in \S~\ref{sec:2.3}.}

The frequency response operator that maps the harmonic forcing term applied to  the above state-space system in Eq.~\eqref{eq:2.2} to the resulting perturbation velocity field is given by \citep{Zhou1995-dl}:
\begin{equation}
\label{eq:2.8}
\begin{aligned}
\mathscr{H}(y;k_x,k_z,\omega)=[\mathscr{C}(k_x,k_z)[i\omega \mathbf{I} - \mathscr{A}(k_x,k_z)]^{-1}\mathscr{B}(k_x,k_z)](y).
\end{aligned} 
\end{equation}
Here, $\mathbf{I}$  is the  identity matrix. As can be seen in equations \eqref{eq:2.4} and \eqref{eq:2.5}, the operators $\mathscr{B}$ and $\mathscr{C}$ have three components each corresponding to different inputs and outputs, respectively.

The frequency response operator is used extensively in the input-output analysis of \citet{jovanovic2005componentwise}. Herein, we define an alternative operator, as proposed by \cite{liu2021}, and used in several structured input-output works \citep[such as][]{mushtaq2023,shuai2023structured}:
\begin{equation}
\label{eq:2.9}
\mathscr{H}_{\nabla}(y;k_x,k_z,\omega) \equiv
\text{diag} (\nabla,\nabla,\nabla)  \mathscr{H}(y;k_x,k_z,\omega).
\end{equation}
This frequency response operator of the system $\mathscr{H}_{\nabla}$ relates the forcing term to the resulting gradients of velocity perturbations. 
This operator will be useful for the derivation of a perturbation-based stability criterion that can incorporate nonlinear interactions via a structured input-output formulation, as will be shown in \S~\ref{sec:2.3}.
 
\subsection{Choice of norms}

The selection of the norm plays a critical role in the analysis of input-output behaviors in flow systems. Norms serve as the mathematical infrastructure for quantifying the magnitude of the response to a specific input and provide a basis for defining and computing the amplification between an input and its corresponding output. It is important to note that different choices of norms are associated with the computation of different quantities. For example, the $\mathscr{H}_2$ norm is specifically used to represent perturbation kinetic energy, as referenced in \citep{jovanovic2005componentwise}. In our research, we utilize the $\mathscr{H}_{\infty}$ norm instead of the $\mathscr{H}_2$ norm, which is related to the amplification of one specific mode (which is the most amplified) in the flow, as shown next.  
 
Given the operator $\mathscr{H}(y;k_x,k_z,\omega)$ after discretisation in $y$, we define the $\mathscr{H}_\infty$ system norm as follows \citep{Zhou1995-dl}:
\begin{equation}
 \label{eq:2.10}
\norm{\mathscr{H}}_{\infty}(k_x,k_z) \equiv \sup_{\omega \in \mathbb{R}}{\overline{\sigma}[ \mathscr{H}(k_x,k_z,\omega)]},
\end{equation}
and similarly,
\begin{equation}
 \label{eq:2.11}
\norm{\mathscr{H}_\nabla}_{\infty}(k_x,k_z) \equiv \sup_{\omega \in \mathbb{R}}{\overline{\sigma}[\mathscr{H}_\nabla(k_x,k_z,\omega)]}.
\end{equation}
Here, $\overline{\sigma}[ \cdot]$ represents the largest singular value under the standard Euclidean inner product, and $\sup_{\omega \in \mathbb{R}}{}$ represents the supremum operation over all temporal frequencies. 
Hence, the $\mathscr{H}_\infty$ norm does not quantify the energy amplification of the whole field, but rather the \ most amplified mode of the system due to harmonic temporal forcing.

\subsection{Non-linear frequency response}
\label{sec:2.3}
\begin{figure}
 \centering
 \includegraphics[width = \textwidth]{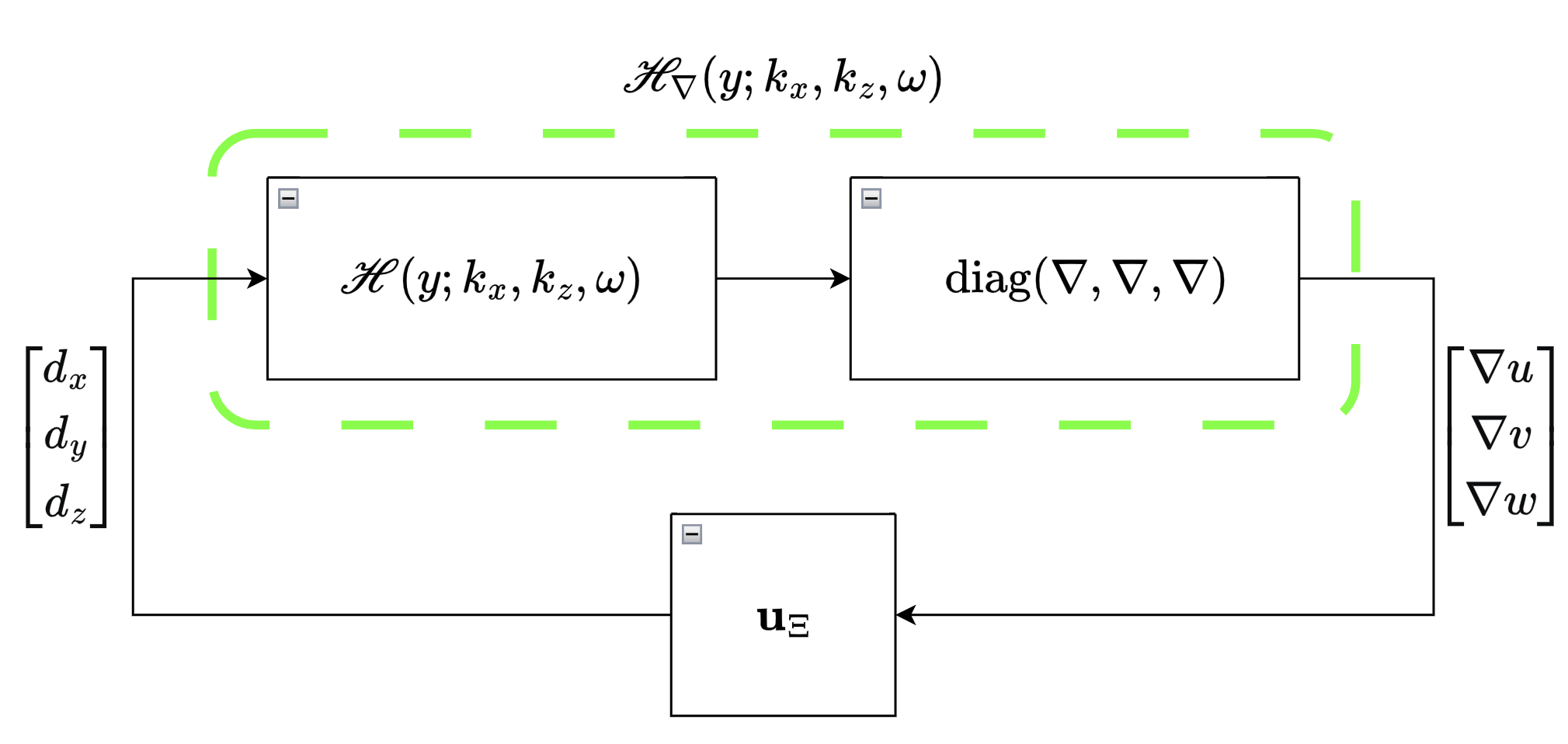}
 \caption{Interconnection loop block diagram for the nonlinear interactions. Adapted from \citet[][Fig. 3]{liu2021}.}
 \label{fig:1}
\end{figure}

In this section, we use structured input-output analysis to model the nonlinear response associated with the full NS system. This modeling of nonlinear interactions will be used in our stability analysis. 
We consider the full NS equations for velocity and pressure perturbations as we seek to incorporate nonlinear behaviors in our analysis. The NSE without a forcing term are as follows:

\begin{equation}
\label{eq:2.12}
\begin{matrix}
 \frac{\partial \mathbf{u}}{\partial t} = -\mathbf{\bar{u}} \cdot \nabla \mathbf{u} - \mathbf{u} \cdot \nabla \mathbf{\bar{u}} - \nabla p + \frac{1}{\Rey} \Delta \mathbf{u} - \mathbf{u} \cdot \nabla \mathbf{u}, \\
\nabla \cdot \mathbf{u} = 0.
\end{matrix}
\end{equation}
Eq. \eqref{eq:2.12} is very similar to Eq. \eqref{eq:2.1}, with the only difference being that the forcing term is swapped by the nonlinear advection term {$\mathbf{u}\cdot \nabla \mathbf{u}$}. Thus, we can model nonlinear interactions as a forcing input to the LNS system. We model the nonlinear advection term in the NSE using the block diagram representation in Fig.~\ref{fig:1}.
Here, $\mathbf{u}_{\Xi}$ is the gain applied to the vectorized velocity gradient $\begin{bmatrix} \nabla^T u & \nabla^T v & \nabla^T w \end{bmatrix}^T$ that retains the component-wise structure of the advection term $\mathbf{u} \cdot \nabla \mathbf{u}$ \citep{liu2021}. To interpret the feedback interconnection in Fig.~\ref{fig:1} physically, we view the nonlinear advection term as an internal forcing mechanism acting on the LNS dynamics. In a standard 'unstructured' analysis, the feedback block ${\textbf{u}}_\Xi$ is treated as a generic operator bounded only by its magnitude. Thus, unstructured analysis can be thought of as a worst-case model of the nonlinearity, as there are no limitations on the individual terms in ${\textbf{u}}_\Xi$. By contrast, the 'structured' feedback connection explicitly imposes constraints on ${\textbf{u}}_\Xi$, such that the value of some of the terms in it is fixed to be zero. Structure constraints are imposed to limit feedback pathways for the goal of replicating the structure of the nonlinearity in the NSE \citep{liu2021}. As will be shown in \S~\ref{sec:2.4}, the unstructured analysis results in more conservative, lower bounds on velocity perturbations.

The exact expression for ${\textbf{u}}_\Xi$ that captures the nature of the nonlinear term in the NSE is (see \citet{liu2021}, equation 2.9):
\begin{equation}
\label{eq:2.13}
 {{\textbf{u}}}_\Xi \equiv \begin{bmatrix}
    -{\textbf{u}}^T & \mathbf{0} & \mathbf{0} \\
    \mathbf{0} & -{\textbf{u}}^T & \mathbf{0} \\
    \mathbf{0} & \mathbf{0} & -{\textbf{u}}^T
\end{bmatrix}.
\end{equation}
For the rest of this work, we consider a discretised version of the operator $\mathbf{u}_\Xi$, which we denote as $\mathbf{U_\Xi}$. The discretisation in the wall-normal direction causes $\mathbf{U}_\Xi$ to be a constant-valued matrix. The exact expression for $\mathbf{U}_\Xi$ can be found in Appendix~\ref{sec:App_theory}.
To incorporate this nonlinearity, we calculate SSVs of the system. The SSVs are a generalization of the concept of singular values for the case of a transfer function or a frequency response operator that is feedback-interconnected to an uncertain matrix, which has a predetermined structure \citep{Zhou1995-dl}.
In detail, similarly to the unstructured case, we calculate SSVs for the frequency response operator $\mathscr{H}_{\nabla}$ by solving the following minimization problem \citep{packard1993,Zhou1995-dl}:

\begin{equation}
\label{eq:2.14}
\mu_{\mathbf{\Delta}}[\mathscr{H}_{\nabla}(k_x,k_z,\omega)] = \frac{1}{\text{min}\{ \overline{\sigma} ({\textbf{U}}_\Xi) : {\textbf{U}}_\Xi \in \mathbf{\Delta}, \text{det}[I - \mathscr{H}_{\nabla}(k_x,k_z,\omega) {\textbf{U}}_\Xi] = 0 \}}.
\end{equation}
In Eq.~\eqref{eq:2.14}, $\overline{\sigma}(\cdot)$ represents the largest singular value, $\text{det}[\cdot]$ is the determinant of a matrix, and $\boldsymbol{\Delta}$ is a set of matrices, that includes all the matrices that have a certain structure. Solving the SSV problem under the same (or similar) structure as ${{\textbf{U}}}_\Xi$ allows modeling the "worst case" nonlinear response to external velocity perturbations. In this framework, the matrix ${{\textbf{U}}}_\Xi$ is not known a priori. Rather, it is computed by solving Eq.~\eqref{eq:2.14} such that the smallest structured uncertainty (in terms of the largest singular value) that will cause instability of the system. While in the current study ${{\textbf{U}}}_\Xi$ models the nonlinear interactions, it is treated in the SSV framework as modeling uncertainty. Thus, we refer to ${{\textbf{U}}}_\Xi$ for the rest of this work as the \emph{structured uncertainty}. A detailed discussion about uncertainty structures is provided in~{\S}\ref{sec:2.5}.

We quantify amplification with nonlinear interactions that are modeled by a structured uncertainty ${\textbf{U}}_\Xi \in \mathbf{\Delta}$ using the supremum over all temporal frequencies, i.e.

\begin{equation}
\label{eq:2.15}
\norm{\mathscr{H}_{\nabla}}_{\mu_{\mathbf{\Delta}}}(k_x,k_z) \equiv \sup_{\omega \in \mathbb{R}}{\mu_{\mathbf{\Delta}}[ \mathscr{H}_{\nabla}(k_x,k_z,\omega)]}, 
\end{equation}
this is the "worst case" amplification of perturbation gradients under the uncertainty structure $\mathbf{\Delta}$. This amplification is not an amplification of the energy of the whole flow field, but rather the gain applied to the most amplified mode. Thus, we note that $\norm{\mathscr{H}_{\nabla}}_{\mu_{\mathbf{\Delta}}}$ is a generalization of $\norm{\mathscr{H}_{\nabla}}_{\infty}$, for the case of interconnection of {discretised} $\mathscr{H}_{\nabla}$ with a structured uncertainty ${\textbf{U}}_\Xi$. Additionally, we use the notation $\norm{\cdot}_{\mu}$ for consistency with previous works \citep{packard1993,liu2021}, despite this operator not being a proper norm since it does not satisfy the triangle inequality. 

\subsection{Small gain theorem formulation of stability}
\label{sec:2.4}
In this section, we use the \textit{small gain theorem} to develop an understanding of the stability of flow systems represented by the NSE. This theorem is used to define the stability of structured feedback interconnections in the structured input-output formulation, providing stability margins for wall-bounded shear flows \citep{liu2021,shuai2023structured,mushtaq2023,mushtaq2024structured}. Herein, we utilize the small gain theorem to establish a threshold on velocity disturbance magnitude that triggers flow instability. 
The interconnection in Fig.~\ref{fig:1} is structured in a way compatible with the small gain theorem {(see chapter 11, \citet{Zhou1995-dl})}, which for a fixed-in-time $\mathbf{U}_\Xi$ is phrased in the following manner:
\begin{theorem}
\label{theo:smg}
Suppose $\mathscr{H}_{\nabla}(k_x,k_z)$ is stable for a given wavenumber pair $(k_x,k_z)$ and let $0 < \gamma < \infty$. The following is a sufficient condition for the feedback loop in Fig.~\ref{fig:1} to be stable:
\begin{equation}
\label{eq:2.16}
\norm{\mathbf{U}_\Xi}_\infty < 1/\gamma
\end{equation}
if and only if:
\begin{equation}
\label{eq:2.17}
\norm{\mathscr{H}_{\nabla}}_{\mu_{\mathbf{\Delta}}}(k_x,k_z) \leq \gamma.
\end{equation}
\end{theorem}
Here, 
$1/\gamma$ constitutes a finite bound on $\norm{\mathbf{U}_\Xi}_\infty$ and $\gamma$ on $\norm{\mathscr{H}_{\nabla}}_{\mu_{\mathbf{\Delta}}}$. Using singular value decomposition, it can be shown that (see Appendix~\ref{sec:App_theory}) that $\norm{{\mathbf{U}}_\Xi}_\infty$ is an approximation of the quantity $\max_{y\in\mathcal{Y}}\norm{\mathbf{u}(y)}_2$ under the discretisation in the wall-normal direction.
Here, $\mathcal{Y}$ is the relevant domain in the wall-normal direction corresponding to the flow system. (for channel flows $\mathcal{Y}=[-1,1]$, for boundary layer flows $\mathcal{Y}=[0,\infty)$);
{$\max_{y\in\mathcal{Y}}\norm{\mathbf{u}(y)}_2$}  denotes the largest velocity perturbation magnitude {that is present in the flow}. 
{Combining equations \eqref{eq:2.16} and \eqref{eq:2.17} leads to the following condition for system stability:
{
\begin{equation}
\label{eq:2.18}
\norm{\mathbf{U}_\Xi}_\infty  < \norm{\mathscr{H}_{\nabla}}_{\mu_{\mathbf{\Delta}}}^{-1}(k_x,k_z).
\end{equation}
}}
\setcounter{theorem}{0}
Therefore, Eq.~\eqref{eq:2.18} provides a bound of value $\norm{\mathscr{H}_{\nabla}}_{\mu_{\mathbf{\Delta}}}^{-1}(k_x,k_z)$ on the magnitude of perturbations that are allowed in the flow to maintain stability. If the perturbation exceeds this bound, {stability is no longer guaranteed}, and a transition to turbulence may occur. 
This formulation is capable of capturing the non-modal instability, i.e., transient growth mechanisms and flow instability triggered via bypass routes in the case of the presence of finite-size disturbances surpassing the threshold set by the small gain theorem criterion.
Alternatively, for a given magnitude of external velocity perturbations, Eq.~\eqref{eq:2.18} can be used to bound the gain applied by the flow system, quantified by $\norm{\mathscr{H}_{\nabla}}_{\mu_{\mathbf{\Delta}}}(k_x,k_z)$. This interpretation relates disturbance size and the magnitude of non-modal growth allowed in the flow.
Following Eq.~\eqref{eq:2.18}, the wave-number pair that determines the stability threshold bound  for a given Reynolds number can be defined as
\begin{equation}
    \label{eq:2.19}
    [(k^*_x,k^*_z)_{\mu_{\mathbf{\Delta}}}](\Rey)\equiv \arg \min \{ [\norm{\mathscr{H}_{\nabla}}_{\mu_{\mathbf{\Delta}}}^{-1}(k_x,k_z)](\Rey) \}.
\end{equation}
Here, we denote this mode as the~\emph{most dominant mode} for a particular Reynolds number that arises from our structured analysis.  
We note that in Eq.~\eqref{eq:2.17} in Theorem~\ref{theo:smg}, the norm-like quantity $\norm{\mathscr{H}_{\nabla}}_{\mu_{\mathbf{\Delta}}}$ can be replaced by $\norm{\mathscr{H}_{\nabla}}_{\infty}${, representing a version of the small gain theorem in which the structure of the uncertainty is not considered (see chapter 9, \citet{Zhou1995-dl})}. This operation yields the following relation:
\begin{equation}
\label{eq:2.20}
\norm{\mathbf{U}_\Xi}_\infty < {\norm{\mathscr{H}_{\nabla}}^{-1}_{\infty}}(k_x,k_z). 
\end{equation}
The use of $\norm{\mathscr{H}_{\nabla}}_{\infty}$ instead of $\norm{\mathscr{H}_{\nabla}}_{\mu_{\mathbf{\Delta}}}$ is equivalent to replacing the set $\mathbf{\Delta}$ with the set of all complex-valued matrices of compatible size \citep{packard1993,Zhou1995-dl,liu2021}, meaning that no particular feedback structure is considered.
As denoted earlier, $\norm{\mathbf{U}_\Xi}_\infty$  corresponds to the largest velocity perturbation magnitude {(i.e., corresponds to $\max_{y\in\mathcal{Y}}\norm{\mathbf{u}(y)}_2$)  that is found in the flow}.The inequality in Eq. \eqref{eq:2.20} provides a stability criterion, which determines the maximal perturbation value allowed in the flow to guarantee the flow system stability. 
 
Equivalently to the structured case, the \emph{most dominant mode} for a given Reynolds number for the unstructured case can be defined as
\begin{equation}
\label{eq:2.21}
    [(k^*_x,k^*_z)_{\infty}](\Rey)\equiv \arg \min \{ [\norm{\mathscr{H}_{\nabla}}_{\infty}^{-1}(k_x,k_z)](\Rey) \}.
\end{equation}
In general, we note that the following inequality always holds for any complex-valued matrix $\mathbf{M}$ and some set of matrices $\mathbf{\Delta}$ \citep{Zhou1995-dl}:

\begin{equation}
\label{eq:2.22}
\forall{\mathbf{\Delta},\mathbf{M}}: \mu_{\mathbf{\Delta}}(\mathbf{M}) \leq \overline{\sigma}(\mathbf{M}).
\end{equation}
By replacing $\mathbf{M}$ with $\mathscr{H}_{\nabla}$ and taking the supremum of both terms in Eq.~\eqref{eq:2.22} we obtain
\begin{equation}
\label{eq:2.23}
\forall{\mathbf{\Delta}}: \norm{\mathscr{H}_{\nabla}}_{\mu_{\mathbf{\Delta}}}(k_x,k_z) \leq \norm{\mathscr{H}_{\nabla}}_{\infty}(k_x,k_z). 
\end{equation}
 From Eq. \eqref{eq:2.23} it follows that:

\begin{equation}
\label{eq:2.24}
\forall{\mathbf{\Delta}}: \norm{\mathscr{H}_{\nabla}}_{\mu_{\mathbf{\Delta}}}^{-1}(k_x,k_z) \geq \norm{\mathscr{H}_{\nabla}}_{\infty}^{-1}(k_x,k_z).
\end{equation}
Both $\norm{\mathscr{H}_{\nabla}}_{\mu_{\mathbf{\Delta}}}^{-1}$ and $\norm{\mathscr{H}_{\nabla}}_{\infty}^{-1}$ are bounds on the magnitude of the maximal velocity perturbation via structured and unstructured approaches, respectively.
Our approach provides a threshold on disturbance magnitude to trigger instability.
In both (structured and unstructured) cases, the wave-number domain is the same, allowing for the same disturbances to be considered in both cases.  
Here, the structured case is a subset of an unstructured case. This, in turn, implies that there are more amplification routes/scenarios for the unstructured case than for structured cases that consider only a subset of such scenarios, limited by the imposed structure on the gain block. 
The structured approach takes into account the structure of ${\textbf{U}}_\Xi$ to model the physics of the nonlinear interactions that are an integral part of the realistic transitional flow physics. Therefore, {we expect} $\norm{\mathscr{H}_{\nabla}}_{\mu_{\mathbf{\Delta}}}^{-1}$ {to be} a more {representative} criterion for stability, while $\norm{\mathscr{H}_{\nabla}}_{\infty}^{-1}$ will be overly strict, and can predict lower bounds on velocity perturbation magnitude than what the actual flow allows.

\subsection{Structures of uncertainty}
\label{sec:2.5}

In this section, we discuss several possible structures for the structured uncertainty ${\textbf{U}}_\Xi$, that are used for approximating the nonlinear term in the NSE and obtaining the best approximation to ${\norm{\mathscr{H}_{\nabla}}_{\mu_{\mathbf{\Delta}}}}$. 
These structures arise after a suitable discretisation in the wall-normal direction is applied to the system. 
The uncertainty structure that approximates the product of ${\textbf{u}}_\Xi$ and the vector of velocity perturbation gradient reproducing the structure of the advection term in the Navier-Stokes equation is:
\begin{equation}
\label{eq:2.25}
\mathbf{\Delta}_{\mathbf{u}} = \left\{\begin{bmatrix}
    \Delta_u & \mathbf{0} & \mathbf{0} \\
    \mathbf{0} & \Delta_u & \mathbf{0} \\
    \mathbf{0} & \mathbf{0} & \Delta_u 
\end{bmatrix} , \Delta_u = [\text{diag}(u_{\xi}), \text{diag}(v_{\xi}), \text{diag}(w_{\xi})]: u_{\xi},v_{\xi},w_{\xi} \in \mathbb{C}^{N_y \times 1} \right\},
\end{equation}
where $u_{\xi},v_{\xi},w_{\xi}$ can have complex entries because they represent Fourier-transformed velocity components; $N_y$ is the number of points that are used in the discretisation of the wall-normal direction. {Eq. \eqref{eq:2.25} represents the structure that arises after a discretisation of Eq. \eqref{eq:2.13}}. This structure ensures that the feedback interconnection loop in Fig.~\ref{fig:1} accurately recreates the structure of the nonlinear advection term. 
Any additional terms that are not included in the set $\mathbf{\Delta}_{\mathbf{u}}$ may be regarded as fictitious nonlinear interactions that are not part of the term $\mathbf{u}\cdot \nabla\mathbf{u}$ and, thus, do not represent physical flow behaviors.

\begin{figure}
\begin{subfigure}{0.28\textwidth}
\includegraphics[width=0.9\linewidth]{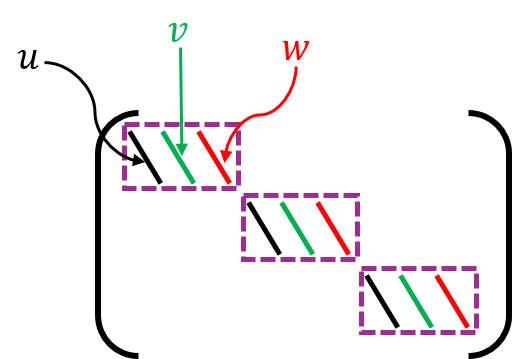}
\caption{${\textbf{U}}_\Xi \in \mathbf{\Delta}_{\mathbf{u}}$}
\label{fig:2a}
\end{subfigure}
\hfill
\begin{subfigure}{0.28\textwidth}
\includegraphics[width=0.9\linewidth]{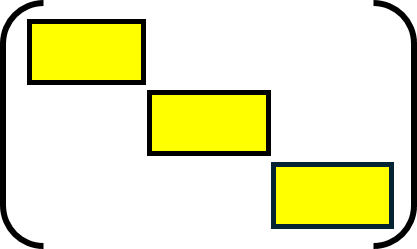}
\caption{${\textbf{U}}_\Xi \in \mathbf{\Delta}_{r}$}
\label{fig:2b}
\end{subfigure}
\hfill
\begin{subfigure}{0.28\textwidth}
\includegraphics[width=0.9\linewidth]{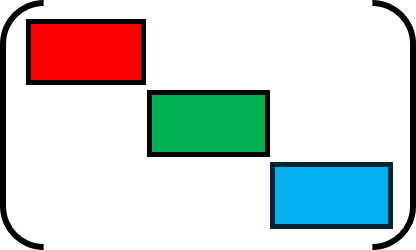} 
\caption{${\textbf{U}}_\Xi \in \mathbf{\Delta}_{nr}$}
\label{fig:2c}
\end{subfigure}
\hspace{0.8cm}

\caption{{Illustration of uncertainty matrices with relevant structures: (a) uncertainty structure that is based on the feedback interconnection as defined in Eq.~\eqref{eq:2.25}, (b) repeated block structure as defined in Eq.~\eqref{eq:2.26}, (c) non-repeated block structure as defined in Eq.~\eqref{eq:2.27}. Visualizations of the uncertainty matrices are shown for (a) ${\textbf{U}}_\Xi \in \mathbf{\Delta}_{\mathbf{u}}$, (b) ${\textbf{U}}_\Xi \in \mathbf{\Delta}_{r}$, (c) ${\textbf{U}}_\Xi \in \mathbf{\Delta}_{nr}$.}}
\label{fig:2}
\end{figure}

The structure of the uncertainty should ideally be similar to $\mathbf{\Delta}_{\mathbf{u}}$ for an accurate model of the physics of nonlinear interactions. 
The solution for uncertainties in $\mathbf{\Delta}_{\mathbf{u}}$ involves the solution of the optimization problem in Eq. \eqref{eq:2.14} with a very complex set of constraints that are required for preserving the structure in Eq. \eqref{eq:2.25}. Such a solution has not yet been found, as far as we know. Alternatively, approximating this structure of the uncertainties is considered, which we detail next. 
In this work, we use two different sets of matrices that are simpler in structure and are meant to approximate the uncertainty structure from Eq.~\eqref {eq:2.14}.
These sets are as follows. 
\begin{enumerate}
\item Set $\mathbf{\Delta}_{r}$: the set of block-diagonal matrices with three repeating full blocks with compatible sizes as was used in the works of \citet{mushtaq2023,mushtaq2024structured}, 
\begin{equation}
\label{eq:2.26}
\mathbf{\Delta}_{r} = \left\{ \begin{bmatrix}
    \Delta_r & \mathbf{0} & \mathbf{0} \\
    \mathbf{0} & \Delta_r & \mathbf{0} \\
    \mathbf{0} & \mathbf{0} & \Delta_r 
\end{bmatrix} : \Delta_r \in \mathbb{C}^{N_y \times 3N_y}\right\}.
\end{equation}
Here, $\Delta_r$ can be any matrix of compatible dimensions ($N_y \times 3N_y$).

\item Set $\mathbf{\Delta}_{nr}$: the set of block-diagonal matrices with three full, independent \textbf{non-repeating} blocks with compatible sizes, {as was used in the works of \citet{liu2021,shuai2023structured}}: 

\begin{equation}
\label{eq:2.27}
\mathbf{\Delta}_{nr} = \left\{\begin{bmatrix}
    \Delta_1 & \mathbf{0} & \mathbf{0} \\
    \mathbf{0} & \Delta_2 & \mathbf{0} \\
    \mathbf{0} & \mathbf{0} & \Delta_3 
\end{bmatrix} : \Delta_1,\Delta_2,\Delta_3 \in \mathbb{C}^{N_y \times 3N_y}\right\}.
\end{equation}
The difference here between $\mathbf{\Delta}_{nr}$ and $\mathbf{\Delta}_{r}$ lies in the fact that that $\mathbf{\Delta}_{nr}$ does not include the constraint of equal blocks, as $\Delta_1$, $\Delta_2$, and $\Delta_3$ can be different matrices.

\end{enumerate}
Visualization of the uncertainty matrices that correspond to the 
structures $\mathbf{\Delta}_{\mathbf{u}}$, $\mathbf{\Delta}_{r}$ and $\mathbf{\Delta}_{nr}$ are illustrated in Fig.~\ref{fig:2}. We note that solving the minimization problem for $\mathbf{\Delta}_{r}$ is computationally more demanding than for $\mathbf{\Delta}_{nr}$. While $\mathbf{\Delta}_{nr}$ imposes a single constraint on the uncertainty structure to be a set of block-diagonal matrices, $\mathbf{\Delta}_{r}$ imposes an additional constraint by requiring these blocks to be equal. 

The definitions in Eq.~\eqref{eq:2.25}, Eq.~\eqref{eq:2.26} and \eqref{eq:2.27} yield that
\begin{equation}
 \label{eq:2.28}
\mathbf{\Delta}_{\mathbf{u}} \subset \mathbf{\Delta}_{r} \subset \mathbf{\Delta}_{nr}\subset \mathbb{C}^{3N_y \times 9N_y}. 
\end{equation}
These relations between sets are manifested in the values of the SSVs for each set. Based on the monotonicity of the SSV with respect to set inclusion  \cite[see][section 4.10]{scherer2001theory}, solving the minimization problem (Eq.~\eqref{eq:2.14}) over these sets of matrices that satisfy the relations in Eq.~\eqref{eq:2.28}, yields, 
\begin{equation}
\label{eq:2.29}
\norm{\mathscr{H}_{\nabla}}_{\mu_{\mathbf{\Delta}_{u}}}(k_x,k_z) \leq \norm{\mathscr{H}_{\nabla}}_{\mu_{\mathbf{\Delta}_{r}}}(k_x,k_z) \leq \norm{\mathscr{H}_{\nabla}}_{\mu_{\mathbf{\Delta}_{nr}}}(k_x,k_z) \leq \norm{\mathscr{H}_{\nabla}}_{\infty}(k_x,k_z). 
\end{equation}
As we will show next, this hierarchical relationship allows us to explain the difference in the predictions of the most dominant mode by unstructured and structured input-output methods. Moreover, the norm-like property of the smallest set $\mathbf{\Delta}_{\mathbf{u}}$ is $\norm{\mathscr{H}_{\nabla}}_{\mu_{\mathbf{\Delta}_{u}}}(k_x,k_z)$, which inverse ($\norm{\mathscr{H}_{\nabla}}_{\mu_{\mathbf{\Delta}_{u}}}^{-1} $) would provide the least tight and most accurate bound to perturbation magnitude to keep the system stable. In particular, the inequality in Eq.~\eqref{eq:2.20}, can serve as a generalized stability criterion, which is more physically justified to study pathways to transition, in contrast to classical stability theory, in which the mathematical formulation is restricted to infinitesimally small perturbations. However, as mentioned previously, the solution to such a set requires the solution of the optimization problem in Eq. \eqref{eq:2.14}, which is elaborate and has not yet been found. The best next approximation of this set would be to use 
$\norm{\mathscr{H}_{\nabla}}_{\mu_{\mathbf{\Delta}_{r}}}^{-1}$, which is the most representing bound on velocity perturbation magnitude that we can compute, while $\norm{\mathscr{H}_{\nabla}}_{\mu_{\mathbf{\Delta}_{nr}}}^{-1}$ and $\norm{\mathscr{H}_{\nabla}}_{\infty}^{-1}$ provide more restricting bounds, respectively. 

\section{Numerical analysis settings}
\label{sec:3} 

To compute the operators in Eq.~\eqref{eq:2.2} and the flow field they produce, we apply discretisation to the wall-normal direction by using Chebyshev collocation points. We implement this discretisation process by using the pseudo-spectral differentiation matrices from \citet{weideman2000matlab} for differentiation in the wall-normal direction. Boundary conditions were implemented based on \citet{trefethen2000spectral}. {In particular, we employ a quasi-parallel assumption and use the change of variables described in \citet{schmid2002stability} to transform the bounded domain $\begin{bmatrix} -1, & 1 \end{bmatrix}$ to the semi-infinite domain $\begin{bmatrix} 0, & \infty \end{bmatrix}$ of the flat plate boundary layer (for details see Appendix A.4 in \citet{schmid2002stability}). 
We set the wall-normal domain range to $[0, L_y]=\begin{bmatrix} 0, & 15 \end{bmatrix}$ as this height is well within the free stream, capturing the entire base flow profile variation. We verified that the selected domain provides converged results, with no significant effect on the results in our study. Specifically, the maximum difference between the contour plots of the  $\norm{\mathscr{H}_{\nabla}}_{\infty}$ generated with increased domain  $L=20$  was less than 4~\%. 
 
All of the calculations are performed in $\text{MATLAB}^\text{®}$ R2023b. We denote by $N_y$ the number of Chebyshev collocation points chosen for the discretisation process. For the wave-number domain, we have used a $50\times90$ grid of logarithmically spaced values with the following boundaries: $(k_{x_{min}}=10^{-4},k_{x_{max}}=3.02)$ and $(k_{z_{min}}=10^{-2},k_{z_{max}}=15.84)$. This is the same grid used in \citep{jovanovic2005componentwise}. A Chebyshev collocation method with $N_y={61}$ points was used for discretisation along the wall-normal direction in all computations and for all base flows. This spacing was determined to be adequate for obtaining reliable results, as doubling the number of source points did not affect the trends and values of the curves, nor did it cause any noticeable changes in the contour plots presented in \S\ref{sec:4}. 
 
In terms of temporal frequencies, an adaptive approach was applied using MATLAB, utilizing the highly computationally efficient \textit{sigma} and \textit{mussv} MATLAB functions. The results obtained from this adaptive method were compared with those obtained using a constant grid of frequency values. It was demonstrated that the results from the constant grid approach were converging to those achieved by adaptively selecting temporal frequencies.

The values of the $\mathscr{H}_{\infty}$ norm were calculated by using the MATLAB function \textit{hinfnorm}, which is part of the robust control toolbox \citep{balas2007robust}. Additionally, we calculate {the estimated value of $\norm{ \mathscr{H}_{\nabla}}_{\mu}$} for the case of ${\textbf{U}}_\Xi \in \mathbf{\Delta}_{nr}$ using the function \textit{mussv} in the robust control toolbox \citep{balas2007robust}, and we use the numerical algorithm of generalized power iteration for finding lower bounds on the SSV \citep{mushtaq2023} for the case of ${\textbf{U}}_\Xi \in \mathbf{\Delta}_{r}$.

\section{Effect of Reynolds number on disturbance threshold}
\label{sec:4}

This section presents a detailed analysis of the effect of the Reynolds number on disturbance threshold, and the dominant structures that determine the stability threshold bound for plane Couette, plane Poiseuille, and Blasius flows.
For each base flow, we study the flow system stability characteristics using our stability criterion, which was derived in \S~\ref{sec:2.4} by evaluating the following.
\begin{enumerate}
\item[(a)] Structured norm-like quantities {(and their inverse)} defined in Eq.~\eqref{eq:2.15} using:
\begin{itemize}
\item[(i)] repeated blocks approach, denoted as $\norm{\mathscr{H}_{\nabla}}_{\mu_{\mathbf{\Delta}_{r}}}(k_x,k_z)$;
\item[(i)] non-repeated blocks approach, denoted as $\norm{\mathscr{H}_{\nabla}}_{\mu_{\mathbf{\Delta}_{nr}}}(k_x,k_z)$.
\end{itemize}
\item[(b)] Unstructured linear norm, denoted as $\norm{\mathscr{H}_{\nabla}}_{\infty}(k_x,k_z)$ in Eq.~\eqref{eq:2.11}. 
\end{enumerate}

We examine and analyze here the contour plots of the quantities mentioned above over the $k_x,k_z$ domain at specific Reynolds numbers.} With this approach, we identify modes of interest that determine the stability threshold bound in our analysis.
Next, we analyze the behavior of pre-selected (three to four) modes across a wide range of Reynolds numbers that we identified as dominant modes that emerged during the system excitation of each base flow. We focus on several specific modes since it would not be feasible to determine the most dominant mode $(k^*_x,k^*_z)$---that determines the stability threshold bound---and the corresponding stability threshold for the above norm quantities, over a wide range of wave-number pairs and for a wide range of Reynolds numbers. {The modes we study} are listed in Table~\ref{tbl:1}, where each mode is represented by a unique wave number pair $(k_x,k_z)$. 

Specifically, the considered modes for each flow case are as follows.

\begin{table}
\centering
\begin{tabular}{p{4cm} c c c}

& Couette & {plane Poiseuille} & Blasius \\

DLR & $(0.196,0.628)$ & $(1.561,0.692)$ & $(0.241,0.274)$ \\

TS & $(0.2,10^{-2})$ & $(1.02,10^{-2})$ & $(0.303,10^{-2})$ \\

SPS & {$(10^{-4},1.436)$} & {$(10^{-4},1.841)$} & {$(10^{-4},0.274)$} \\

DHR & $(0.01, 0.13)$ & $(-)$ & $(-)$ \\

\end{tabular}
\caption{Selected modes {(DLR - dominant low Reynolds, TS - Tollmien–Schlichting, SPS - spanwise periodic streak, DHR - dominant high Reynolds)} for each base flow, represented by a pair of wave numbers $(k_x,k_z)$. 
  }
\label{tbl:1}
\end{table}

\begin{enumerate}
\item The dominant low Reynolds (DLR) mode -- corresponds to the {most dominant} mode observed in the {contour map} of {$\norm{\mathscr{H}_{\nabla}}_{\mu_{\mathbf{\Delta}_{nr}}}^{-1}$} at a sub-critical Reynolds number for each base flow. These Reynolds numbers are $\Rey=358$ for Couette, $\Rey=400$ for Blasius, and $\Rey=690$ for {plane Poiseuille} base flows. {These particular Reynolds numbers were chosen for the cases of Couette and plane Poiseuille flows to equal the Reynolds numbers that were used in \citet{liu2021} and \citet{mushtaq2023}, to provide a validation of our results with the results of these works.}
\item The TS (Tollmien–Schlichting) mode -- The TS mode is set using the smallest $k_z$ in our numerical domain, which is $k_{z_{min}}=10^{-2}$. {For the case of Couette flow}, the $k_x$ value {is} taken to be {$k_x=\arg \max \norm{\mathscr{H}_\nabla}_{\infty}(k_x,k_{z,min})$}, i.e., the streamwise wave number argument of the highest value of {$\norm{\mathscr{H}_\nabla}_{\infty}(k_x,k_z)$}. For the cases of {plane Poiseuille} and Blasius base flows, the $k_x$ component of the TS mode associated
with the critical mode (according to LST) that becomes unstable at the lowest Reynolds number according to LST \citep{schmid2002stability}. Monitoring this mode is very useful to discuss different aspects of our analysis with respect to LST.
\item The spanwise periodic streak (SPS) mode -- corresponds to the smallest $k_x$ value in our numerical domain, which is $k_{x, min}=10^{-4}$, coupled with the $k_z$ values are observed to be {most dominant} on the {contour maps} of {$\norm{\mathscr{H}_{\nabla}}_{\infty}^{-1}$}.  {The unstructured case is selected following the extensive literature that utilized unstructured-like input-output methodologies  where streaks were reported to be dominant  \citep{reddy1993energy,schmid1994optimal,schmid2000linear,bamieh2001energy,jovanovic2004modeling,jovanovic2004unstable,jovanovic2005componentwise}.} 
\item The dominant high Reynolds (DHR) mode --  observed only for Couette base flow and corresponds to the {most dominant} mode of {$\norm{\mathscr{H}_{\nabla}}_{\mu_{\mathbf{\Delta}_{nr}}}^{-1}$} at a {high} Reynolds number, which is selected to be  $\Rey=8000$. 
\end{enumerate}

\subsection{Couette base flow}
\label{sec:4.1}
\begin{figure}
\centering
\begin{subfigure}[a]{1\textwidth}
\centering
\includegraphics[width=\textwidth]{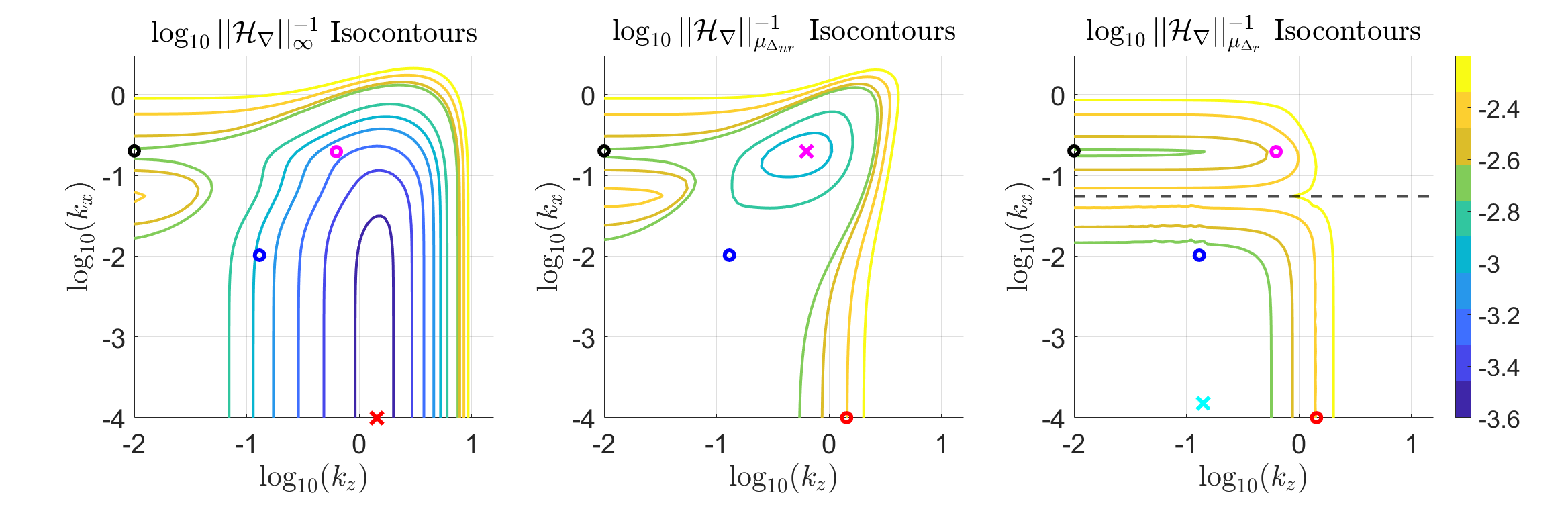}
\caption{$\Rey=358$}
\label{fig:3a}
\end{subfigure}
\hfill
\begin{subfigure}[b]{1\textwidth}
\centering
\includegraphics[width=\textwidth]{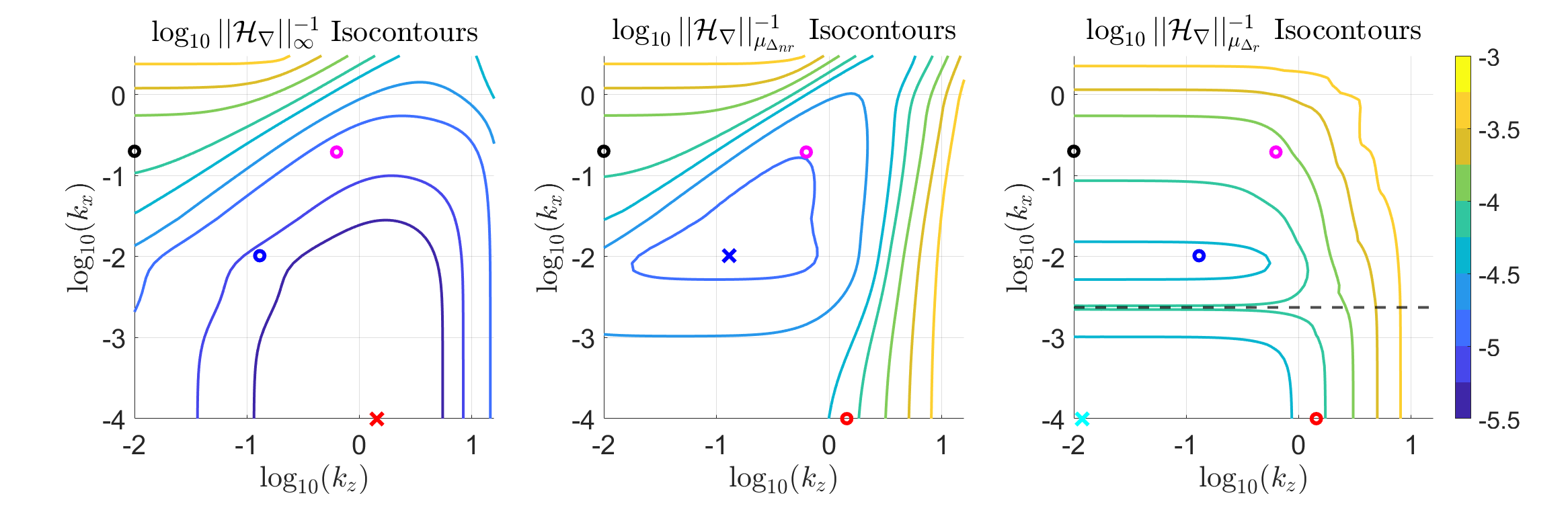}
\caption{$\Rey=8000$}
\label{fig:3b}
\end{subfigure}
\caption{Contour plots in logarithmic scale of (from left to right): $\norm{\mathscr{H}_{\nabla}}_{\infty}^{-1}$, $\norm{\mathscr{H}_{\nabla}}_{\mu_{\mathbf{\Delta}_{nr}}}^{-1}$, and $\norm{\mathscr{H}_{\nabla}}_{\mu_{\mathbf{\Delta}_{r}}}^{-1}$ at (a) $\Rey=358$ and (b) $\Rey=8000$ for Couette flow. The most dominant mode is marked by an X (cyan color indicates that the mode's $(k_x,k_z)$ value  is not denoted in the table), whereas the other modes from Table~\ref{tbl:1} are marked by colored circles (magenta - DLR mode, black - TS mode, red - SPS mode, blue
- DHR mode). Results are showed for (a) $Re=358$ and (b) $Re=8000$.}
\label{fig:3}
\end{figure}

In this section, we present the results of our analysis for the Couette base flow profile of the form $U=y$.
In Fig.~\ref{fig:3}, {we show the {most dominant} flow structures in terms of $\norm{\mathscr{H}_{\nabla}}_{\infty}^{-1}(k_x,k_z)$,  $\norm{\mathscr{H}_{\nabla}}_{\mu_{\mathbf{\Delta}_{nr}}}^{-1}(k_x,k_z)$ and $\norm{\mathscr{H}_{\nabla}}_{\mu_{\mathbf{\Delta}_{r}}}^{-1}(k_x,k_z)$,}  for both of the Reynolds numbers $\Rey=358$ (Fig.~\ref{fig:3a}) and $\Rey=8000$ (Fig.~\ref{fig:3b}). The Reynolds number $\Rey=358$ was chosen to be the same as used in \citet{liu2021} and \citet{mushtaq2024structured}, to validate our results with these works. 
The value $Re=8000$ was selected arbitrarily after the post-critical value observed in experiments. We chose a very high value to demonstrate that the stability characteristics do not change drastically when changing the Reynolds number for Couette flow. 
Over these {contour maps}, we overlay the modes from Table~\ref{tbl:1}. {The most dominant mode is marked by an X symbol. If the most dominant mode is not denoted in Table \ref{tbl:1}, it is colored in cyan.} 
We note that the relationship between the three types of amplification satisfy the relation in Eq.~\eqref{eq:2.29} {for both Reynolds numbers}, i.e., $\norm{\mathscr{H}_{\nabla}}_{\mu_{\mathbf{\Delta}_{r}}}(k_x,k_z) \leq \norm{\mathscr{H}_{\nabla}}_{\mu_{\mathbf{\Delta}_{nr}}}(k_x,k_z) \leq \norm{\mathscr{H}_{\nabla}}_{\infty}(k_x,k_z)$. Here, $\norm{\mathscr{H}_{\nabla}}^{-1}_{\infty}$ reaches values lower than $\norm{\mathscr{H}_{\nabla}}^{-1}_{\mu_{\mathbf{\Delta}_{nr}}}$ and $\norm{\mathscr{H}_{\nabla}}^{-1}_{\mu_{\mathbf{\Delta}_{r}}}$, respectively, as discussed in section \ref{sec:2.5}. 

The use of a structured approach for both cases (repeated and non-repeated approaches) leads to an increased stability threshold. The dominant structures that arise in the unstructured case are streamwise elongated structures, whereas oblique modes become the most dominant structures in structured cases, which is more clearly evident in the non-repeated blocks case. The latter result agrees better with DNS and nonlinear optimal perturbations (NLOP)-based predictions than with traditional linear input-output methods \citep{prigent2003long,duguet2010formation}.
The contour map of {$\norm{\mathscr{H}_{\nabla}}_{\mu_{\mathbf{\Delta}_{nr}}}^{-1}$} shows a {most dominant} oblique wave mode of $(k_x,k_z) \approx (0.1,1)$, {which overpowers} the streamwise elongated structures that are also evident on the same color map by an order of magnitude. It was stated in \cite{liu2021}, "this result is consistent with findings of \citet{reddy1998stability} showing that oblique waves require less perturbation energy to trigger turbulence in plane Couette flow than streamwise vortices". As we will show next, the observed dominance of oblique waves is affected by the type of uncertainty that is considered and the Reynolds number. In the case of {$\norm{\mathscr{H}_{\nabla}}_{\mu_{\mathbf{\Delta}_{r}}}^{-1}$}, where repeated uncertainty is considered, the lowest stability threshold region that is evident in  $\norm{\mathscr{H}_{\nabla}}_{\mu_{\mathbf{\Delta}_{r}}}^{-1}$  {spreads} over a wide range of modes in the bottom left quadrant of the domain, corresponding to a wide range of dominant oblique modes. This result is consistent with the results of \citet{mushtaq2024structured} for $\Rey=358$. 

We note that while the relationship in Eq.~\eqref{eq:2.29} shows that $\norm{\mathscr{H}_{\nabla}}_{\mu_{\mathbf{\Delta}_{r}}}^{-1}$  provides a closer bound than  $\norm{\mathscr{H}_{\nabla}}_{\mu_{\mathbf{\Delta}_{nr}}}^{-1}$  to the accurate value $\norm{\mathscr{H}_{\nabla}}_{\mu_{\mathbf{\Delta}_{\mu}}}^{-1}$  on velocity perturbations from a stability viewpoint, there is no guarantee that it represents more accurately the contour plots in the wavenumber domain.
Approximations of the structure of the nonlinear term in the NSE by either non-repeating or repeating blocks still may include additional feedback pathways that are not physically associated with nonlinear flow physics, but to a significantly smaller extent than in the unstructured case. Therefore, considering two structured cases (repeated and non-repeated blocks) is important in our analysis, where common features may be associated with flow physics, whereas the differences require further investigation.
One such common feature is the shift of dominant modes to lower $k_x$ values with Reynolds number for unstructured and both structured cases,  which means this trend is associated with a linear mechanism.   This trend agrees with results obtained from stability analysis of optimal disturbance growth \citep{schmid2002stability,butler1992three}, where it is shown that for Couette flow, the maximum growth is achieved for $k_x Re \sim \text{C}$, where $C$ is some constant value, which means that for a given non-zero $k_z$,  an increase in the Reynolds number results in a decrease of $k_x$, which corresponds to oblique and $\Lambda$-shaped structures that transform to streaks at higher Reynolds numbers  \citep{hwang2010linear,dokoza2023reynolds}.
Observing {$\norm{\mathscr{H}_{\nabla}}_{\mu_{\mathbf{\Delta}_{nr}}}^{-1}$} (with non-repeated uncertainty) for both Reynolds numbers, we see that there is a change {of} the {most dominant} mode with a Reynolds number increase, where the DLR mode (indicated by a magenta colored circle) of $(k_x,k_z)=(0.196,0.628)$ for $\Rey=358$ in Fig.~\ref{fig:3a}, {has a larger stability threshold associated with it than the} DHR mode (indicated by blue colored circle) of $(k_x,k_z)=(0.01, 0.13)$ for $\Rey=8000$ in Fig.~\ref{fig:3b}. 
 Additionally, in both plots, we added a horizontal dashed black line overlaid on the {contour maps} of $\norm{\mathscr{H}_{\nabla}}_{\mu_{\mathbf{\Delta}_{r}}}$. This horizontal line represents the border between two regions of {relatively unstable modes} in the {contour maps} of {$\norm{\mathscr{H}_{\nabla}}_{\mu_{\mathbf{\Delta}_{r}}}^{-1}$} for both Reynolds numbers, \Rey = 358 and \Rey = 8000, respectively. This line migrates towards a lower $k_x$ value with the Reynolds number increase, due to the shift of dominant modes to lower $k_x$ values. Thus, it is evident that there is a set of oblique modes with this streamwise wavenumber (of the dashed line) and spanwise wavenumber values of $k_z<1$, that will exhibit increased stability compared with neighboring modes. 
 
In the results for the higher Reynolds number, lower stability threshold values are observed for all of the quantities {$\norm{\mathscr{H}_{\nabla}}_{\mu_{\mathbf{\Delta}_{r}}}^{-1}(k_x,k_z)$}, {$\norm{\mathscr{H}_{\nabla}}_{\mu_{\mathbf{\Delta}_{nr}}}^{-1}(k_x,k_z)$} and {$\norm{\mathscr{H}_{\nabla}}_{\infty}^{-1}(k_x,k_z)$}. In Fig.~\ref{fig:3a}, the smallest perturbations that cause instability have a magnitude of $10^{-3.6}$ for the unstructured case. In Fig.~\ref{fig:3b}, significantly lower values appear, with the smallest perturbations, which can cause instability being of magnitude $10^{-5.5}$  for the unstructured case.
 
\begin{figure}
\centering
\begin{overpic}[width=1\linewidth]{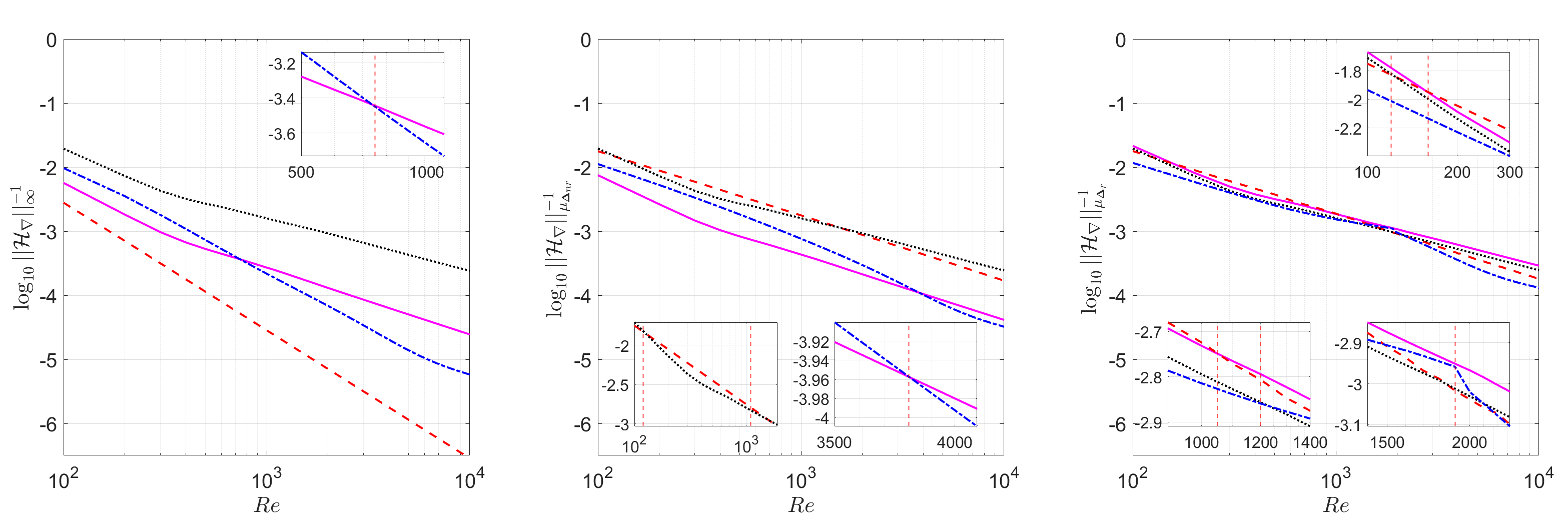}
\put(0, 28){\small (a) \hspace{4.05cm} (b) \hspace{4.05cm} (c)}
\end{overpic}
 
\caption{Evolution curves of imposed thresholds due to flow structures associated with pre-selected modes of interest denoted in Table~\ref{tbl:1} (magenta solid - DLR mode, black dotted - TS mode, red dashed - SPS mode, blue dashed-dotted - DHR mode) for Couette base flow as a function of the Reynolds number in terms of (a) $\norm{\mathscr{H}_{\nabla}}_{\infty}^{-1}$, (b) $\norm{\mathscr{H}_{\nabla}}_{\mu_{\mathbf{\Delta}_{nr}}}^{-1}$ and (c) $\norm{\mathscr{H}_{\nabla}}_{\mu_{\mathbf{\Delta}_{r}}}^{-1}$.} 
\label{fig:4}
\end{figure}

Comparing only two Reynolds numbers is insufficient to reveal the interplay between the dominant structures that are evident in these {contour maps} with Reynolds number variation. Therefore, next, we focus our analysis on key pre-selected modes of interest from these {contour maps} and monitor their behavior for a wide range of Reynolds numbers. In Fig.~\ref{fig:4}, we show the evolution of the pre-selected modes that are listed in Table~\ref{tbl:1} as a function of Reynolds number in the range $\Rey \in [100,10000]$. The trends from each plot are accompanied by insets of zoomed-in views for certain limited Reynolds number ranges at different locations, revealing the interesting interplay between the mode curves for {$\norm{\mathscr{H}_{\nabla}}_{\mu_{\mathbf{\Delta}_{r}}}^{-1}$}, {$\norm{\mathscr{H}_{\nabla}}_{\mu_{\mathbf{\Delta}_{nr}}}^{-1}$} and {$\norm{\mathscr{H}_{\nabla}}_{\infty}^{-1}$}. In particular, the following insights are gained from these plots. 

First, for {$\norm{\mathscr{H}_{\nabla}}_{\infty}^{-1}$} (Fig.~\ref{fig:4}a), we observe that the SPS mode ({red} curve) remains {the most dominant, i.e., providing the lowest threshold,} through the entire range of Reynolds numbers. A higher threshold is shown for the DLR mode (magenta curve) that overpowers the DHR mode (blue curve) for Reynolds numbers up to $\Rey=750$, and vice versa afterward. The SPS modes are extensively reported in the literature to be the dominant structures for a variety of base flows at different Reynolds numbers when using linear input-output analysis \citep{bamieh2001energy,jovanovic2004modeling,jovanovic2004unstable,jovanovic2005componentwise,schmid2007nonmodal}. Thus, it is not surprising to us that the dominance of the SPS mode for a wide range of Reynolds numbers in the linear analysis represented by $\norm{\mathscr{H}_{\nabla}}_{\infty}$ is in agreement with previous works.
In contrast, this mode is not always dominant if {structured analysis} is considered. 
In particular, in the non-repeated case, for {$\norm{\mathscr{H}_{\nabla}}_{\mu_{\mathbf{\Delta}_{nr}}}^{-1}$} (Fig.~\ref{fig:4}b), the {threshold that is associated to the SPS mode is always higher} than the DLR mode  (magenta) and the DHR mode (blue) thresholds. The DLR and DHR modes {are most dominant} for the entire Reynolds number range, where the oblique DLR mode (magenta curve) {sets {a lower} threshold on disturbance to make the flow unstable than} the DHR mode (blue curve) for Reynolds number up to $\Rey\approx 3800$ (denoted by a red vertical dashed line), and vice versa afterward. 
In the repeated case, for {$\norm{\mathscr{H}_{\nabla}}_{\mu_{\mathbf{\Delta}_{r}}}^{-1}$} (Fig.~\ref{fig:4}c) we observe first that the oblique DLR mode (magenta curve) {provides the highest threshold} (out of the four modes that were tested) for the entire range of Reynolds numbers that were considered. Second, the oblique DHR mode (blue curve) is the {most dominant} mode for $\Rey<900$ and $\Rey>3500$, whereas in the range $900<\Rey<3500$, the SPS mode ({red} curve)  { imposes a slightly lower threshold on perturbation magnitudes for instability to occur than} the DHR mode. 
The blue curve trend is shown to be {decreasing}, but with a subtle kink at $\Rey\approx 1900$, where the DHR mode drops close to the DLR mode value and then grows again. 
This kink can be explained by inspecting the contour maps of {$\norm{\mathscr{H}_{\nabla}}_{\mu_{\mathbf{\Delta}_{r}}}^{-1}$}. In Fig.~\ref{fig:3a} and \ref{fig:3b}, we mark dashed horizontal lines representing {modes that are stable for larger perturbations,} dividing two distinct regions of {relative instability} for both Reynolds numbers, $\Rey=358$ and $\Rey=8000$, respectively.  {This horizontal line is located at $k_x\approx 0.05$ for $\Rey=358$, i.e,   above the $k_x$ value of the DHR mode (blue circle)--- which is defined by the wavenumber pair $(k_x,k_z)=(0.01, 0.13)$---whereas it shifts towards  $k_x\approx 0.0025$, {below} the $k_x$ value of the DHR mode at $\Rey=8000$. This means that when this line migrates towards lower streamwise wave number with a Reynolds number increase, it intersects the  $k_x=0.01$ value at $Re\approx 1900$ where this line matches the same $k_x$ value of the DHR mode. 

Second, the TS mode (black curve) imposed threshold is the highest in the unstructured case for all Reynolds numbers, whereas for both {$\norm{\mathscr{H}_{\nabla}}_{\mu_{\mathbf{\Delta}_{r}}}^{-1}$} (Fig.~\ref{fig:4}c) and {$\norm{\mathscr{H}_{\nabla}}_{\mu_{\mathbf{\Delta}_{nr}}}^{-1}$ (Fig.~\ref{fig:4}b), the TS mode imposed threshold is higher than the SPS mode threshold (red curve) for almost the entire range of Reynolds numbers that were considered, besides a limited range between $200<\Rey<700$. When using the structured analysis, the most dominant mode changes with the Reynolds number, demonstrating the importance of looking at a range of Reynolds numbers as opposed to a single one to understand the full physical behavior and stability characteristics of the flow system at hand.

Third, we note that for all Reynolds numbers and modes tested, the curve trends of the {most dominant} modes agree with the relationship in Eq.~\eqref{eq:2.29} that holds for all Reynolds numbers as expected. In particular, one can observe that the graphs of all of the chosen modes for both of the structured quantities, {$\norm{\mathscr{H}_{\nabla}}_{\mu_{\mathbf{\Delta}_{nr}}}^{-1}$} and {$\norm{\mathscr{H}_{\nabla}}_{\mu_{\mathbf{\Delta}_{r}}}^{-1}$}, remain less spread out than in the case of $\norm{\mathscr{H}_{\nabla}}_{\infty}^{-1}$ through the whole range of Reynolds numbers, which is mostly apparent for $\norm{\mathscr{H}_{\nabla}}_{\mu_{\mathbf{\Delta}_{r}}}^{-1}$. This observation correlates with the number of constraints imposed on the amplification routes that yield the hierarchical relation of subsets in Eq.~\eqref{eq:2.28}. Here, all of the modes impose similar thresholds on the magnitude of perturbations that can be supported by the flow without losing stability. The stability margin represented by $\norm{\mathscr{H}_{\nabla}}_{\infty}^{-1}$ is significantly more strict (with the lowest threshold) than in the structured case with repeated-block uncertainty, which is represented by $\norm{\mathscr{H}_{\nabla}}_{\mu_{\mathbf{\Delta}_{r}}}^{-1}$. 
For example, at $\Rey=1000$, the system represented by $\norm{\mathscr{H}_{\nabla}}_{\infty}^{-1}$ can remain stable up to perturbations of a magnitude of $\norm{\vc{u}} \sim O(10^{-4.5})$, whereas for the system represented by $\norm{\mathscr{H}_{\nabla}}_{\mu_{\mathbf{\Delta}_{nr}}}^{-1}$, the perturbations magnitude increase to a magnitude of $\norm{\vc{u}} \sim O(10^{-3.5})$, and the system represented by $\norm{\mathscr{H}_{\nabla}}_{\mu_{\mathbf{\Delta}_{r}}}^{-1}$ can support perturbations up to a magnitude of $\norm{\vc{u}} \sim O(10^{-3})$. We consider here the results of previous works \citep{reddy1998stability,duguet2010towards,duguet2013minimal}, which used optimization methods and DNS to predict the minimal perturbation kinetic energy that induces transition to turbulence. These works predict a minimal kinetic energy between $O(10^{-6})$ and $O(10^{-5})$ to cause transition at $\Rey=1000$. Kinetic energy is proportional to the square of velocity magnitude; therefore, it corresponds to the disturbance of  $O(10^{-3})$ and $O(10^{-2.5})$, where it is clear that the predictions by $\norm{\mathscr{H}_{\nabla}}_{\infty}^{-1}$ are overly strict whereas $\norm{\mathscr{H}_{\nabla}}_{\mu_{\mathbf{\Delta}_{nr}}}^{-1}$ and $\norm{\mathscr{H}_{\nabla}}_{\mu_{\mathbf{\Delta}_{r}}}^{-1}$, produce stability thresholds that better agree with DNS results. Since the main cause for the large difference in stability thresholds between the unstructured and structured approaches is associated with the SPS mode, we conclude that the unstructured analysis exaggerates the effect of SPS compared with the real flow physics. A more thorough comparison between the predictions of our stability criterion and the results from numerical simulations and experimental works is provided in~\S\ref{sec:5}.

\subsection{Plane Poiseuille base flow}
\label{sec:4.2}
 
In this section, we present {stability analysis} results obtained using the plane Poiseuille base flow profile of the form $U=1-y^2$ {for three-dimensional perturbations}. In Fig.~\ref{fig:5}, {we show the {most dominant} flow structures in terms of $\norm{\mathscr{H}_{\nabla}}_{\infty}^{-1}(k_x,k_z)$,  $\norm{\mathscr{H}_{\nabla}}_{\mu_{\mathbf{\Delta}_{nr}}}^{-1}(k_x,k_z)$ and $\norm{\mathscr{H}_{\nabla}}_{\mu_{\mathbf{\Delta}_{r}}}^{-1}(k_x,k_z)$,} for the Reynolds numbers $\Rey=690$ (Fig.~\ref{fig:5a}), $\Rey=5700$ (Fig.~\ref{fig:5b}) and $\Rey=6000$ (Fig.~\ref{fig:5c}). 
The Reynolds number $\Rey=690$ is the same as used in \citet{liu2021} and \citet{mushtaq2023} and is chosen to validate our results. {The Reynolds numbers $\Rey=5700$ and $\Rey=6000$ are chosen as one is below and one is above the critical Reynolds number, $\Rey_{c}=5772$, predicted by LST for plane Poiseuille flow.}
Over these {contour maps}, we overlay the modes from Table~\ref{tbl:1}. 
The most dominant mode is marked by an X symbol. If the most dominant mode is not denoted in Table \ref{tbl:1}, it is colored in cyan.
\begin{figure}
\centering
\begin{subfigure}[a]{1\textwidth}
\centering
\includegraphics[width=\textwidth]{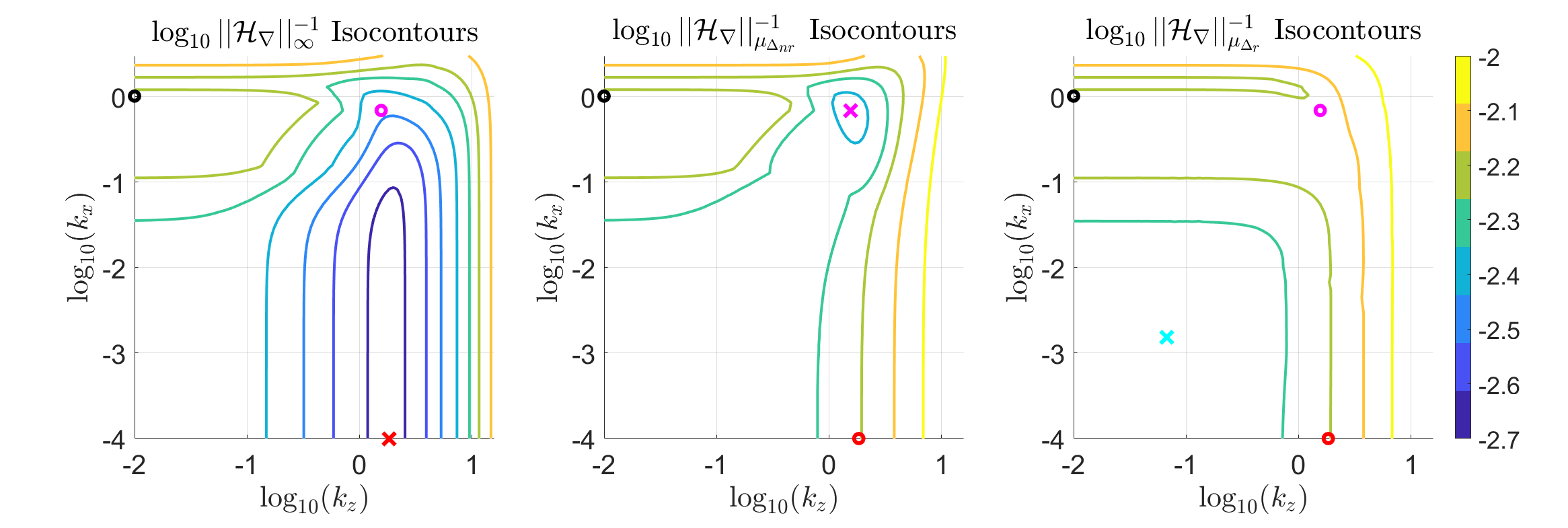}
\caption{$\Rey=690$}
\label{fig:5a}
\end{subfigure}
\hfill
\begin{subfigure}[b]{1\textwidth}
\centering
\includegraphics[width=\textwidth]{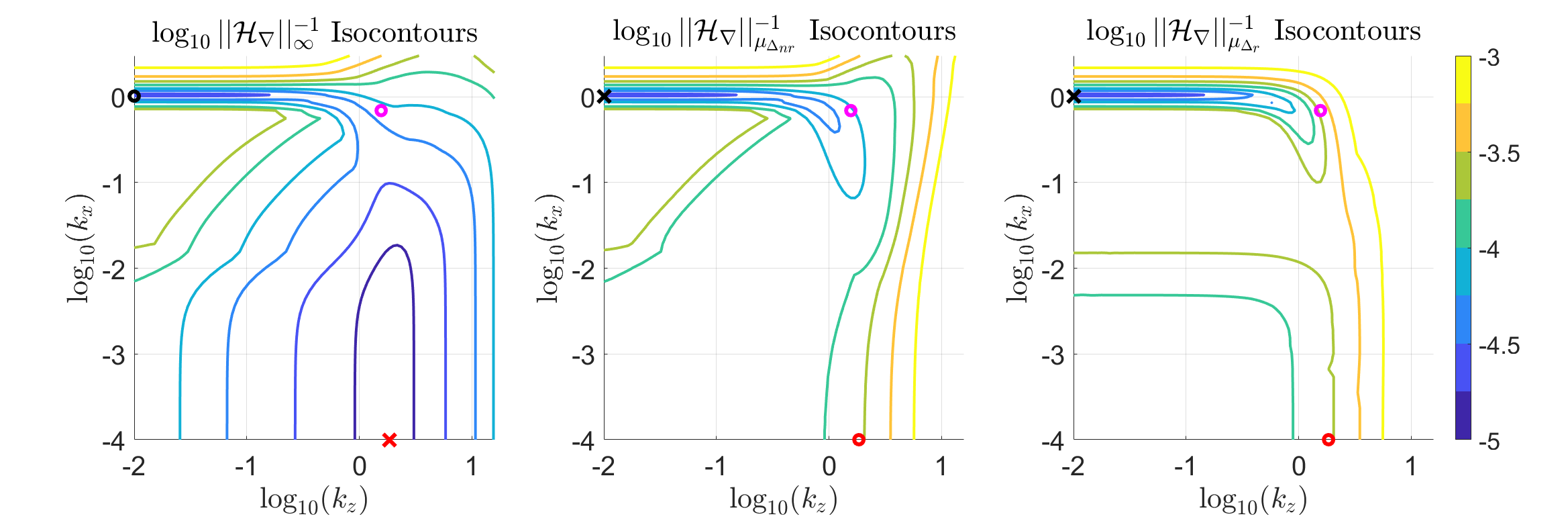}
\caption{$\Rey=5700$}
\label{fig:5b}
\end{subfigure}
\begin{subfigure}[c]{1\textwidth}
\centering
\includegraphics[width=\textwidth]{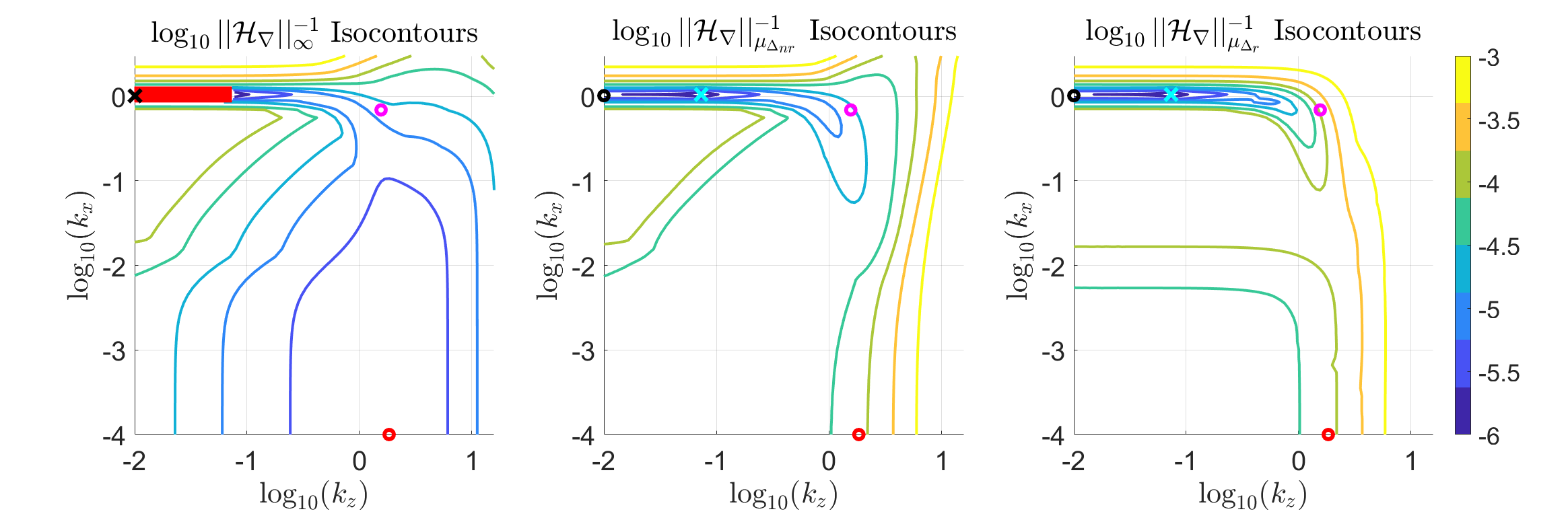}
\caption{$\Rey=6000$}
\label{fig:5c}
\end{subfigure}
\caption{Contour plots in logarithmic scale of (from left to right): $\norm{\mathscr{H}_{\nabla}}_{\infty}^{-1}$, $\norm{\mathscr{H}_{\nabla}}_{\mu_{\mathbf{\Delta}_{nr}}}^{-1}$, and $\norm{\mathscr{H}_{\nabla}}_{\mu_{\mathbf{\Delta}_{r}}}^{-1}$ at (a) $\Rey=690$, (b) $\Rey=5700$, and (c) $\Rey=6000$ for plane Poiseuille flow. {The {most dominant} mode is marked by an X {(cyan color indicates that the mode's $(k_x,k_z)$ value  is  not denoted in the table)}, whereas the other modes from Table~\ref{tbl:1} are marked by colored circles (magenta - DLR mode, black - TS mode, red - SPS mode). Results are shown for (a) $Re=690$, (b) $Re=5700$, (c) $Re=6000$.}}
 \label{fig:5}
\end{figure}
For the three Reynolds numbers that were tested, the relationship of the norm-like quantities in Eq. \eqref{eq:2.29}, i.e., {$\norm{\mathscr{H}_{\nabla}}_{\mu_{\mathbf{\Delta}_{r}}}(k_x,k_z) \leq \norm{\mathscr{H}_{\nabla}}_{\mu_{\mathbf{\Delta}_{nr}}}(k_x,k_z) \leq \norm{\mathscr{H}_{\nabla}}_{\infty}(k_x,k_z)$}, is satisfied as in the Couette case.

The use of structured approaches shows that the lowest threshold is governed by dominance of oblique modes rather than the streamwise elongated streaky structures in the unstructured case, similarly to the Couette base flow. Structured case results were found to better agree with the NLOP-based predictions study of   
\citep{farano2015hairpin}, in which it was shown that hairpin vortex structures (associated with oblique modes) can be the outcome of a nonlinear optimal growth process in plane Poiseuille flow, in a similar way as streaky structures can be the result of a linear optimal growth mechanism.

It was stated in \cite{liu2021}, "This result is consistent with findings of Reddy et al. \citep{reddy1998stability} showing that oblique waves require slightly less perturbation energy to trigger turbulence in plane Poiseuille flow than streamwise vortices." Similarly to the case of Couette base flow, herein we show that this conclusion is highly affected by the type of uncertainty that is considered and the Reynolds number. In the case of {$\norm{\mathscr{H}_{\nabla}}_{\mu_{\mathbf{\Delta}_{r}}}^{-1}$} {the smallest perturbation magnitudes} are associated to a wide range of modes, manifested as a uniform region {in the bottom left quadrant of the $k_x,k_z$ plane} without any obvious dominant oblique mode. Moreover, observing {$\norm{\mathscr{H}_{\nabla}}_{\mu_{\mathbf{\Delta}_{nr}}}^{-1}$} (non-repeated uncertainty) for $\Rey=5700$ (Fig.~\ref{fig:5b}) and $\Rey=6000$ (Fig.~\ref{fig:5c}) reveals that as in the case of Couette base flow, the {most dominant mode} is highly dependent on the Reynolds number. While in Fig.~\ref{fig:5a} the {most dominant mode} is the DLR mode (indicated by a magenta-colored circle) of $(k_x,k_z)=(1.561,0.692)$, in Fig.~\ref{fig:5b} the {most dominant mode} is the TS mode (indicated by a black-colored circle) of $(k_x,k_z)=(1.02,10^{-2})$. In Fig.~\ref{fig:5c}, where the Reynolds number is increased further to $\Rey=6000$, the {most dominant mode in the sense} of {$\norm{\mathscr{H}_{\nabla}}_{\mu_{\mathbf{\Delta}_{nr}}}^{-1}$} shifts again to {be an} oblique mode located between the DLR mode {and the} TS mode on the $k_x,k_z$ domain.

Increasing the Reynolds number from the subcritical value $\Rey=690$ to $\Rey=5700$ (as shown in Fig.~\ref{fig:5b}), leads to the dominance of the TS mode and slightly oblique TS modes (represented by a streamwise wave number of {$k_x \approx 1$} and a spanwise wave number of $k_z \ll 1$), which impose a very low threshold for instability to occur for unstructured and both structured cases. Further increasing the Reynolds number to $\Rey=6000$ (as shown in Fig.~\ref{fig:5c}) causes a difference in the {stability} characteristics between the structured approaches ($\norm{\mathscr{H}_{\nabla}}_{\mu_{\mathbf{\Delta}_{r}}}^{-1}$ and $\norm{\mathscr{H}_{\nabla}}_{\mu_{\mathbf{\Delta}_{nr}}}^{-1}$) and the unstructured ($\norm{\mathscr{H}_{\nabla}}_{\infty}^{-1}$) approaches. 
While in the contour maps of $\norm{\mathscr{H}_{\nabla}}_{\mu_{\mathbf{\Delta}_{r}}}^{-1}$ and $\norm{\mathscr{H}_{\nabla}}_{\mu_{\mathbf{\Delta}_{nr}}}^{-1}$, TS modes and slightly oblique modes govern the lowest threshold on perturbations up to a magnitude of $O(10^{-6})$, in the contour map of $\norm{\mathscr{H}_{\nabla}}_{\infty}^{-1}$ the imposed threshold governed by the same modes drops to an infinitesimally small value, i.e., the flow will lose stability for perturbations of any size (even asymptotically small), indicated by the red rectangle in Fig.~\ref{fig:5c}, which corresponds to $\norm{\mathscr{H}_{\nabla}}_{\infty}^{-1}\rightarrow 0$. 

Next, we study the effect of  Reynolds numbers on stability threshold over a wide range in plane Poiseuille flow.  In Fig.~\ref{fig:6} we show the evolution of  $\norm{\mathscr{H}_{\nabla}}_{\mu_{\mathbf{\Delta}_{r}}}^{-1}$}, $\norm{\mathscr{H}_{\nabla}}_{\mu_{\mathbf{\Delta}_{nr}}}^{-1}$ and $\norm{\mathscr{H}_{\nabla}}_{\infty}^{-1}$ for the chosen modes of interest as a function of the Reynolds number in the interval $\Rey \in [100,10000]$. The trends in each graph are accompanied by insets of zoomed-in views for a limited Reynolds number range at different locations, revealing the interesting interplay between the dominance of modes for $\norm{\mathscr{H}_{\nabla}}_{\mu_{\mathbf{\Delta}_{r}}}^{-1}$, $\norm{\mathscr{H}_{\nabla}}_{\mu_{\mathbf{\Delta}_{nr}}}^{-1}$ and $\norm{\mathscr{H}_{\nabla}}_{\infty}^{-1}$. In particular, four insights are gained from these plots.
\begin{figure}
\centering
\begin{overpic}[width=1\linewidth]{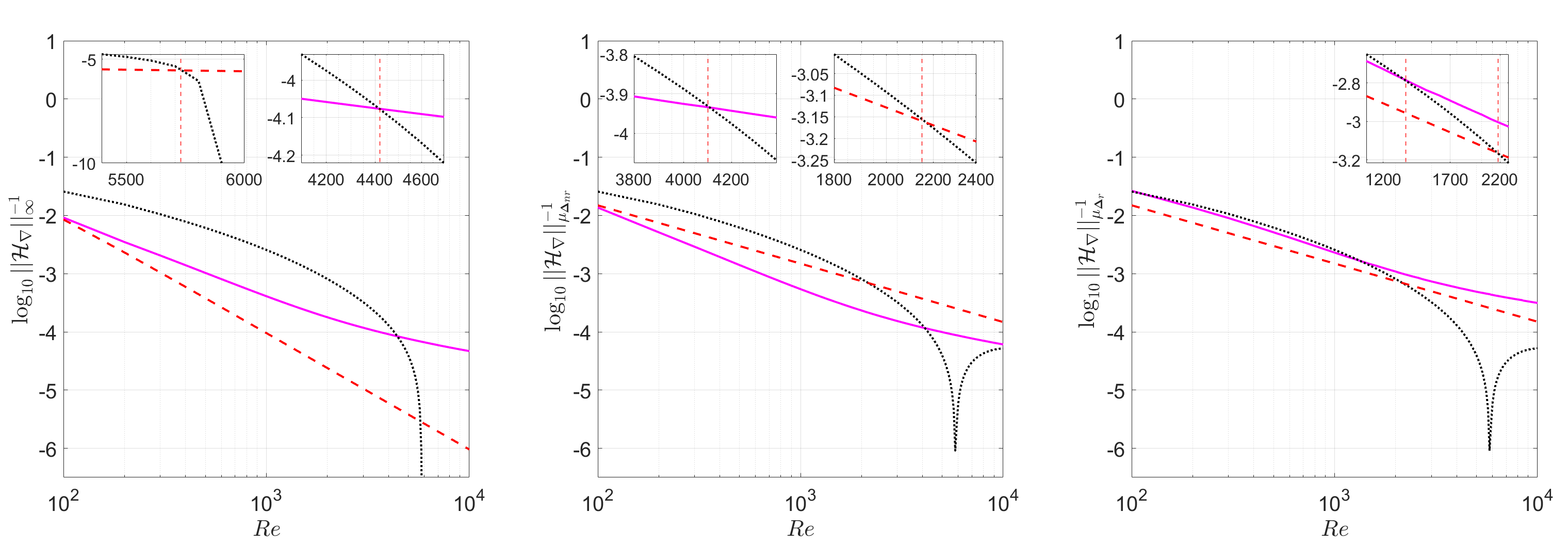}
\put(0, 28){\small (a)
\hspace{4.05cm} (b) \hspace{4.05cm} (c)}
\end{overpic}
 
\caption{Evolution curves of imposed thresholds due to flow structures associated to pre-selected modes of interest denoted in Table~\ref{tbl:1} (magenta solid - DLR mode, black dotted - TS mode, {red} dashed - SPS mode) for plane Poiseuille base flow as a function of the Reynolds number in terms of (a) $\norm{\mathscr{H}_{\nabla}}_{\infty}^{-1}$, (b) $\norm{\mathscr{H}_{\nabla}}_{\mu_{\mathbf{\Delta}_{nr}}}^{-1}$ and (c) $\norm{\mathscr{H}_{\nabla}}_{\mu_{\mathbf{\Delta}_{r}}}^{-1}$.} 
\label{fig:6}
\end{figure}
 
First, for the unstructured case of $\norm{\mathscr{H}_{\nabla}}_{\infty}^{-1}$, we observe that the SPS mode (red) is the most dominant up until very close to $\Rey=5772$, the critical Reynolds number that is predicted by LST  \citep{orszag1971accurate,schmid2002stability}. The TS mode (black) becomes most dominant, imposing the lowest threshold at proximity and after the critical Reynolds number. When using both of the structured approaches ($\norm{\mathscr{H}_{\nabla}}_{\mu_{\mathbf{\Delta}_{nr}}}^{-1}$ and $\norm{\mathscr{H}_{\nabla}}_{\mu_{\mathbf{\Delta}_{r}}}^{-1}$), the TS mode becomes the most dominant mode at a lower Reynolds number compared with the unstructured analysis. In the case of $\norm{\mathscr{H}_{\nabla}}_{\mu_{\mathbf{\Delta}_{nr}}}^{-1}$, the TS mode becomes the most dominant after $\Rey \approx 4100$ (marked by a dashed red line in Fig.~\ref{fig:6}b) and in the case of {$\norm{\mathscr{H}_{\nabla}}_{\mu_{\mathbf{\Delta}_{r}}}^{-1}$}, it becomes dominant at $\Rey \approx 2175$ (marked by a dashed red line in Fig.~\ref{fig:6}c). A recent study by \citep{Huang_Gao_Gao_Xi_2024} showed that in the case of two-dimensional plane Poiseuille flow, TS waves are the optimal perturbation and can cause a sustained transition to turbulence for $Re \geq 2400$, in agreement with the results obtained for the case of repeated blocks. considering structured analysis methods.
The non-repeated case ($\norm{\mathscr{H}_{\nabla}}_{\mu_{\mathbf{\Delta}_{nr}}}^{-1}$) and repeated case ($\norm{\mathscr{H}_{\nabla}}_{\mu_{\mathbf{\Delta}_{r}}}^{-1}$) differ mainly in the {stability threshold that {are} imposed by the DLR mode (magenta)}, which represents an oblique wave mode. The DLR mode {imposes a lower stability threshold in the case of $\norm{\mathscr{H}_{\nabla}}_{\mu_{\mathbf{\Delta}_{nr}}}^{-1}$ compared with $\norm{\mathscr{H}_{\nabla}}_{\mu_{\mathbf{\Delta}_{r}}}^{-1}$} for the entire range of Reynolds numbers that were tested, similarly to the Couette base flow. 

Second, the TS mode curve of $\norm{\mathscr{H}_{\nabla}}_{\infty}^{-1}$ (in Fig.~\ref{fig:6}a) exhibits an asymptotic behavior where it approaches {zero} at $\Rey=5772$. The fact that $\norm{\mathscr{H}_{\nabla}}_{\infty}^{-1}$ approaches zero for the critical mode of actuation and the critical Reynolds number predicted by LST is, of course, not a coincidence. Similarly to Fig.~\ref{fig:12}, the results of $\norm{\mathscr{H}_{\nabla}}_{\infty}^{-1} \rightarrow 0$ correspond to the flow being unstable even for asymptotically small perturbations, and thus, results that match the predictions of LST are to be expected.
It is important to note that even though close to the critical Reynolds number and beyond, a TS wave sets the lowest threshold for disturbance to trigger transition, it does not necessarily translate to transition being caused by a TS mode. Any mode or combination of modes that surpasses the corresponding stability threshold will destabilize the flow system, which consequently may trigger transition. For example, due to a transient growth scenario or the presence of finite-size disturbances in a noisy flow environment. 

Third, in the graphs of $\norm{\mathscr{H}_{\nabla}}_{\mu_{\mathbf{\Delta}_{nr}}}^{-1}$ (Fig.~\ref{fig:6}b) and $\norm{\mathscr{H}_{\nabla}}_{\mu_{\mathbf{\Delta}_{r}}}^{-1}$ (Fig.~\ref{fig:6}c), the TS mode curve {reaches a finite value minimum} {exactly} at the critical Reynolds number value that is predicted by LST. After this {minimum}, the values of both $\norm{\mathscr{H}_{\nabla}}_{\mu_{\mathbf{\Delta}_{nr}}}^{-1}$ and $\norm{\mathscr{H}_{\nabla}}_{\mu_{\mathbf{\Delta}_{r}}}^{-1}$ increase. This {minimum}  has a finite value in contrast to the behavior of $\norm{\mathscr{H}_{\nabla}}_{\infty}^{-1}$ - where the TS mode threshold curve {asymptotes towards zero}. {We explain this observation in more detail in{~\S}\ref{sec:5}, where we will show the associated stability diagrams for both unstructured and structured cases.
At higher post-critical Reynolds numbers, the most dominant mode shifts to an oblique mode as the Reynolds number increases.} 
Lastly, the curve trends of the dominant modes that arise in our analysis agree with the relationship in Eq.~\eqref{eq:2.29}, for all Reynolds numbers.

\subsection{Blasius base flow}
\label{sec:4.3}

Here, we analyze the effect of the Reynolds number on the stability and the dominant structures in Blasius flow. The Blasius base flow profile is obtained numerically by solving the Blasius equation: $f''' + ff'' = 0$ \citep{schlichting2016boundary}, where $f'$ is the normalized base flow velocity varying with a function of the similarity variable $\eta=y/\delta^*$, and $\delta^*$ is a displacement thickness. The boundary conditions used for the solution of the Blasius equation are $f(0)=f'(0)=0$ and $\lim_{\eta\to\infty} f'(\eta) = 1$. In Fig.~\ref{fig:7}, we show the most dominant flow structures in terms of $\norm{\mathscr{H}_{\nabla}}_{\infty}^{-1}(k_x,k_z)$,  $\norm{\mathscr{H}_{\nabla}}_{\mu_{\mathbf{\Delta}_{nr}}}^{-1}(k_x,k_z)$ and $\norm{\mathscr{H}_{\nabla}}_{\mu_{\mathbf{\Delta}_{r}}}^{-1}(k_x,k_z)$,} for the Reynolds numbers $\Rey=400$ (Fig.~\ref{fig:7a}), $\Rey=500$ (Fig.~\ref{fig:7b}) and $\Rey=530$ (Fig.~\ref{fig:7c}).
The Reynolds numbers $\Rey=500$ and $\Rey=530$ are chosen as one is below and one is above the critical Reynolds number, $\Rey_c=520$, predicted by LST for Blasius flow \citep{schmid2002stability}. The Reynolds number $\Rey=400$ was chosen to observe the stability characteristics of the flow
at a subcritical Reynolds number, which is further away from the critical value.
Over these contour maps, we overlay the modes from Table~\ref{tbl:1}. 
{The most dominant mode is marked by an X symbol. If the most dominant mode is not denoted in Table \ref{tbl:1}, it is colored in cyan.}
For the three Reynolds numbers that were tested the relationship between norm-like quantities in Eq. \eqref{eq:2.29} ($\norm{\mathscr{H}_{\nabla}}_{\mu_{\mathbf{\Delta}_{r}}}(k_x,k_z) \leq \norm{\mathscr{H}_{\nabla}}_{\mu_{\mathbf{\Delta}_{nr}}}(k_x,k_z) \leq \norm{\mathscr{H}_{\nabla}}_{\infty}(k_x,k_z)$) is satisfied, similarly to the cases of Couette flow and {plane Poiseuille flow.  This in turn shows that the relation in Eq. \eqref{eq:2.29}  is invariant to the base flow geometry considered (at least for the cases we tested).} 
\begin{figure}
\centering
\begin{subfigure}[a]{1\textwidth}
\centering
\includegraphics[width=\textwidth]{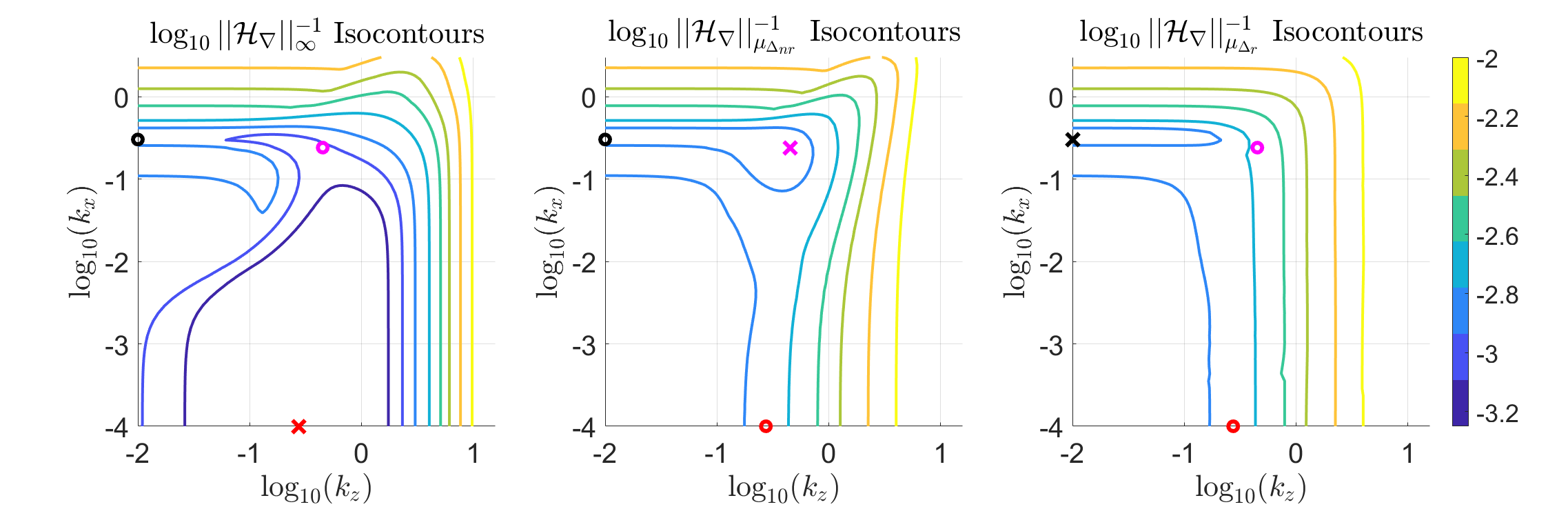}
\caption{$\Rey=400$}
\label{fig:7a}
\end{subfigure}
\hfill
\begin{subfigure}[b]{1\textwidth}
\centering
\includegraphics[width=\textwidth]{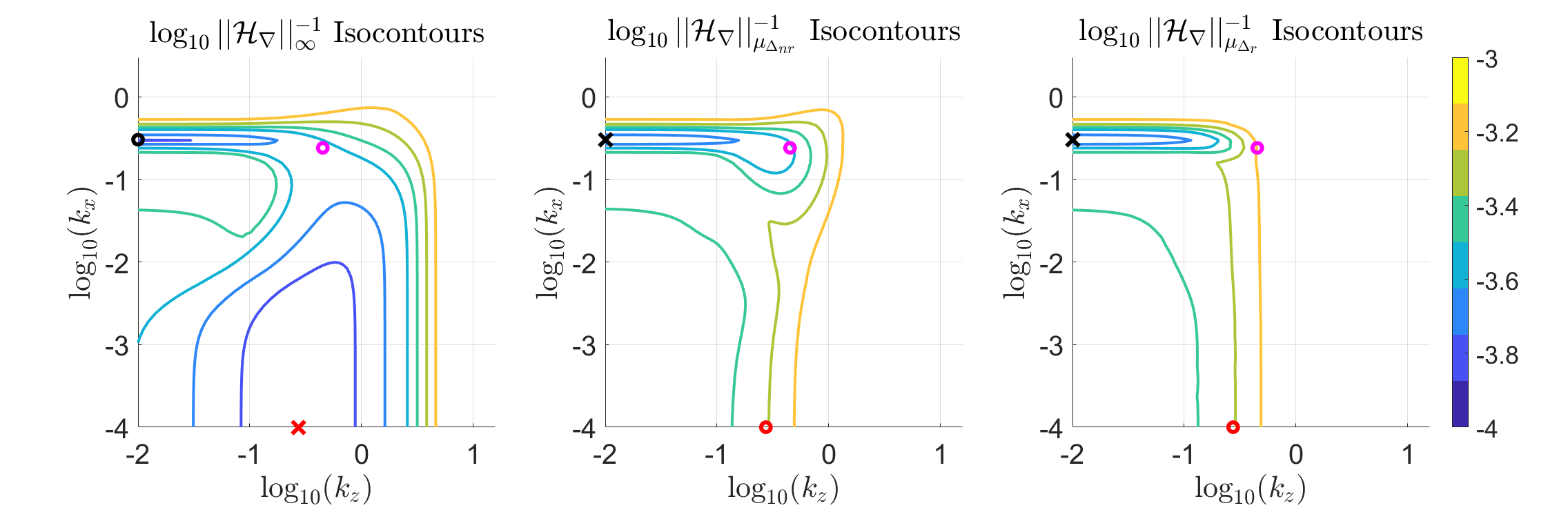}
\caption{$\Rey=500$}
\label{fig:7b}
\end{subfigure}
\hfill
\begin{subfigure}[c]{1\textwidth}
\centering
\includegraphics[width=\textwidth]{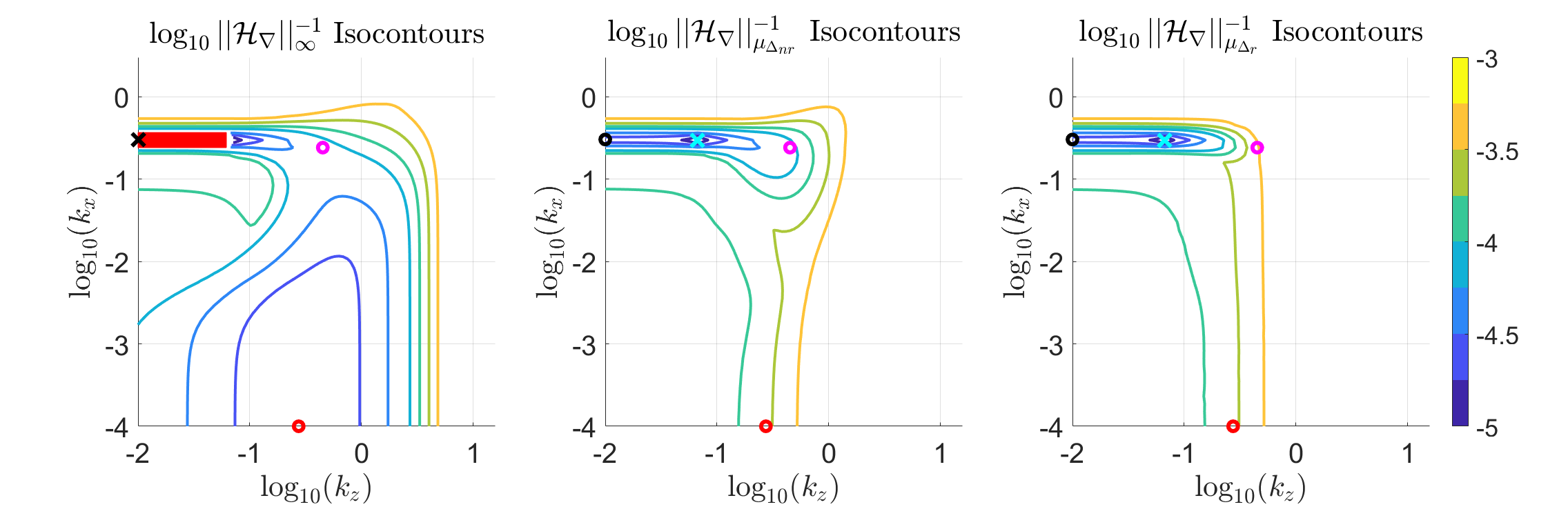}
\caption{$\Rey=530$}
\label{fig:7c}
\end{subfigure}
\caption{Contour plots in logarithmic scale of $\norm{\mathscr{H}_{\nabla}}_{\infty}^{-1}$, $\norm{\mathscr{H}_{\nabla}}_{\mu_{\mathbf{\Delta}_{nr}}}^{-1}$, and $\norm{\mathscr{H}_{\nabla}}_{\mu_{\mathbf{\Delta}_{r}}}^{-1}$ at (a) $\Rey=400$, (b) $\Rey=500$, and (c) $\Rey=530$ for Blasius flow. The {most dominant} mode is marked by an X (cyan color indicates that the mode's $(k_x,k_z)$ value is not denoted in the table), whereas the other modes from Table~\ref{tbl:1} are marked by colored circles (magenta - DLR mode, black - TS mode, red - SPS mode). Results are shown for (a) $Re=400$, (b) $Re=500$, (c) $Re=530$.}
\label{fig:7}
\end{figure}

The most dominant mode of actuation is highly dependent on the Reynolds number and the type of structured uncertainty considered (repeated versus non-repeated). In the case of $\norm{\mathscr{H}_{\nabla}}_{\mu_{\mathbf{\Delta}_{nr}}}^{-1}$, the {most dominant} mode at $\Rey=400$ (Fig.~\ref{fig:7a}) is the DLR mode, which is an oblique mode. However, increasing the Reynolds number causes the {most dominant} mode to change, as can be observed in Fig.~\ref{fig:7b} and Fig.~\ref{fig:7c}. When using repeated uncertainty ({$\norm{\mathscr{H}_{\nabla}}_{\mu_{\mathbf{\Delta}_{r}}}^{-1}$}), the DLR mode and neighboring modes {have} significantly {lower stability thresholds}.

For the Reynolds numbers below $\Rey_c$, the quantities $\norm{\mathscr{H}_{\nabla}}_{\mu_{\mathbf{\Delta}_{r}}}^{-1}$, $\norm{\mathscr{H}_{\nabla}}_{\mu_{\mathbf{\Delta}_{nr}}}^{-1}$ and $\norm{\mathscr{H}_{\nabla}}_{\infty}^{-1}$ all display similar characteristics of TS modes and slightly oblique TS modes (represented by a streamwise wave number of $k_x \approx 0.3$ and a spanwise wave number of $k_z \ll 1$) {being the most dominant}. At $\Rey=400$ (displayed in Fig.~\ref{fig:7a}), there is a band of TS and slightly oblique TS modes with relatively small stability thresholds for both the unstructured and structured approaches. When increasing the Reynolds number to $\Rey=500$ (as shown in Fig.~\ref{fig:7b}), the TS and slightly oblique TS modes become unstable for significantly smaller perturbations and are the most dominant for both $\norm{\mathscr{H}_{\nabla}}_{\mu_{\mathbf{\Delta}_{r}}}^{-1}$ and $\norm{\mathscr{H}_{\nabla}}_{\mu_{\mathbf{\Delta}_{nr}}}^{-1}$. In the case of $\norm{\mathscr{H}_{\nabla}}_{\infty}^{-1}$, these modes have slightly higher stability thresholds than the dominant streamwise elongated streak mode that imposes the lowest threshold. Further increasing the Reynolds number to $\Rey=530$ (as shown in Fig.~\ref{fig:7c}) causes a difference in the {stability characteristics} of TS and slightly oblique TS modes between the structured approaches ($\norm{\mathscr{H}_{\nabla}}_{\mu_{\mathbf{\Delta}_{r}}}^{-1}$ and $\norm{\mathscr{H}_{\nabla}}_{\mu_{\mathbf{\Delta}_{nr}}}^{-1}$) and the unstructured ($\norm{\mathscr{H}_{\nabla}}_{\infty}^{-1}$) approach. In the contour map of $\norm{\mathscr{H}_{\nabla}}_{\infty}^{-1}$, the region of slightly oblique TS modes becomes unstable even for asymptotically small perturbations, as in these regions the stability threshold approaches zero (marked as red regions in the contour map).
In the structured cases, the most dominant mode of $\norm{\mathscr{H}_{\nabla}}_{\mu_{\mathbf{\Delta}_{r}}}^{-1}$ and $\norm{\mathscr{H}_{\nabla}}_{\mu_{\mathbf{\Delta}_{nr}}}^{-1}$ shifts to a larger value of $k_z$, i.e., the mode becomes more oblique, similarly to the case of plane Poiseuille flow. 

\begin{figure}
\centering
\begin{overpic}[width=1\linewidth]{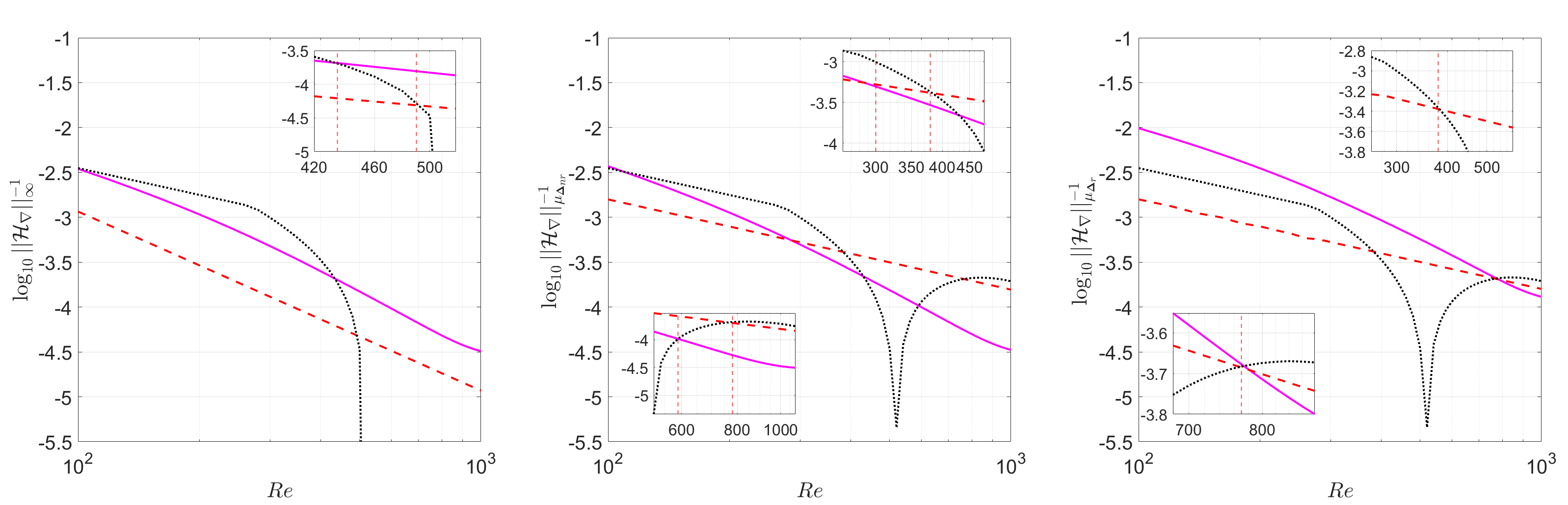}
\put(0, 30){\small (a) \hspace{4.05cm} (b) \hspace{4.05cm} (c)}
\end{overpic}

\caption{Evolution curves of imposed thresholds due to flow structures associated to pre-selected modes of interest denoted in Table~\ref{tbl:1} (magenta solid - DLR mode, black dotted - TS mode, {red} dashed - SPS mode) for Blasius base flow as a function of the Reynolds number in terms of (a) $\norm{\mathscr{H}_{\nabla}}_{\infty}^{-1}$, (b) $\norm{\mathscr{H}_{\nabla}}_{\mu_{\mathbf{\Delta}_{nr}}}^{-1}$ and (c) $\norm{\mathscr{H}_{\nabla}}_{\mu_{\mathbf{\Delta}_{r}}}^{-1}$.}
 
\label{fig:8}
\end{figure}

Next, we explore the stability thresholds variation over a wider range of Reynolds numbers. In Fig.~\ref{fig:8} the evolution of $\norm{\mathscr{H}_{\nabla}}_{\mu_{\mathbf{\Delta}_{r}}}^{-1}$, $\norm{\mathscr{H}_{\nabla}}_{\mu_{\mathbf{\Delta}_{nr}}}^{-1}$ and $\norm{\mathscr{H}_{\nabla}}_{\infty}^{-1}$ for the chosen modes of interest as a function of the Reynolds number in the interval $\Rey \in [100,1000]$ is shown. The trends in each graph are accompanied by insets of zoomed-in views for a certain Reynolds number range at different locations where intersection between different mode curves are observed for $\norm{\mathscr{H}_{\nabla}}_{\mu_{\mathbf{\Delta}_{r}}}^{-1}$, $\norm{\mathscr{H}_{\nabla}}_{\mu_{\mathbf{\Delta}_{nr}}}^{-1}$ and $\norm{\mathscr{H}_{\nabla}}_{\infty}^{-1}$. In particular, these plots reveal the following observations.

First, for the unstructured case of $\norm{\mathscr{H}_{\nabla}}_{\infty}^{-1}$, we observe that the SPS mode (red) is the most dominant up to $\Rey=520$, the critical Reynolds number as predicted by LST \citep{schmid2002stability}. After this critical Reynolds number, the TS mode (black) becomes the most dominant mode. When using both structured analysis methods ($\norm{\mathscr{H}_{\nabla}}_{\mu_{\mathbf{\Delta}_{nr}}}^{-1}$ and $\norm{\mathscr{H}_{\nabla}}_{\mu_{\mathbf{\Delta}_{r}}}^{-1}$), the TS mode becomes the most dominant mode at a lower-Reynolds-number values as we detail next. In the case of $\norm{\mathscr{H}_{\nabla}}_{\mu_{\mathbf{\Delta}_{nr}}}^{-1}$, the TS mode becomes the most dominant mode at $\Rey \approx 430$ (marked by a dashed red line in Fig.~\ref{fig:8}b) and in the case of $\norm{\mathscr{H}_{\nabla}}_{\mu_{\mathbf{\Delta}_{r}}}^{-1}$, it becomes {the most dominant} as early as $\Rey \approx 370$ (marked by a dashed red line in Fig.~\ref{fig:8}c). The non-repeated case ($\norm{\mathscr{H}_{\nabla}}_{\mu_{\mathbf{\Delta}_{nr}}}^{-1}$) and the repeated case ($\norm{\mathscr{H}_{\nabla}}_{\mu_{\mathbf{\Delta}_{r}}}^{-1}$) differ only in the magnitude of perturbations that cause instability of the DLR mode (magenta), which represents an oblique wave mode. The perturbation magnitude that causes instability of the DLR mode is slightly smaller for $\norm{\mathscr{H}_{\nabla}}_{\mu_{\mathbf{\Delta}_{nr}}}^{-1}$ compared with $\norm{\mathscr{H}_{\nabla}}_{\mu_{\mathbf{\Delta}_{r}}}^{-1}$ for all of the tested range of Reynolds numbers, similarly to the cases of Couette and plane Poiseuille flow, and in agreement with Eq.~\eqref{eq:2.29}.

Second, in the graph of $\norm{\mathscr{H}_{\nabla}}_{\infty}^{-1}$ (Fig.~\ref{fig:8}a), there is an asymptotic behavior of the TS mode threshold curve that is approaching {zero} at $\Rey=520$. This asymptotic behavior is explained by LST, which predicts that the chosen TS mode causes the Blasius base flow to become unstable for $\Rey > 520$ \citep{schmid2002stability} for asymptotically small perturbations. 
In the graphs of $\norm{\mathscr{H}_{\nabla}}_{\mu_{\mathbf{\Delta}_{nr}}}^{-1}$ (Fig.~\ref{fig:8}b) and $\norm{\mathscr{H}_{\nabla}}_{\mu_{\mathbf{\Delta}_{r}}}^{-1}$ (Fig.~\ref{fig:8}c), there is a finite amplitude minimum at the critical Reynolds number predicted by LST, reminiscent of the one observed for the case of plane Poiseuille flow. After this minimum, the value of the TS modes in both $\norm{\mathscr{H}_{\nabla}}_{\mu_{\mathbf{\Delta}_{nr}}}^{-1}$ and $\norm{\mathscr{H}_{\nabla}}_{\mu_{\mathbf{\Delta}_{r}}}^{-1}$ increase. As in the case of plane Poiseuille, this minimum corresponds to the flow being extremely sensitive to perturbations and very close to the verge of instability. However, this {minimum} still reaches a finite value in contrast to the behavior of $\norm{\mathscr{H}_{\nabla}}_{\infty}^{-1}$ (that {approaches zero}) due to the imposed structure of the nonlinear term in the Navier-Stokes equation. The existence of this minimum in the graph of the TS mode will be explained similarly to the case of plane Poiseuille flow in more detail in section ~\ref{sec:5}, where we will show the associated stability diagrams for both unstructured and structured cases.

Third, using structured analysis creates a much more complicated interplay between the different modes than in the unstructured analysis. In detail, for $\norm{\mathscr{H}_{\nabla}}_{\mu_{\mathbf{\Delta}_{nr}}}^{-1}$ (Fig.~\ref{fig:8}b), the SPS mode is the most dominant for $\Rey < 255$ and, for higher Reynolds numbers, the most dominant mode is the DLR mode, in the range $255 < \Rey < 430$. The TS mode is {most dominant} for $430 < \Rey < 590$ and, for higher Reynolds numbers, the DLR mode is again the most dominant. For $\norm{\mathscr{H}_{\nabla}}_{\mu_{\mathbf{\Delta}_{r}}}^{-1}$ (Fig.~\ref{fig:8}c), the SPS mode is the most dominant for $\Rey < 367$. The TS mode is most dominant for $367 < \Rey < 770$, and for higher Reynolds numbers, the DLR mode becomes the most dominant. 
The results for the repeated blocks analysis in Fig.~\ref{fig:8}c show that after $Re=800$  the oblique DLR mode dominates the flow, setting a threshold value, which is the lowest but very close to the SPS mode curve for both structured cases. Support for such results may be found in the experimental study by \cite{elofsson2000615}, where it was noted that for the post-critical transition process in the Blasius boundary layer, the combined effect of non-modal growth of streaks and a second stage with exponential growth of oblique waves is responsible for the initiation of the final breakdown stage.
The intricate relationship between the different modes that set the stability threshold, as depicted in Fig.~\ref{fig:8}, highlights the necessity of analyzing a broad range of Reynolds numbers. Limiting the examination to a single Reynolds number would result in overlooking significant physical characteristics of the flow system in this case.
 
Lastly, we note that for all Reynolds numbers, the curve trends of the imposed thresholds that are set by the most dominant modes agree with the relationship we derived in Eq.~\eqref{eq:2.29}. 
As explained in Section \ref{sec:2.5}, $\norm{\mathscr{H}_{\nabla}}_{\mu_{\mathbf{\Delta}_{r}}}^{-1}$ is the best approximation available to us for $\norm{\mathscr{H}_{\nabla}}_{\mu_{\mathbf{\Delta}_{u}}}^{-1}$, which incorporates the accurate structure of the nonlinear term in the NSE. Thus, we expect $\norm{\mathscr{H}_{\nabla}}_{\mu_{\mathbf{\Delta}_{r}}}^{-1}$ to be the bound on velocity perturbations, which is the closest to the accurate value $\norm{\mathscr{H}_{\nabla}}_{\mu_{\mathbf{\Delta}_{u}}}^{-1}$
from a stability viewpoint. As denoted in previous sections, there is no guarantee that it represents more accurately the contour plots in the wavenumber domain compared with the $\norm{\mathscr{H}_{\nabla}}_{\mu_{\mathbf{\Delta}_{nr}}}^{-1}$ case.
Using the formulation of stability with the small gain theorem shows that instability can occur in Blasius flow at any Reynolds number. Specifically, instability can occur at smaller Reynolds numbers than the critical value predicted via LST, $\Rey=520$,  given that perturbations of finite magnitude are present in the flow, surpassing the threshold provided by our stability criterion. 

The results presented here can be used to explain experimental evidence of subcritical transition in flat plate boundary layers. In the work of \cite{asai1995boundary},
subcritical transition is triggered by energetic
hairpin eddies. The boundary layer is found to be stable below $\Rey_x=2.8\times10^{4}$, which is equivalent to $\Rey=288$ (based on the displacement thickness) for a perturbation magnitude of $0.3$~\% (as shown in \citet{asai1995boundary} (Fig.~3) at $x \approx 100$~mm that correspond to this Reynolds number) at the leading edge (for the case of weak forcing). This magnitude corresponds to a value of $10^{-2.5}$, which is in relatively good agreement with our analysis for the structured two cases, where the imposed threshold by the most dominant mode is $O(10^{-3})$, which is the SPS mode at this Reynolds number. {Subcritical transition induced by finite-size perturbations is also observed in DNS studies, such as \cite{cherubini2011minimal} and \cite{vavaliaris2020optimal}, where it was found that a finite threshold value exists for $\Rey=300$ and $\Rey=275$, respectively. In particular, the energy threshold for transition was computed in \cite{vavaliaris2020optimal}, where a value of $E_c=3.64\times 10^{-2}$ was obtained. This value refers to the combined energy of the velocity perturbation field inside the entire spatial domain, whereas in the current work, we focus only on the specific perturbation of the largest magnitude. For that reason, a comparison between the results can be performed by dividing the energy by the volume of the computational domain reported in \cite{vavaliaris2020optimal}, to obtain an energy density value of $1.2\times10^{-8}$. The square root of this value corresponds to the average magnitude of localized perturbations in the flow, which turns out to be $1.1\times10^{-4}$. While the value of the largest perturbation magnitude is not provided in their work, it is reasonable to assume that it is significantly larger than the average value computed here. Thus, a maximal perturbation magnitude of $O(10^{-3})$, which was found to be the stability threshold using our methodology, seems reasonable. 

\section{Critical Reynolds number prediction and comparison with literature}
\label{sec:5}

This section focuses on comparing our stability criteria with {LST} and on validating our analysis of the predicted critical Reynolds number using results from reported experimental and DNS studies. The following two types of analyses are conducted for Couette, plane Poiseuille, and Blasius flows:
\begin{enumerate}
\item We draw a direct comparison between our stability analysis approach and predictions
of {LST}, including the
reproduction of stability diagrams.
\item We determine the critical Reynolds number using our stability criterion and the critical perturbation energy required for transition, and compare them with experimental and DNS results from available literature.
\end{enumerate}

To draw a direct comparison between our stability analysis approach and predictions
of LST, we consider here only two-dimensional modes with $k_z=0$ for two reasons: (i) these modes were shown to be dominant close to critical Reynolds numbers in~\S\ref{sec:2.3}. (ii) to compare our stability diagram with LST results, which are usually obtained for two-dimensional modes,  which, according to LST, yield the lowest critical Reynolds number \citep{schmid2002stability}. In addition, only the non-repeated block case is considered for the structured {analysis}, since we found no significant difference for $k_z=0$ compared with the repeated-blocks case (see appendix~\ref{sec:App_numeric}), where the non-repeated method is more numerically efficient to compute. For each flow case, we also validate our analysis for a wide range of Reynolds numbers with results from literature, by computation of the critical perturbation energy that causes instability, defined as $E_c=A_c^2/2$, where $A_c$ is the perturbation magnitude (herein we used common notation from the literature).

\subsection{Couette base flow}
\label{sec:5.1}
In Fig.~\ref{fig:9} we show an example of the stability diagram in the $\Rey - k_x$ domain, showing only a single isocontour for a threshold value equal to $10^{-2.43}=3.7\times10^{-3}$  for  $\norm{\mathscr{H}_{\nabla}}_{\infty}^{-1}$ at $k_z=0$. 
The $(\Rey,k_x)$ plane is divided into two regions: the stable region is found to the left of the isocontour given the presence of perturbations of magnitude below $3.7\times10^{-3}$, while the unstable region is found to its right, i.e., at higher Reynolds numbers. In our case,  the critical Reynolds number for this perturbation magnitude is defined as the smallest Reynolds number for which the perpendicular line (denoted by the red dashed line in the figure) is tangent to the contour of the threshold $\norm{\mathscr{H}_{\nabla}}_{\infty}^{-1}=3.7\times10^{-3}$.  In this case, the critical Reynolds number turns out to be $\Rey_c=325$ for perturbations that are $0.37\%$ of the characteristic base flow velocity (in the Couette case, it is half the velocity difference between the walls). This critical Reynolds number matches the experimental results of \cite{dauchot1995finite} and simulations by  \cite{kreiss1994bounds}. In \cite{dauchot1995finite},  it was noted that very large perturbations are introduced via transverse jet flow injected into the laminar flow, where the information on the exact magnitude was not available. In  \cite{kreiss1994bounds},  the relation $A_c\sim \Rey^{-1}$ is established for the critical perturbation amplitude in Couette flow. Considering the Reynolds number $\Rey=325$, this relation yields a critical value of $\Rey^{-1}= 3.1\times 10^{-3}$ (upper bound on the disturbance by their analysis), which is similar to the value we predict using $\norm{\mathscr{H}_{\nabla}}_{\infty}^{-1}$ and also for the structured case, which yields the same results as shown in Fig.~\ref{fig:10}.
We note that in \cite{kreiss1994bounds}, there is a large prefactor of $\sqrt{2000}$ that multiplies $Re^{-1}$. As noted by \cite{kreiss1994bounds}, they compute an upper bound for the perturbation magnitude and not necessarily the exact value. In \cite{kreiss1994bounds}, only one initial condition is considered (two counter-rotating vortices added to Stokes' noise), and thus their result can be significantly higher than the bound for the least stable initial condition. This highlights a significant advantage of our analysis, as we do not require exhaustive simulations of many different initial conditions to obtain the optimal bound on the perturbation magnitude.

\begin{figure}
    \centering
    \includegraphics[width=0.7\linewidth]{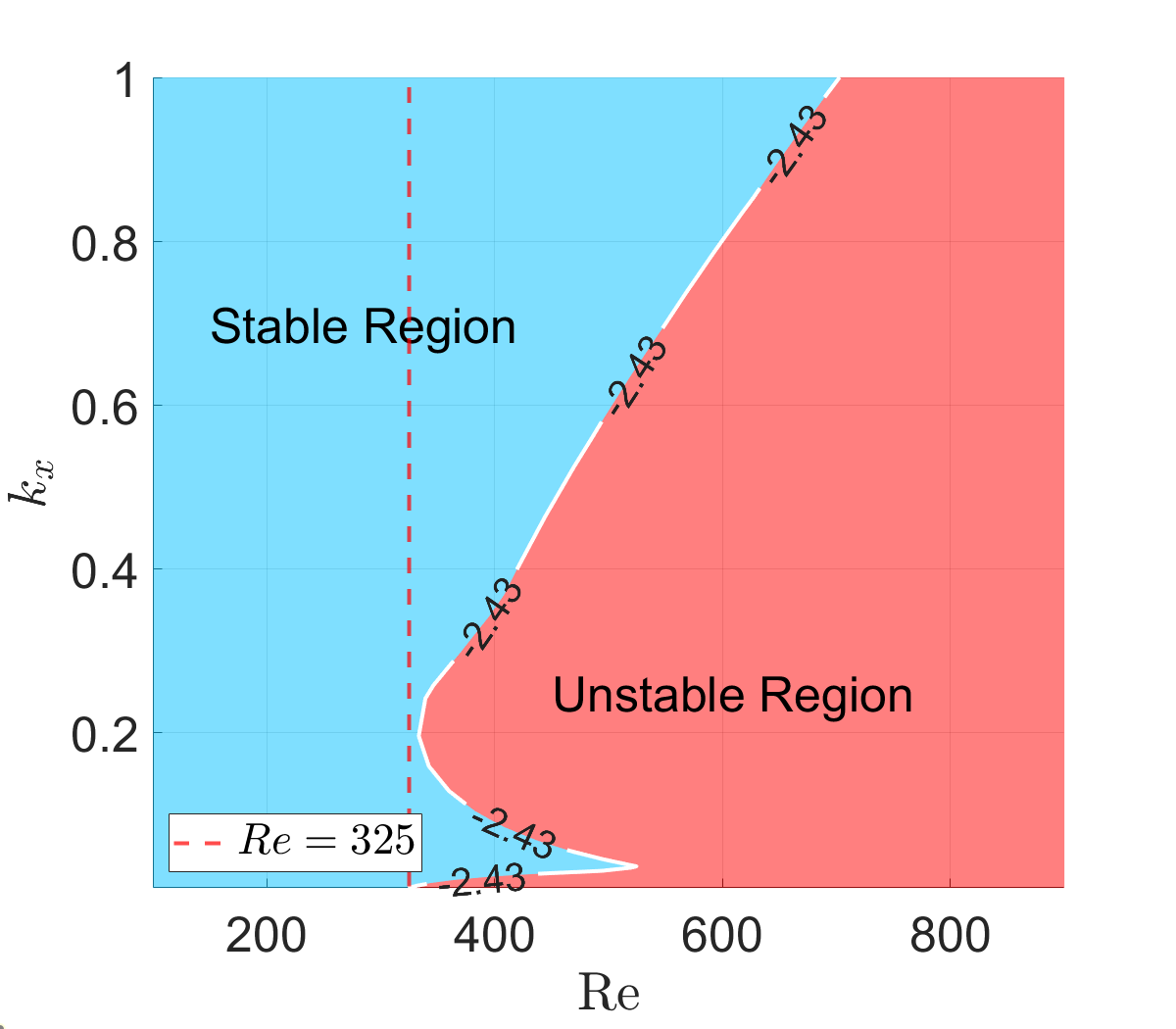}
    \caption{Contour plot of $\norm{\mathscr{H}_{\nabla}}_{\infty}^{-1}=10^{-2.43}$ and stability interpretation based on Eq. \eqref{eq:2.20}  at $k_z=0$.}
    \label{fig:9}
\end{figure}
 
The analysis presented here can be extended to any range of threshold values, as shown in Fig.~\ref{fig:10}, where we show stability diagrams in {the $\Rey-k_x$} domain at $k_z=0$, showing several isocontours for different threshold values for both cases: $\norm{\mathscr{H}_{\nabla}}_{\infty}^{-1}$  and $\norm{\mathscr{H}_{\nabla}}_{\mu_{\mathbf{\Delta}_{nr}}}^{-1}$.  
The stability diagrams in both structured and unstructured cases displays a very similar behavior, showing that the structure of the interconnected nonlinearity does not affect the stability characteristics of Couette flow for $k_z=0$, i.e., if only two-dimensional TS modes are considered. Both subfigures (panels a and b) show that as the Reynolds number increases,  the flow becomes more sensitive to smaller velocity perturbations.  
We observe that Couette flow is more sensitive to perturbations characterized by smaller $k_x$ values, especially at higher Reynolds numbers.

Stability analysis via our small gain theorem approach implies that there exists a perturbation of a finite value that could cause the flow to become unstable. With the increase in Reynolds number, the magnitude of perturbation that will not trigger transitioning to turbulence becomes smaller, and thus,~\emph{there is a higher probability for transition to occur}. In contrast to LST, which predicts that Couette flow is asymptotically stable for any Reynolds number \citep{schmid2007nonmodal}, our small gain theorem approach provides mathematical proof that Couette flow can become unstable for a finite disturbance magnitude that exceeds a particular bound such as $\norm{\mathscr{H}_{\nabla}}_{\infty}^{-1}$, $\norm{\mathscr{H}_{\nabla}}_{\mu_{\mathbf{\Delta}_{nr}}}^{-1}$ or} $\norm{\mathscr{H}_{\nabla}}_{\mu_{\mathbf{\Delta}_{r}}}^{-1}$ for any Reynolds number, which agrees with experimental observations, such as \citet{tillmark1992experiments,dauchot1995finite}, and simulation results \citep{dou2012direct,barkley2005computational} showing transition at $\Rey = 320-370$.
In particular, the vertical dashed red lines in Fig.~\ref{fig:10a} and Fig.~\ref{fig:10b} correspond to $Re=325$, the critical Reynolds number found experimentally by \cite{dauchot1995finite}, and the vertical dashed-dotted red lines correspond to $Re=360$, the critical Reynolds number found experimentally by \cite{tillmark1992experiments}. In their study, the authors introduced 
controlled finite-size disturbances of short duration that were made by a solenoid-activated, 1 mm diameter fluid jet. However, no information is available on the magnitude of the disturbance. 
According to our stability criterion, the stability bound depends on the geometric shape of the disturbance (represented by its wavenumber pair $k_x,k_z$), its temporal frequency $\omega$, and its magnitude. Thus, differences in the magnitude of the introduced disturbances, in geometric shape, and in the time period between disturbances can all cause the discrepancy between the critical Reynolds number predictions of \cite{dauchot1995finite} and \cite{tillmark1992experiments}.

 \begin{figure}
    \centering
    \subfloat[\label{fig:10a}]{\includegraphics[width=0.49\columnwidth]{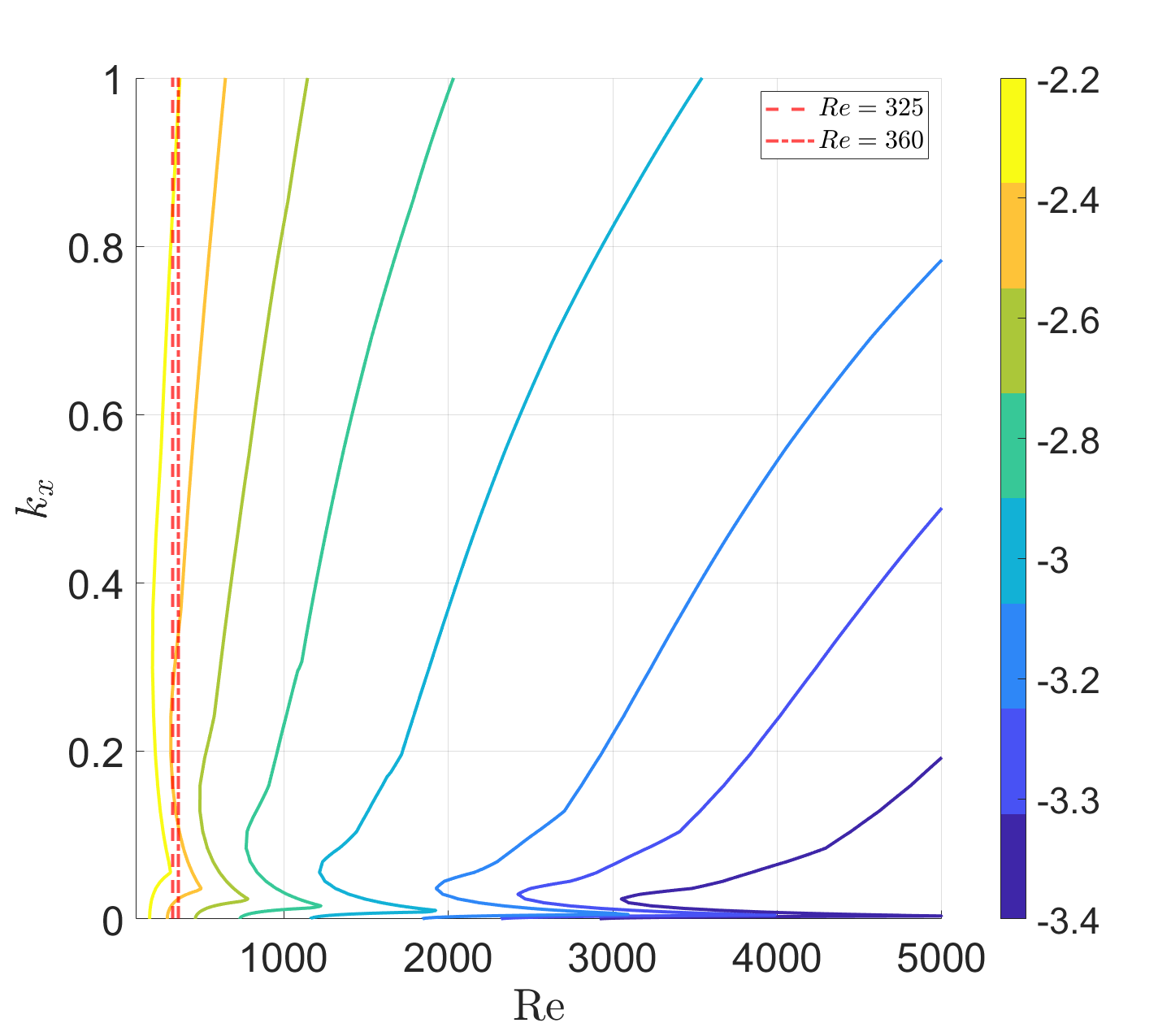}}
    \hfill
    \subfloat[\label{fig:10b}]{\includegraphics[width=0.49\columnwidth]{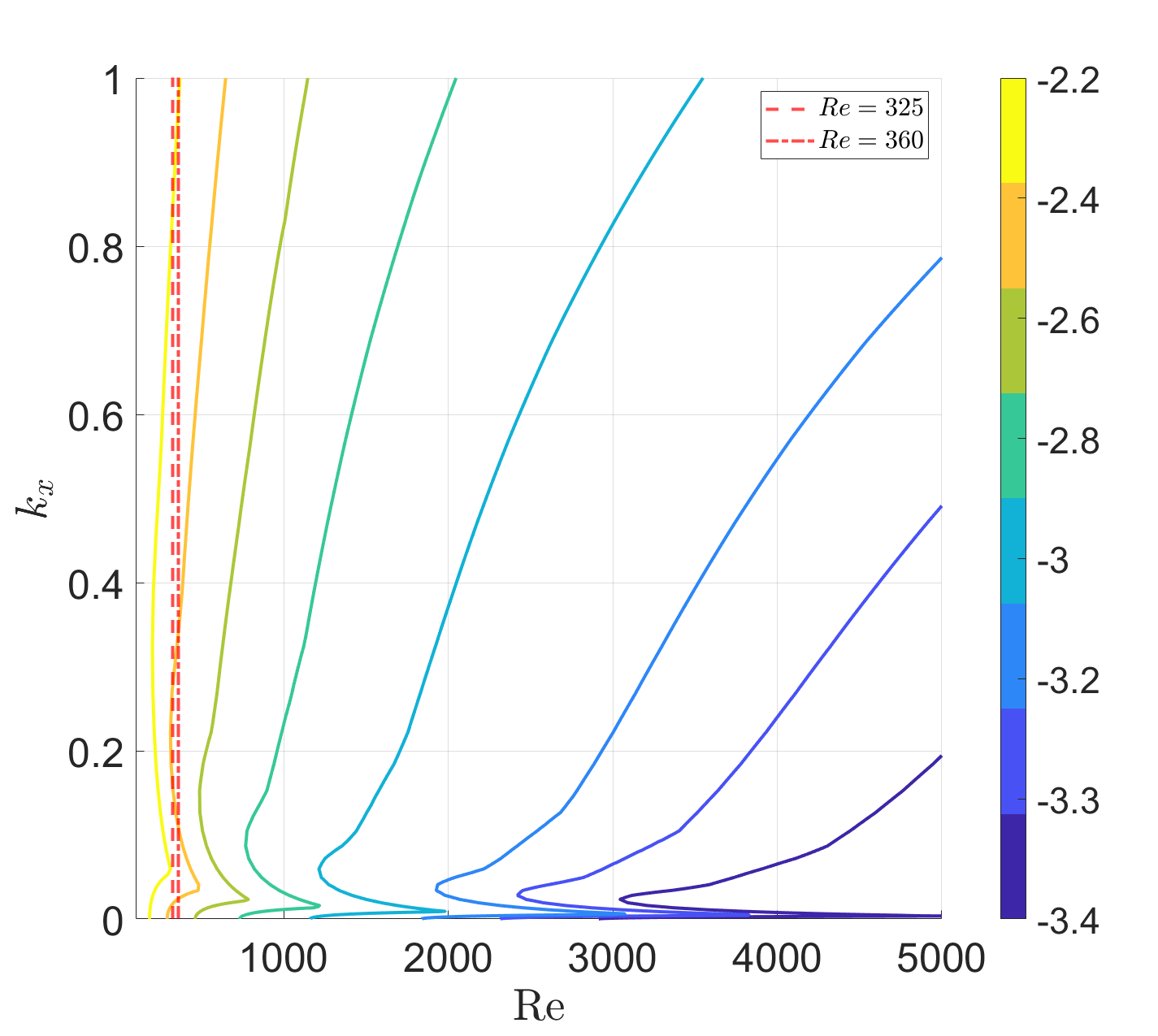}}
    \caption{{Stability diagrams for Couette flow showing contour plots on the $\Rey - k_x$ domain for $k_z=0$  of imposed thresholds of  (a) $\norm{\mathscr{H}_{\nabla}}_{\infty}^{-1}$ and (b) $\norm{\mathscr{H}_{\nabla}}_{\mu_{\mathbf{\Delta}_{nr}}}^{-1}$.}}
    \label{fig:10}
\end{figure}

\begin{figure}
    \centering
    \subfloat[\label{fig:11a}]{\includegraphics[width=0.49\columnwidth]{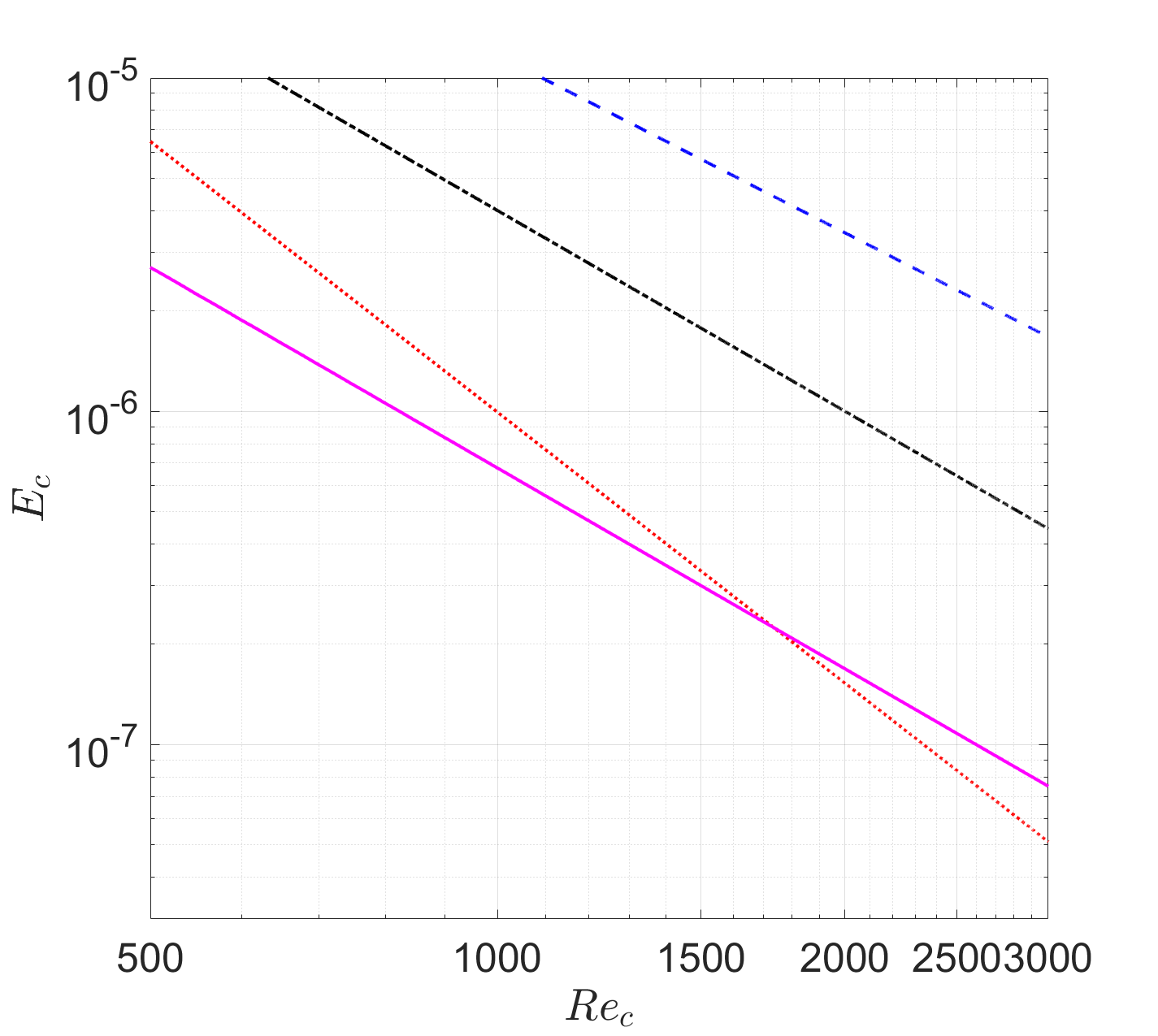}}
    \hfill
    \subfloat[\label{fig:11b}]{\includegraphics[width=0.49\columnwidth]{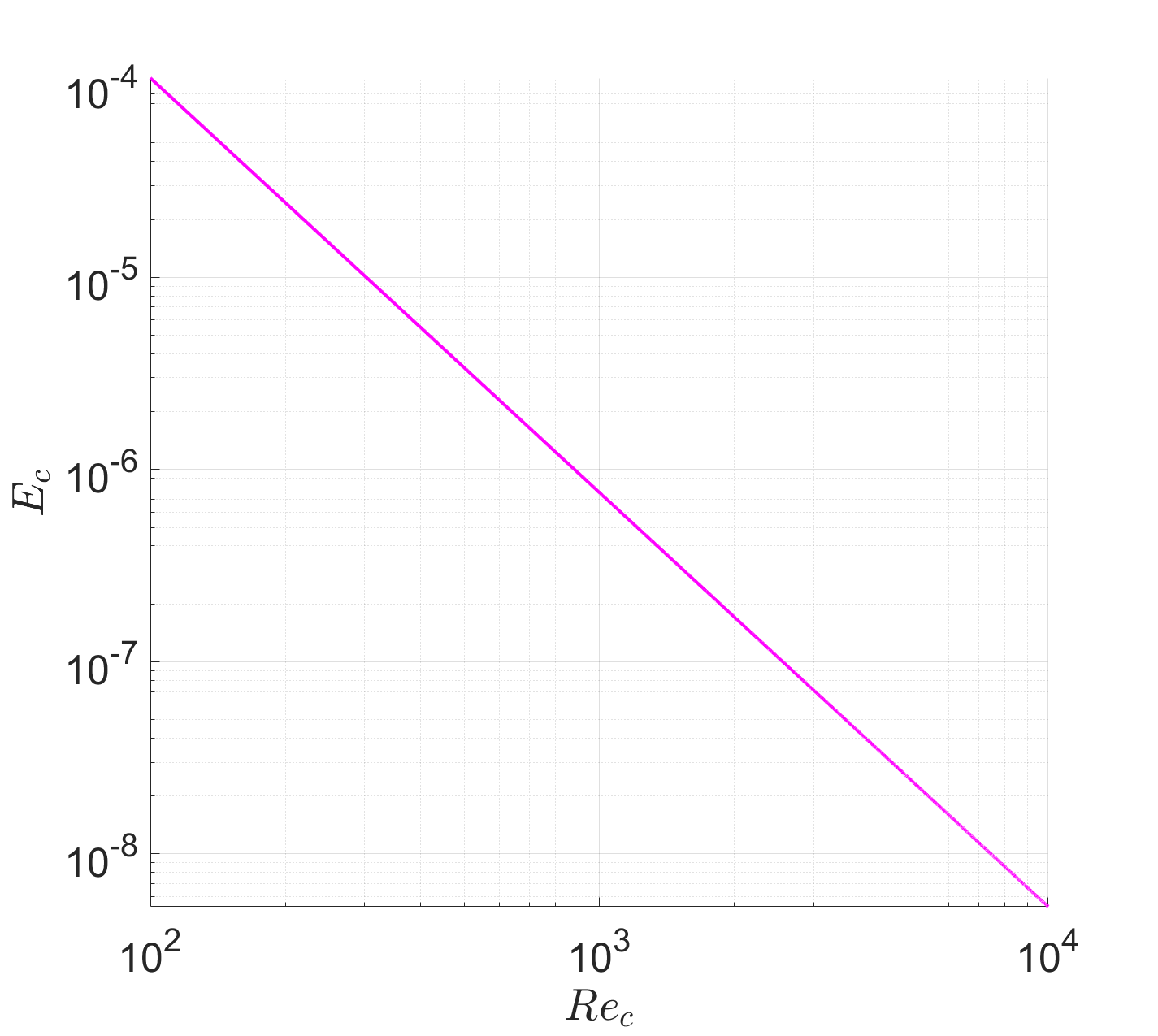}}
    \caption{Critical perturbation energy versus critical Reynolds number for Couette flow: (a) Our analysis versus results from literature:  magenta solid - our results; black dashed-dotted - results from \citet{duguet2010towards}; red dotted - results from \citet{duguet2013minimal}, blue dashed - results from \citet{reddy1998stability}. (b) Our analysis for a wider range of Reynolds numbers, $\Rey \in [100,10000]$.}
    \label{fig:11}
\end{figure}

Next, to validate our analysis for a wide range of Reynolds numbers with results from the literature, we compute the critical perturbation energy that causes instability.  
In Fig.~\ref{fig:11a}, we conduct a comparison between the critical perturbation energy from our analysis versus DNS data from three studies \citep{reddy1998stability,duguet2010towards,duguet2013minimal}. Fig.~\ref{fig:11a} shows that the predictions for the critical Reynolds number and the critical energy for instability are similar to the prediction made by \citet{duguet2013minimal}. While both \citet{reddy1998stability} and \citet{duguet2010towards} consider a specific oblique wave scenario, by employing a nonlinear optimization method, in \citet{duguet2013minimal} a more general transition scenario was considered, leading to lower critical energy predictions. The linear trend (in logarithmic scale) in Fig.~\ref{fig:11a} corresponds to a relationship of the form $E_c\propto Re^{-\gamma}$. Using our methodology, a value of $\gamma\approx2$ is predicted, similar to the results of \citet{reddy1998stability} and \citet{duguet2010towards}. However, a value of $\gamma\approx 2.7$ was reported by \citet{duguet2013minimal}, which better agrees with our results. It is possible that considering modes with a non-zero spanwise wavenumber can lead to even better agreement. Lastly, in Fig.~\ref{fig:11b} we show that the obtained relationship $E_c\propto Re^{-2}$  is sustained for a wide range of Reynolds numbers, by increasing the domain to $\Rey \in [100,10000]$. We note that we only display the results using $\norm{\mathscr{H}_{\nabla}}_{\infty}^{-1}$, as both the unstructured and structured methods yielded the same results, similarly to the results in Fig.~\ref{fig:10}.

\subsection{Plane Poiseuille base flow}
\label{sec:5.2}

\begin{figure}
    
    \centering
    \subfloat[\label{fig:12a}]{\includegraphics[width=0.49\columnwidth]{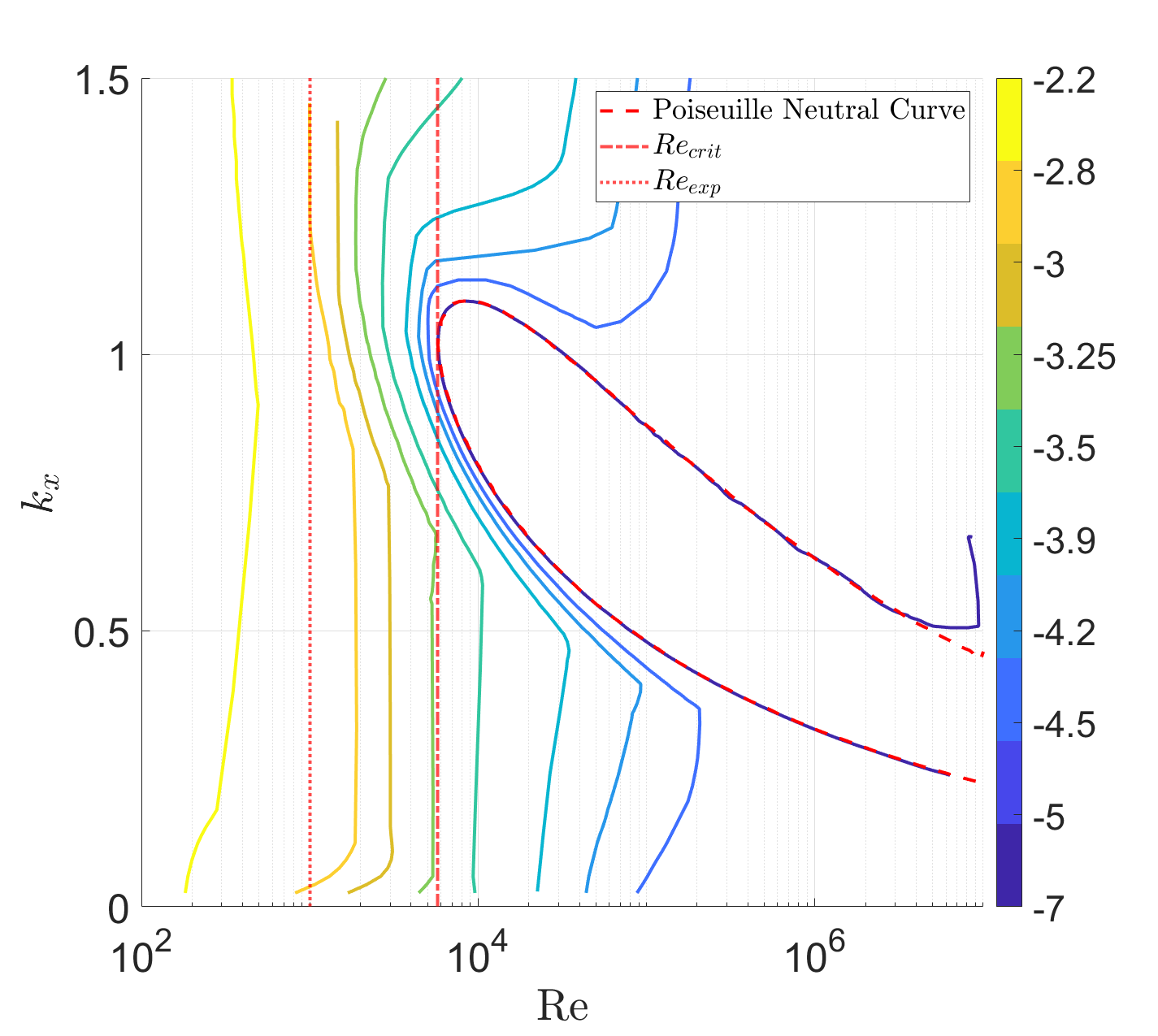}}\hfill
    \subfloat[\label{fig:12b}]{\includegraphics[width=0.49\columnwidth]{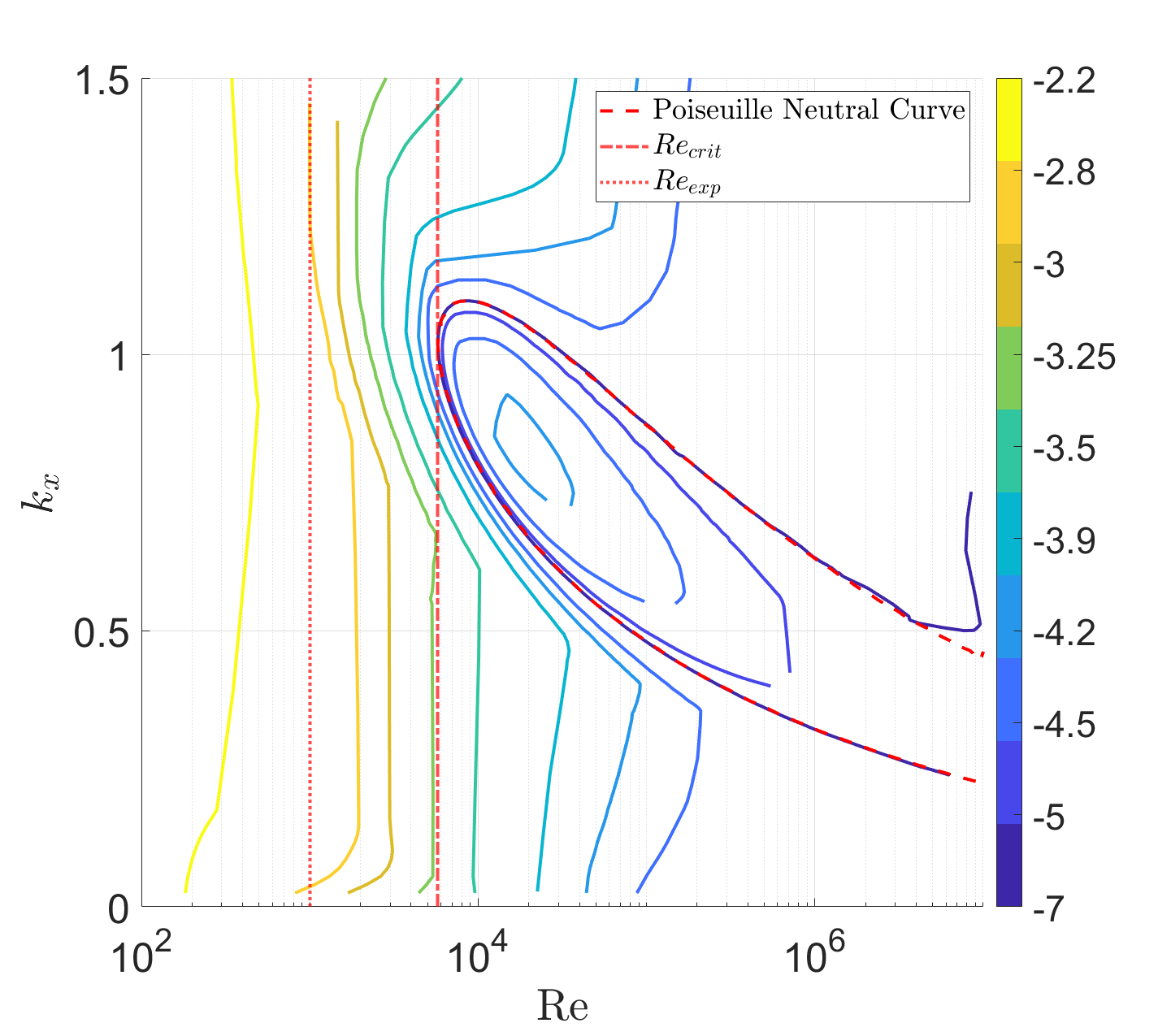}}
    \caption{Stability diagrams for plane Poiseuille flow showing contour plots on the $\Rey - k_x$ domain for $k_z=0$  of imposed thresholds of  (a) $\norm{\mathscr{H}_{\nabla}}_{\infty}^{-1}$ and (b) $\norm{\mathscr{H}_{\nabla}}_{\mu_{\mathbf{\Delta}_{nr}}}^{-1}$. The vertical dotted line corresponds to $Re=1000$ and the dashed-dotted line corresponds to $Re=5772$. The dashed contour is the neutral stability contour computed using LST.}
    \label{fig:12}
\end{figure}

In Fig.~\ref{fig:12}, we show stability diagrams for plane Poiseuille flow in {the $\Rey-k_x$} domain at $k_z=0$, showing several isocontours for different threshold values for both cases: $\norm{\mathscr{H}_{\nabla}}_{\infty}^{-1}$  (Fig.~\ref{fig:12a}) and $\norm{\mathscr{H}_{\nabla}}_{\mu_{\mathbf{\Delta}_{nr}}}^{-1}$ (Fig.~\ref{fig:12b}). 
The dashed red contour in Fig.~\ref{fig:12a} and \ref{fig:12b} {corresponds} to the neutral curve for Poiseuille base flow computed using eigenvalue analysis (LST), which is shown for comparison with our results. 

In Fig.~\ref{fig:12a},  as the Reynolds number increases, the threshold on disturbance that is required to make the flow unstable decreases similarly to the case of Couette flow. However, in the case of plane Poiseuille flow, as the threshold decreases, the resulting contours approach the neutral curve obtained by LST. 
When $\norm{\mathscr{H}_{\nabla}}_{\infty}^{-1}(k_x,k_z) \rightarrow 0$, infinitesimal velocity perturbations will cause the system to become unstable. This is how our small gain formulation of stability relates to LST, according to which, the region within the neutral curve is where the modes are infinitely amplified.
The predictions of LST, including critical Reynolds numbers, are valid only for the case of asymptotically small perturbations. Herein, the small gain theorem formulation provides an extension of LST and allows us to determine stability based on the magnitude of the perturbations in the flow for any Reynolds number. Using our small gain theorem approach to study transition states shows that instability can occur in plane Poiseuille flow at lower Reynolds numbers than the critical value $\Rey=5772$ (denoted by vertical dashed-dotted red lines in the figures) that is predicted via LST and our use of {the bounds} $\norm{\mathscr{H}_{\nabla}}_{\infty}^{-1}$ and $\norm{\mathscr{H}_{\nabla}}_{\mu_{\mathbf{\Delta}_{nr}}}^{-1}$. 
The exact Reynolds number associated with the loss of stability depends on the magnitude of velocity perturbations that exist in the flow field.
The flow will become unstable at earlier Reynolds number if the velocity perturbation exceeds the bound of $\norm{\mathscr{H}_{\nabla}}_{\mu_{\mathbf{\Delta}_{u}}}^{-1}$, which  {can be approximated by $\norm{\mathscr{H}_{\nabla}}_{\infty}^{-1}$, $\norm{\mathscr{H}_{\nabla}}_{\mu_{\mathbf{\Delta}_{nr}}}^{-1}$ or $\norm{\mathscr{H}_{\nabla}}_{\mu_{\mathbf{\Delta}_{r}}}^{-1}$}. Thus, experimental observations of subcritical transition in plane Poiseuille flow, such as in \citet{nishioka1985some} and \citet{sano2016universal}, can be explained by our novel stability criterion. The perturbation magnitude that corresponds to a critical Reynolds number of $Re=1000$ (denoted by the vertical dotted red lines in the figures) is approximately $10^{-2.9}=1.1\times 10^{-3}$ (for both structured and unstructured cases), meaning that perturbations that are $0.11\%$ of the centerline velocity are needed to be present in the flow for instability to occur for this Reynolds number \citep{carlson1982flow,nishioka1985some}. In both studies, finite-size disturbances are introduced to the laminar flow to induce instability, though we could not determine their magnitudes from these studies.

We note that for both unstructured (Fig.~\ref{fig:12a}) and structured (Fig.~\ref{fig:12b}) cases, the stability threshold contours plot outside the neutral curve region are very similar, including the obtained infinitesimal thresholds on the curve of the neutral loop itself. The difference between the two is limited to the stability characteristics for threshold values that lie inside the neutral loop. While in Fig.~\ref{fig:12a}, any perturbation that corresponds to a mode that is found inside the neutral is deemed unstable. However, in Fig.~\ref{fig:12b} there are contours that are found inside the neutral loop, corresponding to the fact that according to the structured analysis using $\norm{\mathscr{H}_{\nabla}}_{\mu_{\mathbf{\Delta}_{nr}}}^{-1}$ modes that are found inside the neutral loop can be stable given small enough perturbations.
The interpretation of contours of the structured quantities $\norm{\mathscr{H}_{\nabla}}_{\mu_{\mathbf{\Delta}_{r}}}^{-1}$ or $\norm{\mathscr{H}_{\nabla}}_{\mu_{\mathbf{\Delta}_{nr}}}^{-1}$ inside the neutral curve should be treated with caution. The small gain theorem is valid only given that the nominal system $\mathscr{H}_\nabla(k_x,k_z)$ is stable (as we denote in Theorem~\ref{theo:smg}). Linear stability theory predicts an exponentially growing mode within the neutral curve region, driven by linear interactions. Thus, our analysis holds only when the system within the neutral curve region is initially stable. Still, the uncertainty matrices in our analysis may provide a stabilizing effect on the interconnected system represented by the loop block diagram in Fig.~\ref{fig:1}, allowing the flow to remain stable even at post-critical Reynolds numbers. This is in line with studies by \citet{cossu2002stabilization,fransson2005experimental} that have shown that finite-amplitude streaks stabilize TS instability. Furthermore, nonlinearity in the NSE has also been suggested to provide a stabilizing effect, leading to the saturation of linear growth \citep{schmid2002stability,mckeon2017engine,jin2021energy}. The extent to which these effects are replicated using the current methodology with constant uncertainty matrices, and the implications for stability within the neutral curve, are subject to future study.
Results from our analysis can be used to explain the experimental observations of the work by \citet{nishioka1975experimental}, in which plane Poiseuille flow is found experimentally to remain stable up to a Reynolds number of $Re=8000$, which cannot be explained using LST. Our analysis suggests that the perturbation magnitude that leads to the loss of stability at a Reynolds number of $Re=8000$ is $10^{-4.2}$, which corresponds to a turbulence intensity of $0.006\%$ of the centerline velocity. This result matches the experimental setup in \citet{nishioka1975experimental}, where the turbulence intensity is reported to be "essentially less than $0.01\%$".

Next, we validate our analysis for a wide range of Reynolds numbers with results from the literature by computing the critical perturbation energy.
\begin{figure}
    \centering
    \subfloat[\label{fig:13a}]{\includegraphics[width=0.49\columnwidth]{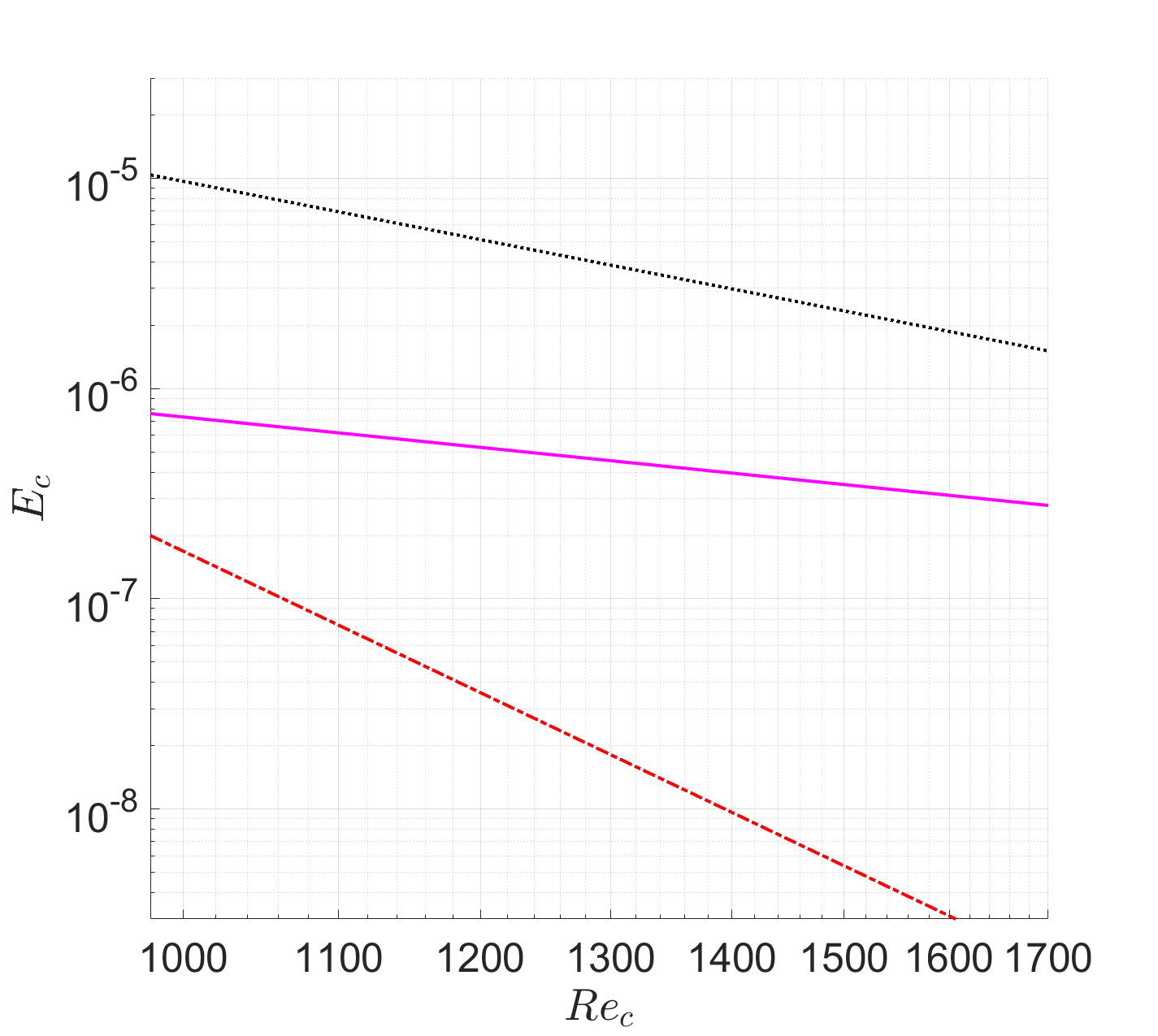}}\hfill
    \subfloat[\label{fig:13b}]{\includegraphics[width=0.49\columnwidth]{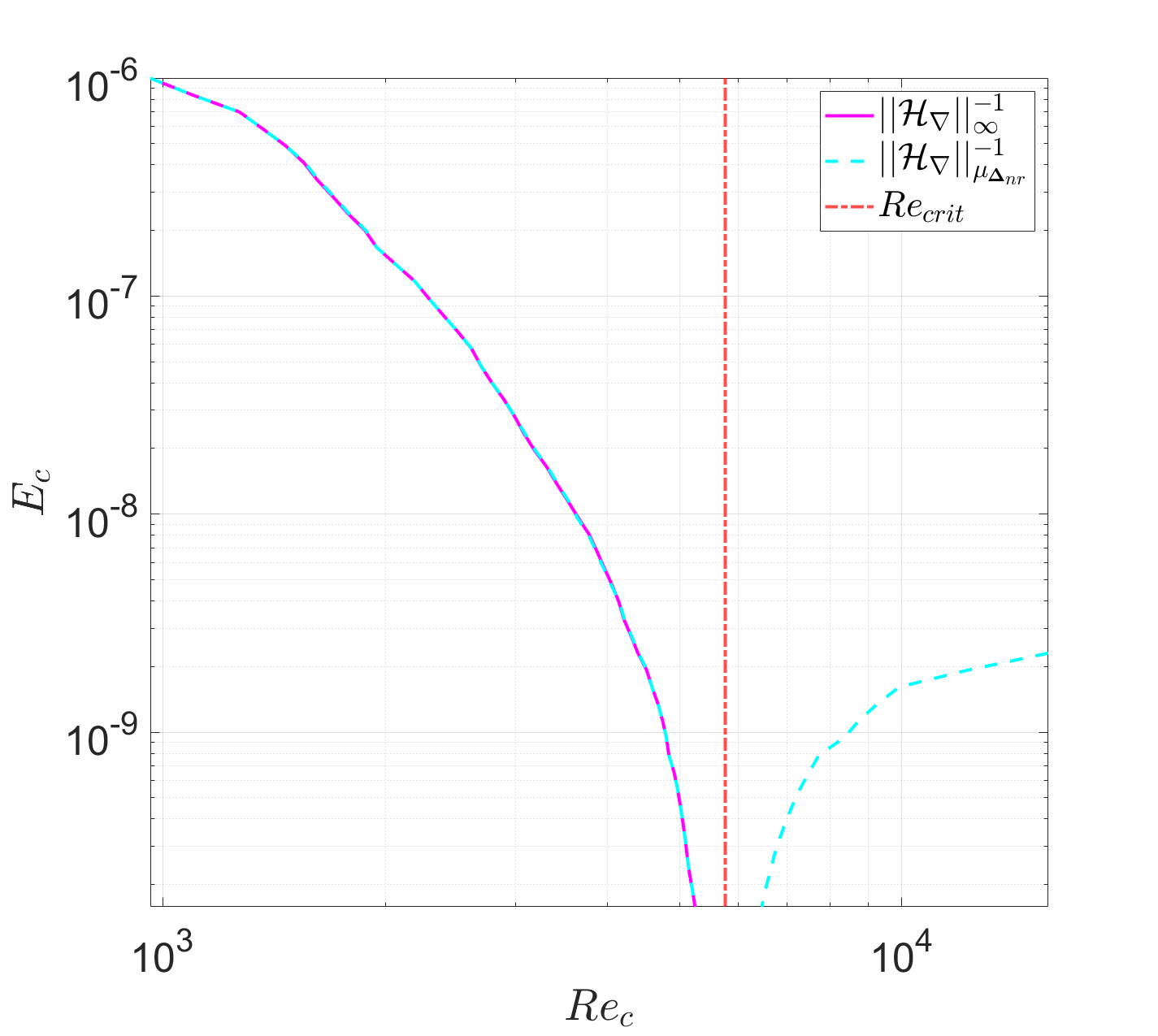}}
    \caption{Critical perturbation energy versus critical Reynolds number for plane Poiseuille flow: (a) Our analysis versus result from literature: magenta solid - our results; black dotted - \citet{reddy1998stability}; red dashed-dotted -\citep{parente2022minimal}  (minimal energy threshold using large domain). (b) Our analysis for the increased range $\Rey\in [100,15,000]$.}
\end{figure}
Fig.~\ref{fig:13a} shows that the predictions of the critical Reynolds number and the critical energy for instability are found between the prediction by \citet{parente2022minimal} and the prediction by \citet{reddy1998stability}. While in \citet{reddy1998stability}, a specific oblique wave scenario is considered, \citet{parente2022minimal} employs a nonlinear variational optimization method to consider a more general transition scenario, leading to lower critical energy predictions. As shown in Fig.~\ref{fig:6}, the TS mode is the least dominant for subcritical Reynolds numbers (according to both structured and unstructured methods). Thus, we expect that considering more general cases that include modes with non-zero spanwise wavenumbers will very likely (as discussed in section~\ref{sec:4.2}) lower the critical perturbation energy predicted by our methodology, making it closer to the results of \citet{parente2022minimal}.

The linear trend in Fig.~\ref{fig:13a} corresponds to a relationship of the form $E_c=O(Re^{-\gamma})$. Similarly to the case of Couette flow, a value of $\gamma\approx 2$ is predicted using our methodology. However, this trend does not hold for all Reynolds numbers as we demonstrate in  Fig.~\ref{fig:13b}, which shows the relationship between the critical energy perturbation and critical Reynolds number for a wider range of Reynolds numbers,  $\Rey \in [100,15,000]$. Fig.~\ref{fig:13b} shows that the linear trend (in logarithmic scale) does not continue for  Reynolds numbers above $1700$. Instead, there is a rapid decrease of the curve, that asymptotes towards zero at critical Reynolds number $\Rey=5772$ for both $\norm{\mathscr{H}_{\nabla}}_{\infty}^{-1}$ and $\norm{\mathscr{H}_{\nabla}}_{\mu_{\mathbf{\Delta}_{nr}}}^{-1}$ at $\Rey=5772$. This result reflects the imposed threshold by the TS mode in Fig.~\ref{fig:6}b, where the perturbation energy that is required to cause instability approaches zero, matching the prediction of the critical Reynolds number by LST and our previous discussion on the results of Fig.~\ref {fig:12}. The difference between the structured and unstructured cases is restricted to post-critical Reynolds numbers (higher than $Re=5772$). In detail, $\norm{\mathscr{H}_{\nabla}}_{\infty}^{-1}$ predicts that any perturbation, even of infinitesimal energy, will cause a transition. However, the structured analysis using $\norm{\mathscr{H}_{\nabla}}_{\mu_{\mathbf{\Delta}_{nr}}}^{-1}$ predicts that while for $Re=5772$ very small perturbations will cause instability, as the Reynolds number increases beyond this value the critical energy also increases. These results are consistent with the discussion regarding Fig.~\ref{fig:12b}, in which thresholds on perturbation magnitude for stability exist inside the neutral curve.

\subsection{Blasius base flow}
\label{sec:5.3}

\begin{figure}
    \centering
    \subfloat[\label{fig:14a}]{\includegraphics[width=0.49\columnwidth]{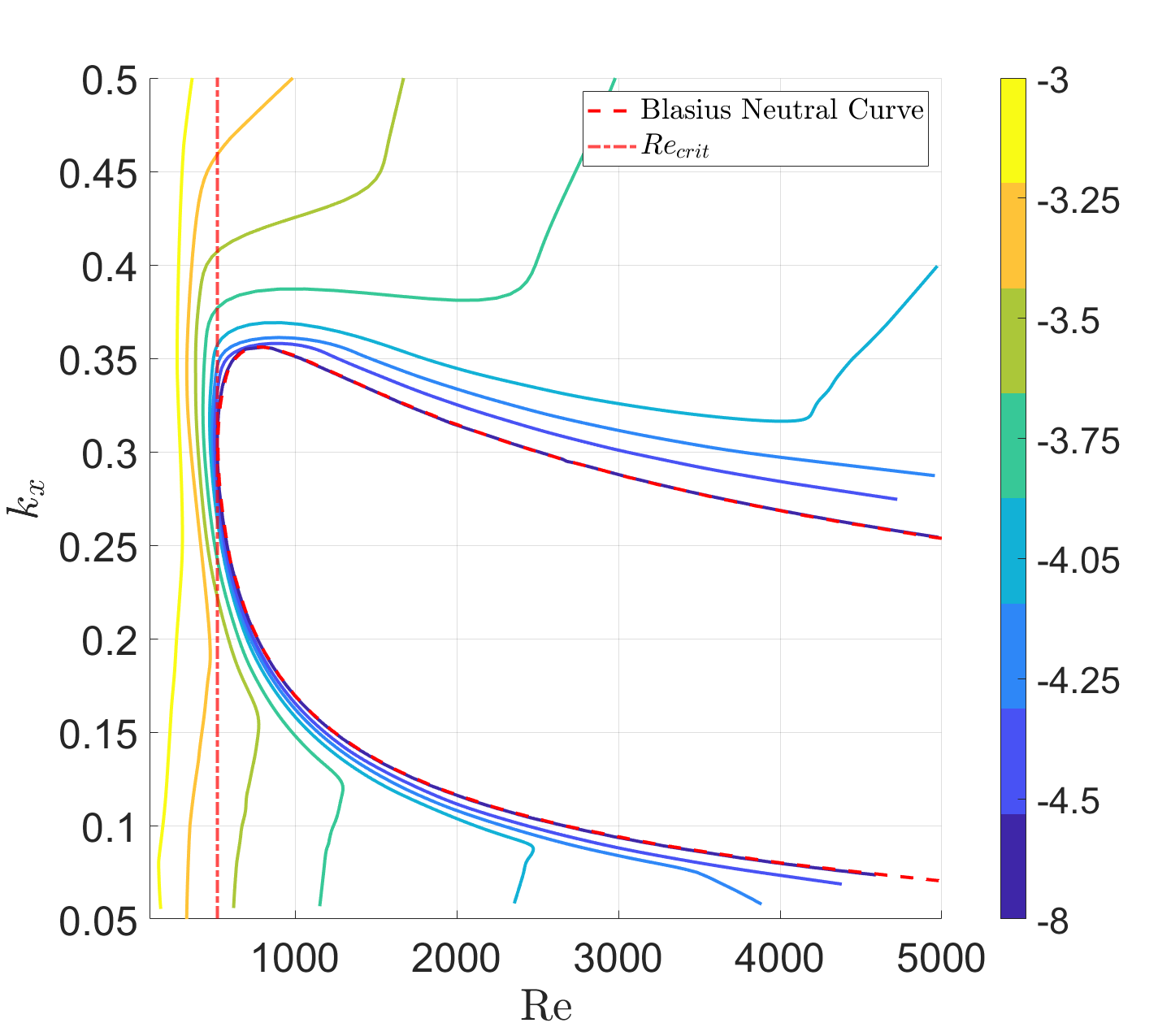}}
    \subfloat[\label{fig:14b}]{\includegraphics[width=0.49\columnwidth]{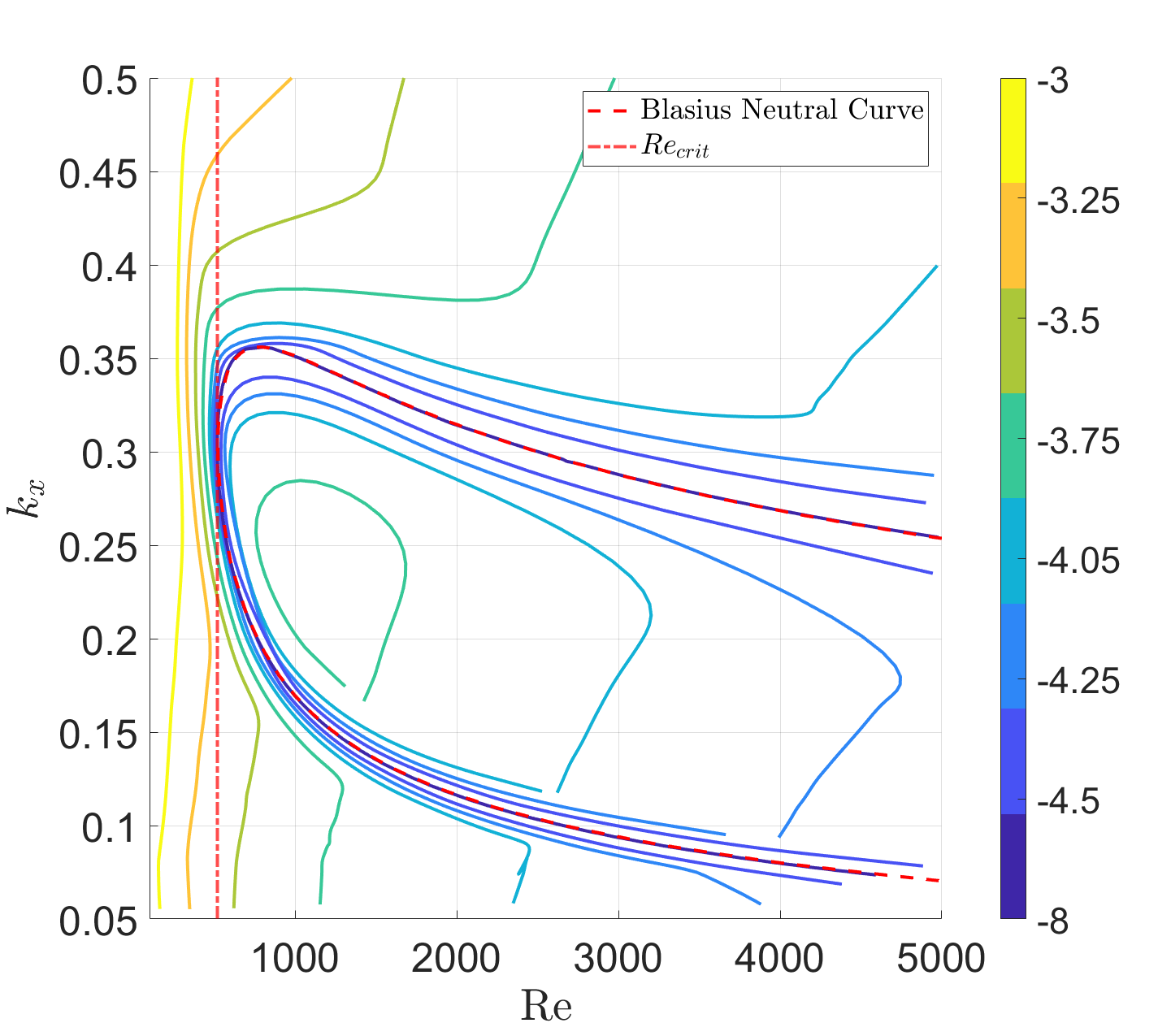}}\caption{Stability diagrams for Blasius flow showing contour plots on the $\Rey - k_x$ domain for $k_z=0$  of imposed thresholds of  (a) $\norm{\mathscr{H}_{\nabla}}_{\infty}^{-1}$ and (b) $\norm{\mathscr{H}_{\nabla}}_{\mu_{\mathbf{\Delta}_{nr}}}^{-1}$. The vertical dashed-dotted line corresponds to $Re=520$, and the dashed contour is the neutral stability contour computed using LST.}
    \label{fig:14}
\end{figure}

Stability diagrams showing contour plots of threshold on disturbance magnitude using $\norm{\mathscr{H}_{\nabla}}_{\infty}^{-1}$ and $\norm{\mathscr{H}_{\nabla}}_{\mu_{\mathbf{\Delta}_{nr}}}^{-1}$ for Blasius flow are shown in Fig.~\ref{fig:14a} and \ref{fig:14b} respectively. The dashed red curve in both figures corresponds to the so-called 'banana' neutral curve for Blasius base flow, as predicted by LST, and the vertical dashed-dotted red lines correspond to $Re=520$, the critical Reynolds number predicted by LST for Blasius flow. These plots show that as the Reynolds number increases, the perturbation magnitude that is required to make the flow unstable decreases, similar to the cases of Couette and plane Poiseuille flows. Additionally, as the threshold magnitude decreases, the resulting threshold contours approach the neutral curve obtained via LST, as observed for the case of plane Poiseuille flow. This result shows once again that imposing an infinitesimally small threshold on velocity perturbations leads our analysis to converge to the results of LST, while extending them by allowing us to assess stability in the presence of finite-amplitude perturbations. 
 
Both diagrams for unstructured and structured cases show a very similar topology of threshold isolines outside and on the neutral loop region. Whereas, inside the neutral loop of the structured case diagram (Fig.~\ref{fig:14b}), there are finite-size threshold contours present. Imposing structure should integrate some aspects of nonlinearity in the computation of $\norm{\mathscr{H}_{\nabla}}_{\mu_{\mathbf{\Delta}_{nr}}}^{-1}$, {causing} the flow to become stable for small finite-{sized} perturbations inside the neutral loop. {This is} in contrast to the unstructured case, which predicts any mode inside the neutral loop to be unstable for any perturbation magnitude.
However, the presence of a stability threshold isoline inside the neutral curve should be treated with caution, as explained in~\S\ref{sec:5.2} due to the inherent assumption of the small gain theorem that the interconnected system should be initially stable. It is shown to be feasible to enter the postcritical Reynolds number region, while keeping the flow stable via a nonlinear route to transition as reported in \cite{cherubini2011minimal,cherubini2012purely},  where a finite perturbation magnitude is determined for causing transition of Blasius flow at a Reynolds number of $\Rey=610$. This result is in agreement with our results for the structured case, as illustrated in Fig.~\ref{fig:14b}.
In particular, in \cite{cherubini2012purely} a critical energy value (of the entire velocity perturbation field) of $E_c=4.4\times 10^{-3}$ is obtained for this Reynolds number. As explained in the last paragraph of~\S~\ref{sec:4.3}, the average magnitude of perturbations in the flow can be computed given the energy, yielding an average magnitude of $3.3\times 10^{-4}=10^{-3.5}$. This value is slightly higher than the value predicted in Fig.~\ref{fig:14b}, which is approximately $10^{-4.05}$. This difference can perhaps be attributed to approximating the structure of the nonlinear term in the NSE by non-repeated blocks, as we expect lower stability margin predictions using the exact structure according to Eq.~\eqref{eq:2.29}, and also may be due to not accounting for non-parallel effects of the boundary layer in the current study.

\begin{figure}
    \centering
    \subfloat[\label{fig:15a}]{\includegraphics[width=0.49\columnwidth]{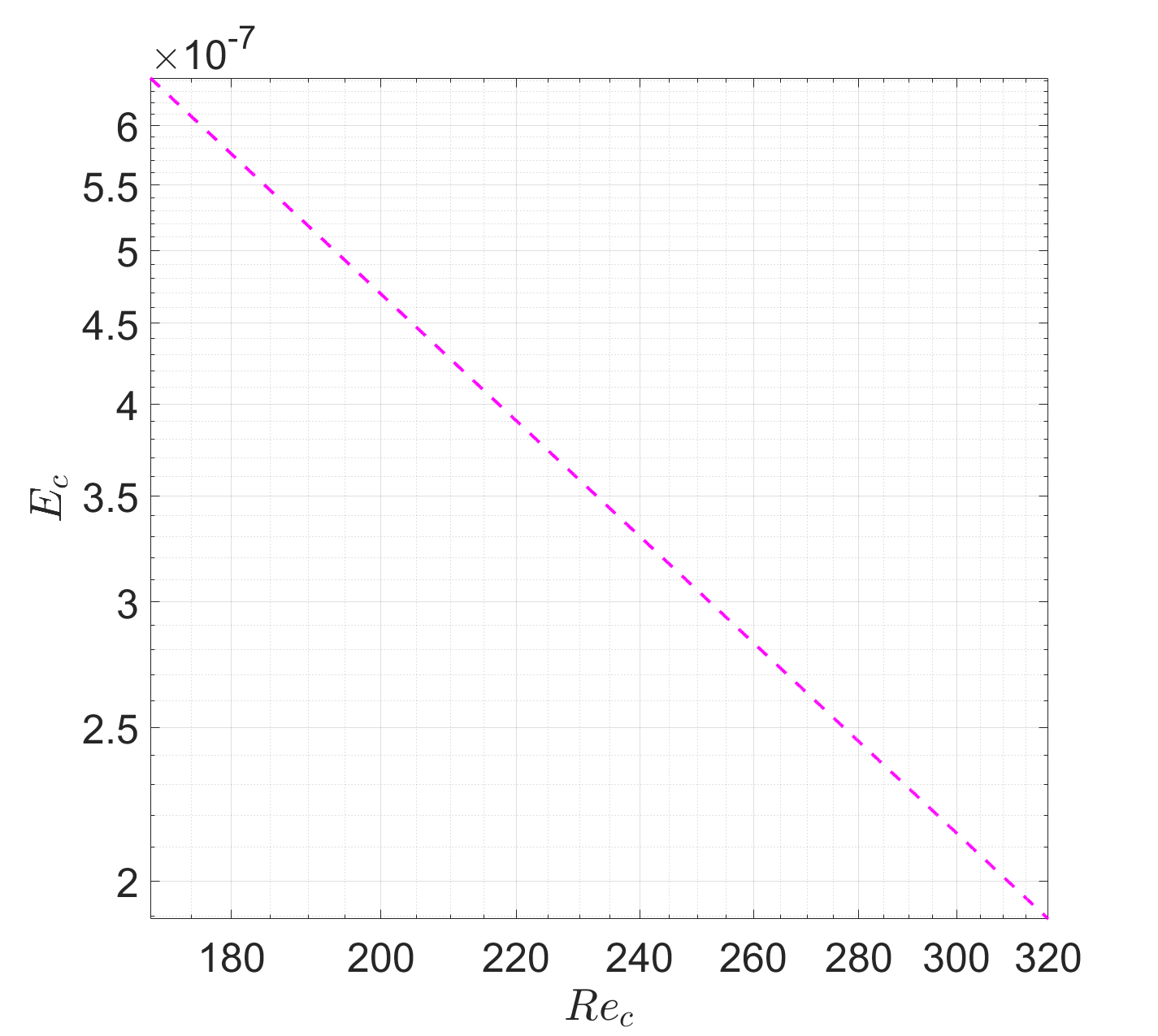}}\hfill
    \subfloat[\label{fig:15b}]{\includegraphics[width=0.49\columnwidth]{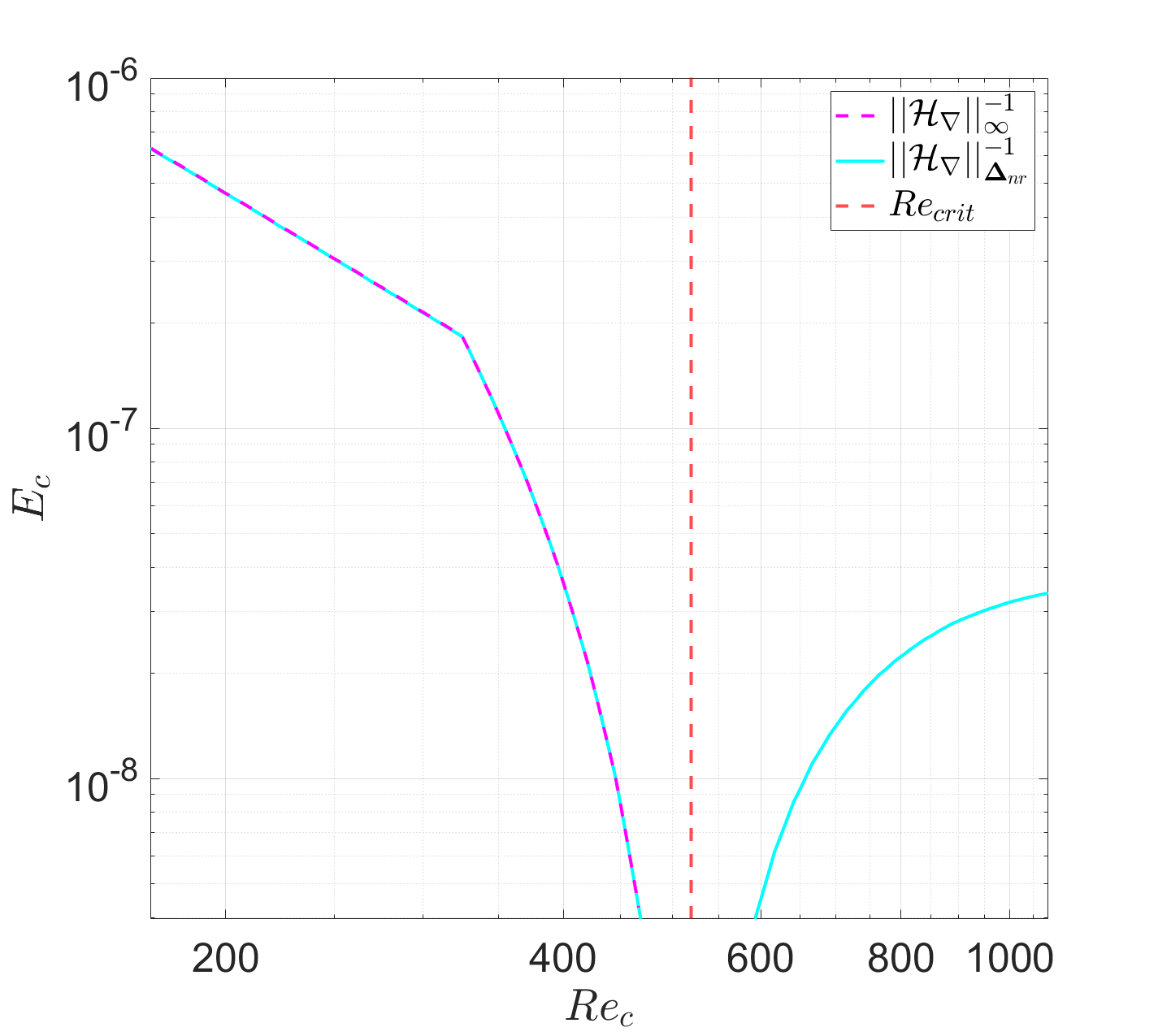}}\hfill 
    \caption{Critical perturbation energy versus critical Reynolds number for Blasius flow via our analysis: (a) for a limited range of $\Rey$ up to 320, (b) for an increased range of Reynolds numbers up to 1100.}
    \label{fig:15}
\end{figure}

Lastly, we now turn our attention to computing the critical energy perturbation for the transition of Blasius flow. Fig.~\ref{fig:15a} displays the relationship between the critical perturbation energy and the critical Reynolds number on a logarithmic scale, defined similarly as in the cases of Couette and Poiseuille flows. 
The linear trend in logarithmic scale in Fig.~\ref{fig:15a} corresponds to a relationship of the form $E_c=O(Re^{-\gamma})$. Similarly to the previous cases, a value of $\gamma\approx 2$ is predicted using our methodology.  This linear trend holds up to about $\Rey_c=320$, after which there is a rapid drop in the critical perturbation energy, as shown in Fig.~\ref{fig:15b} that shows the same relationship between the critical perturbation energy and Reynolds number for a larger range of $\Rey \in [0,1100]$. 
Similar to the observed behavior for plane Poiseuille flow,
the  $E_c$  curve asymptotes towards zero at $\Rey=520$ for both structured and unstructured cases. The observed trend change in the curve after $\Rey=300$ shown in Fig.~\ref{fig:15b} can be explained by a shift in the dominant flow structure that is associated with the least stable mode.

For $\Rey>520$, the energy curve for structured methods departs from the unstructured case, exhibiting similar behavior as the TS mode in Fig.~\ref{fig:8}b, which we discussed in detail for the case of plane Poiseuille flow in section~\ref{sec:5.2}, and will not repeat it here again for brevity.
  
Details on validation of our stability criterion for Blasius flow with some results from experimental studies from literature are provided in the previous section~\ref{sec:4.3} (last paragraph), where we discuss the general case for our analysis with non-zero $k_z$, which is associated with an oblique transition scenario. 

\section{Conclusions}
\label{sec:6}

Herein, we propose a novel stability criterion based on the \emph{small gain theorem} coupled with unstructured and structured input-output analyses. The criterion provides an upper bound for velocity perturbations that ensures stability. Our stability criterion does not guarantee a transition, but provides an estimate of a threshold for disturbance magnitude that will trigger flow instability. 
We employ three different input-output approaches for modeling nonlinearity (unstructured, structured with non-repeating blocks, and structured with repeating blocks). In particular, we show that the imposed threshold obtained by these three methods complies with a hierarchical relationship based on set inclusion that we obtained in Eq.~\eqref{eq:2.29}. According to this hierarchy, the unstructured case is the most conservative, imposing the lowest bound on disturbance magnitude. Moreover, structured input-output with repeating uncertainty analysis provides the closest bound to $\norm{\mathscr{H}_{\nabla}}_{\mu_{\mathbf{\Delta}_{u}}}^{-1}$, which incorporates the accurate structure of the nonlinear term in the NSE (as defined in Eq.~\eqref{eq:2.20}).

This work provides a complementary point of view to nonmodal analysis, quantifying the threshold that nonmodal growth has to reach to cause flow instability.
It allows studying complex transition scenarios that are associated with bypass transition scenarios in the presence of finite-size disturbances and transient growth scenarios caused by a combination of decaying modes. 
In detail, \textit{our framework addresses the discrepancy between the computed critical Reynolds number via LST and the documented Reynolds numbers at which transition is observed in experiments.} Stability analysis with our approach is in agreement with previous experimental work, which found that transition can occur at lower Reynolds numbers than the critical Reynolds number predicted by LST, and due to finite amplitude exceeding a certain bound. Nonmodal nonlinear stability analysis provides similar thresholds on perturbation magnitude, via computing the evolution of the flow field, iterating over different initial conditions, converging to the one with the smallest magnitude that leads to transition, which is a numerically costly procedure.
Our approach is significantly numerically cheaper, allowing us to
conduct a more detailed study compared with previous works using nonmodal nonlinear approaches, considering a large range of Reynolds numbers and multiple modes in the $k_x,k_z$ domain. 

For the limit of infinitesimally small perturbations, we show that our analysis reproduces LST stability diagrams. It is not a coincidence, but it will not always be true. Linear stability theory is based on eigenvalue analysis, which identifies the TS mode as the least stable eigenvalue of the system, and the critical Reynolds number is the Reynolds number at which the growth rate of the TS mode becomes positive. 
In our analysis, the dominant mode for a given Reynolds number is the mode that sets the lowest threshold on velocity perturbation magnitude to trigger instability. This mode is not necessarily identical to the least stable mode predicted by LST, defined by the largest growth rate. At Reynolds numbers below the critical value, our analysis shows that the TS mode is not the dominant mode, as the threshold is initially set by SPS (in unstructured case) or SPS and later DLR mode in structured cases. However, being the least stable,  the TS mode experiences explosive growth, which in turn, makes it the dominant mode that sets the threshold for disturbance magnitude at the proximity of the critical Reynolds number for both plane Poiseuille and Blasius flows, according to structured and unstructured cases.

When approximating nonlinearity using unstructured uncertainty, our results indicate that the region inside the neutral loop is unstable for any perturbation magnitude. In contrast, when using structured uncertainty, our analysis predicts a stable region inside the neutral stability curve predicted by LST, given sufficiently small perturbations for both plane Poiseuille and Blasius flows. The obtained threshold is in agreement with the disturbance level observations in experimental and simulation studies \cite{nishioka1975experimental,cherubini2011minimal,cherubini2012purely} for which the flows remained laminar at post-critical Reynolds numbers in plane channel flow. However, this result should be treated with caution.
The small gain theorem guarantees stability only for cases where the nominal system (i.e., the linear interactions represented by $\mathscr{H}_\nabla$) is stable. Linear stability theory predicts that the linear system is unstable inside the neutral loop, due to an exponentially growing mode. Thus, our analysis holds only in the case that the system inside the neutral curve region is initially stable. 
Nonlinear effects may provide a stabilizing effect, explaining our results. The validity of this claim and the accuracy with which our model for nonlinearity represents nonlinear effects in the NSE remain open questions.
Furthermore, in spatially evolving, i.e., non-parallel, base flows such as the Blasius boundary layer, the Reynolds number increases as the disturbance propagates downstream. This might cause a disturbance of any size to eventually cross the unstable region, where an infinitesimal threshold around the perimeter of the neutral curve will trigger instability of the flow.

We also use these three input-output approaches to elucidate the interplay between the dominant modes that set the threshold in our stability criterion on disturbance magnitude. We conducted our analysis on three canonical flows---Couette,  plane Poiseuille, and Blasius for a wide range of Reynolds numbers, which in turn allowed us to obtain new insights on the interplay between flow structures that govern the transition physics of wall-bounded shear flows, which were elusive in past studies.  
We demonstrate that there is an interplay between flow structures such as streaks, oblique modes and TS waves that govern the flow physics for different Reynolds numbers, where either oblique \citep{liu2021,shuai2023structured} and streak modes \citep{reddy1993energy,schmid1994optimal,schmid2000linear,bamieh2001energy,jovanovic2004modeling,jovanovic2004unstable,jovanovic2005componentwise} govern the flow at subcritical Reynolds number, setting a finite-size threshold on disturbance magnitude,  whereas upon the critical value the TS mode dominates and sets the threshold such that any infinitesimal disturbance would trigger transition at the critical Reynolds value predicted by LST. This is due to the fact, as explained before,  that at the proximity of the critical Reynolds number, LST shows that the TS mode becomes unstable close to the critical Reynolds number based on eigen mode analysis, it grows rapidly, and overpowers other modes that in turn sets the threshold on disturbance magnitude based on our stability approach.

For a given magnitude of external disturbances, our framework can quantify the threshold below which finite-size disturbances or nonmodal growth are allowed to maintain stability. Insights from our stability analysis can lead to improved, efficient control strategies by identifying the forcing that achieves the largest impact on the flow and by predicting flow responses to finite-size actuation forcing. 
An interesting extension of our study for the future could be to incorporate base flow modifications into our analysis by implementing a structured uncertainty model with a time-varying model for the linearized flow system \citep{dullerud2002new}.
\\
\\
{\bf Declaration of Interests}. \\The authors report no conflict of interest.

\appendix

\section{Derivation of bound on perturbation magnitude}
\label{sec:App_theory}

In this appoendix we explicitly derive an expression for the norm $\norm{\mathbf{U}_{\Xi}}_{\infty}$, to show that it is equivalent to the largest perturbation magnitude in the flow, as stated in~\S\ref{sec:2.4}. Here $\mathbf{U}_{\Xi}$ is the structured uncertainty applied to the vectorized velocity gradient $\begin{bmatrix} \nabla^T u & \nabla^T v & \nabla^T w \end{bmatrix}^T$ that retains the component-wise structure of the advection term $\boldsymbol{u} \cdot \nabla \boldsymbol{u}$ \citep{liu2021}.
The exact expression for ${\textbf{U}}_\Xi$ that captures the nature of the nonlinear term in the NSE after discretisation is: 

\begin{equation}
\label{eq:A1}
 {\textbf{U}}_\Xi \equiv \begin{bmatrix}
    -{\text{}\textbf{u}_{\xi}}^T & \mathbf{0} & \mathbf{0} \\
    \mathbf{0} & -{\textbf{u}_{\xi}}^T & \mathbf{0} \\
    \mathbf{0} & \mathbf{0} & -{\textbf{u}_{\xi}}^T 
\end{bmatrix}.
\end{equation} 
Here, ${\textbf{u}_{\xi}}^T \equiv \begin{bmatrix}
    \text{diag}(u_\xi) & \text{diag}(v_\xi) & \text{diag}(w_\xi)
\end{bmatrix}$ is a frozen-in-time approximation of the discretised perturbation velocity field. As in Eq.~\eqref{eq:2.25}, $u_\xi, v_\xi, w_\xi \in \mathbb{C}^{N_y\times 1}$ represent discretised versions of the perturbations velocity components. To retain concise notation, we denote $\text{diag}(u_\xi)$ by $u_{\xi}^d$, and similarly for the other velocity components. Based on Eq.~\eqref{eq:A1}, the explicit expression for $\mathbf{U}_{\Xi}$ is:
\begin{equation}
    \label{eq:A2}
    \mathbf{U}_{\Xi} = 
    \begin{bmatrix}
        -u_{\xi}^d & -v_{\xi}^d & -w_{\xi}^d & 0 & 0 & 0 & 0 & 0 & 0 \\
        0 & 0 & 0 & -u_{\xi}^d & -v_{\xi}^d & -w_{\xi}^d & 0 & 0 & 0 \\
        0 & 0 & 0 & 0 & 0 & 0 & -u_{\xi}^d & -v_{\xi}^d & -w_{\xi}^d
    \end{bmatrix}.
\end{equation}
Using the definition in Eq.~\eqref{eq:A2}, we obtain
\begin{equation}
    \label{eq:A3}
    \mathbf{U}_{\Xi}^* = 
    \begin{bmatrix}
        -{u^d_{\xi}}^* & 0 & 0 \\
        -{v^d_{\xi}}^* & 0 & 0 \\
        -{w^d_{\xi}}^* & 0 & 0 \\
        0 & -{u^d_{\xi}}^* & 0 \\
        0 & -{v^d_{\xi}}^* & 0 \\
        0 & -{w^d_{\xi}}^* & 0 \\
        0 & 0 & -{u^d_{\xi}}^* \\
        0 & 0 & -{v^d_{\xi}}^* \\
        0 & 0 & -{w^d_{\xi}}^*
    \end{bmatrix}.
\end{equation}
Here, $^*$ denotes the conjugate transpose. We multiply the results of Eq.~\eqref{eq:A2} and Eq.~\eqref{eq:A3} to obtain the following:
\begin{equation}
    \label{eq:A4}
    (\mathbf{U}_{\Xi}\mathbf{U}_{\Xi}^*)_{i,j} = \begin{cases}
        |{u_{\xi}}_i|^2 + |{v_{\xi}}_i|^2 + |{w_{\xi}}_i|^2, \quad i = j \\
        0, \quad\quad\quad\quad\quad\quad\quad\quad\quad\,\,\,\,\,\, \text{else}
    \end{cases}
\end{equation}
The notation ${u_{\xi}}_i$ (and similarly for the two other velocity components) denotes the $i$-th entry of ${u_{\xi}}$ in the discretised domain, which contains $N_y$ entries.

By the definition of the largest singular value:
\begin{equation}
    \label{eq:A5}
    \bar{\sigma}(\mathbf{U}_{\Xi}) = \sqrt{\bar{\lambda}(\mathbf{U}_{\Xi}\mathbf{U}_{\Xi}^*)}.
\end{equation}
Here, $\bar{\lambda}(\cdot)$ represents the largest eigenvalue of a matrix. Since the eigenvalues of a diagonal matrix (such as $\mathbf{U}_{\Xi}\mathbf{U}_{\Xi}^*$) are the diagonal entries, using Eq.~\eqref{eq:A4} and Eq.~\eqref{eq:A5} it is clear that the following is true:
\begin{equation}
    \label{eq:A6}
        \bar{\sigma}(\mathbf{U}_{\Xi}) = \max_i \sqrt{|{u_{\xi}}_i|^2 + |{v_{\xi}}_i|^2 + |{w_{\xi}}_i|^2} \equiv \max_i \norm{{\mathbf{u}_{\xi}}_i}_2.
\end{equation}
Here,  ${\textbf{u}_{\xi}}_i \equiv \begin{bmatrix}
    {u_\xi}_i & {v_\xi}_i & {w_\xi}_i
\end{bmatrix}^T$. By the definition of the $\mathscr{H}_\infty$ norm (Eq.~\eqref{eq:2.10}) and using $\mathbf{U}_{\Xi}$ to approximate the continuous matrix operator $\mathbf{u}_\Xi$, we obtain the following result:
\begin{equation}
    \label{eq:A7}
    \norm{\mathbf{U}_{\Xi}}_{\infty} = \max_i \norm{{\mathbf{u}_{\xi}}_i}_2 \approx \max_{y\in \mathcal{Y}} \norm{\mathbf{u}(y)}_2.
\end{equation}
Here, there is no supremum over temporal frequencies since $\mathbf{U}_{\Xi}$ is constant in the frequency domain; $\mathcal{Y}$ is the relevant domain in the wall-normal direction corresponding to the flow system (for channel flows, $\mathcal{Y}=[-1,1]$; for boundary layer flows, $\mathcal{Y}=[0,\infty)$);
{$\max_{y\in\mathcal{Y}}\norm{\mathbf{u}(y)}_2$} denotes the largest velocity perturbation magnitude that is present in the flow, which we approximate by the maximum velocity perturbation in the discretised domain. 


\section{Comparison of structured approaches for $k_z=0$.}
\label{sec:App_numeric}

\begin{figure}
    \centering
     \begin{overpic}[width=1\linewidth]{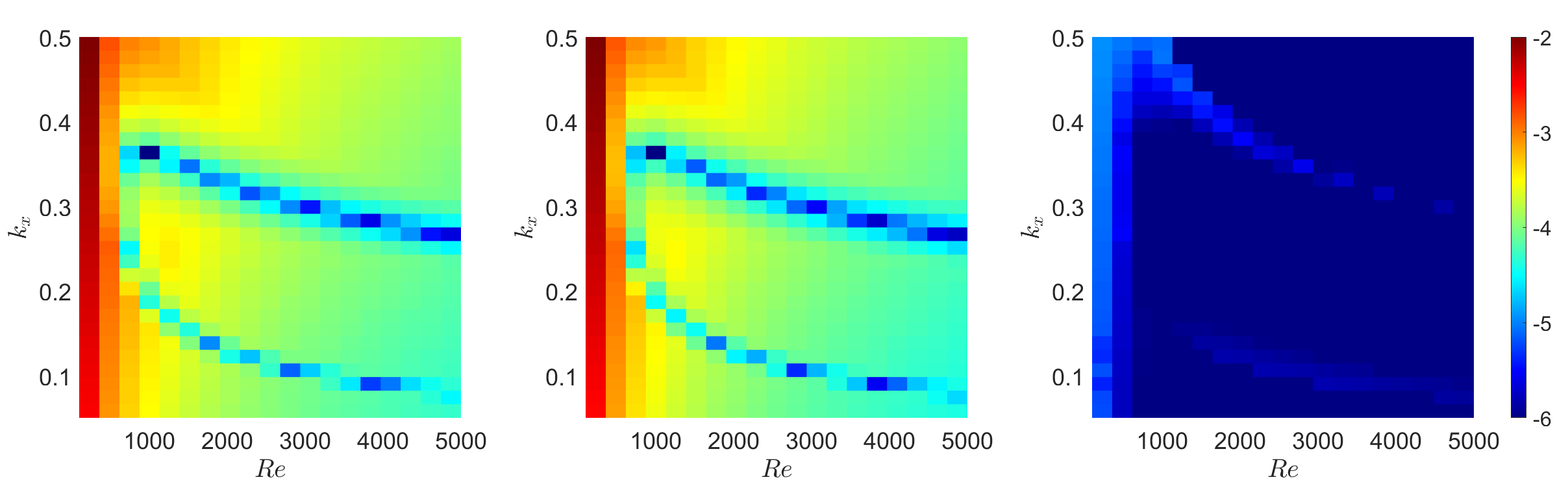}
     \put(-1, 28){\small (a) \hspace{3.8cm} (b) \hspace{3.8cm} (c)}
 \end{overpic}
    \caption{A comparison between the structured analysis methods for Blasius base flow with $k_z=0$.  Colormaps of: (a) $\norm{\mathscr{H}_{\nabla}}_{\mu_{\mathbf{\Delta}_{nr}}}^{-1}$, (b) $\norm{\mathscr{H}_{\nabla}}_{\mu_{\mathbf{\Delta}_{r}}}^{-1}$, and (c) the difference between $\norm{\mathscr{H}_{\nabla}}_{\mu_{\mathbf{\Delta}_{r}}}^{-1}$ and  $\norm{\mathscr{H}_{\nabla}}_{\mu_{\mathbf{\Delta}_{nr}}}^{-1}$. All colormaps are presented on a logarithmic scale.}
    \label{fig:16}
\end{figure}

Here, we compare results of $\norm{\mathscr{H}_{\nabla}}_{\mu_{\mathbf{\Delta}_{nr}}}$ (non-repeated blocks) and $\norm{\mathscr{H}_{\nabla}}_{\mu_{\mathbf{\Delta}_{r}}}$ (repeated blocks) for TS waves with a spanwise wavenumber of $k_z=0$. This comparison is made to justify our decision to consider only the non-repeated structured approach in our analysis for two-dimensional modes, presented in~\S\ref{sec:5}. In Fig.~\ref{fig:16}, we present colormap plots of $\norm{\mathscr{H}_{\nabla}}_{\mu_{\mathbf{\Delta}_{nr}}}^{-1}$, $\norm{\mathscr{H}_{\nabla}}_{\mu_{\mathbf{\Delta}_{r}}}^{-1}$, and the difference $(\norm{\mathscr{H}_{\nabla}}_{\mu_{\mathbf{\Delta}_{r}}}^{-1}-\norm{\mathscr{H}_{\nabla}}_{\mu_{\mathbf{\Delta}_{nr}}}^{-1}$, for Blasius base flow and for different $k_x$ and $Re$ values. This computation was made on a low-resolution grid, since it is very numerically demanding to obtain results using the repeated blocks structure. Thus, we consider $29$ values of $k_x$ in the domain $k_x \in [0.05,0.5]$ and $20$ Reynolds number values in the domain $Re\in [100,5000]$.

Fig.~\ref{fig:16}(a) and Fig.~\ref{fig:16}(b) show very similar qualitative behavior of $\norm{\mathscr{H}_{\nabla}}_{\mu_{\mathbf{\Delta}_{nr}}}^{-1}$ and $\norm{\mathscr{H}_{\nabla}}_{\mu_{\mathbf{\Delta}_{r}}}^{-1}$ producing similar thresholds on velocity perturbations across the entire computational domain. The difference in the results between the two structured approaches, which is shown in Fig.~\ref{fig:16}(c), is very small across the entire computational domain, with a maximal value of $1.16\times 10^{-5}$. This difference is significantly smaller than the threshold predicted by either method for almost all of the computational domain (apart from very close to the neutral stability curve predicted by LST) and thus has no significant effect on any of the results presented in~\S\ref{sec:5}.

\bibliographystyle{jfm}
\bibliography{jfm}

\end{document}